\documentclass[shortAfour,times,11pt]{sagej}      
\usepackage[utf8]{inputenc}
\usepackage{moreverb,url}
\usepackage{subcaption}
\usepackage{threeparttable}
\usepackage{makecell}
\usepackage{multirow}
\usepackage{multicol}
\usepackage{ragged2e}
\usepackage{stfloats}
\usepackage{float}
\usepackage{cite}
\usepackage{csquotes}
\makeatletter
\newcommand*{\rom}[1]{\expandafter\@slowromancap\romannumeral #1@}
\makeatother
\usepackage{setspace}
\onecolumn
\usepackage{tabu}
\usepackage{bbm}
\usepackage{pifont}
\DeclareMathAlphabet{\msfsl}{T1}{cmvtt}{m}{n}
\usepackage{tabularx}

\newcolumntype{Y}{>{\centering\arraybackslash}X}

\newcommand{\RN}[1]{%
  \textup{\uppercase\expandafter{\romannumeral#1}}%
}
\usepackage[colorlinks,bookmarksopen,bookmarksnumbered,citecolor=red,urlcolor=red]{hyperref}
\newcommand\BibTeX{{\rmfamily B\kern-.05em \textsc{i\kern-.025em b}\kern-.08em
T\kern-.1667em\lower.7ex\hbox{E}\kern-.125emX}}

\begin{document}
\runninghead{Wang et al.}

\title{{Bridging simulation and reality in subsurface radar-based sensing: physics-guided hierarchical domain adaptation with deep adversarial learning}}

\author{Zixin Wang\affilnum{1}; Ishfaq Aziz\affilnum{2}; Mohamad Alipour\affilnum{3}}

\affiliation{\affilnum{1}Postdoctoral Research Associate, Ph.D., Department of Civil and Environmental Engineering, University of Illinois Urbana-Champaign, Urbana, IL, 61801, USA (corresponding author). Email: zixinw@illinois.edu\\
\affilnum{2}Ph.D. Candidate, Department of Civil and Environmental Engineering, University of Illinois Urbana-Champaign, Urbana, IL, 61801, USA. Email: ishfaqa2@illinois.edu\\
\affilnum{3}Research Assistant Professor, Ph.D., Department of Civil and Environmental Engineering, University of Illinois Urbana-Champaign, Urbana, IL, 61801, USA. Email: alipour@illinois.edu\\}

\maketitle

\noindent \textbf {\fontsize{11}{12}\fontfamily{qhv}\selectfont Abstract}\\
\noindent Accurate estimation of subsurface material properties, such as soil moisture, is critical for wildfire risk assessment and precision agriculture. Ground-penetrating radar (GPR) is a non-destructive geophysical technique widely used to characterize subsurface conditions. Data-driven parameter estimation methods typically require large amounts of labeled training data, which is expensive to obtain from real-world GPR scans under diverse subsurface conditions. A physics-based GPR model using the finite-difference time-domain (FDTD) method can be employed to generate large synthetic datasets through simulations across varying material parameters, which are then utilized to train data-driven models. A key limitation, however, is that simulated data (source domain) and real-world data (target domain) often follow different distributions, which can cause data-driven models trained on simulations to underperform in real-world scenarios. To address this challenge, this study proposes a novel physics-guided hierarchical domain adaptation framework with deep adversarial learning for robust subsurface material property estimation from GPR signals. The proposed framework is systematically evaluated through the laboratory tests for single- and two-layer materials, as well as the field tests for single- and two-layer materials, and is benchmarked against state-of-the-art methods, including the one-dimensional convolutional neural network (1D CNN) and domain adversarial neural network (DANN). The results demonstrate that the proposed framework achieves higher correlation coefficients R and lower Bias between the predicted and measured parameter values, along with smaller standard deviations in the estimations, thereby validating their effectiveness in bridging the domain gap between simulated and real-world radar signals and enabling efficient subsurface material property retrieval.

\vspace{4mm}

\noindent \textbf {\fontsize{11}{12}\fontfamily{qhv}\selectfont Keywords}\\
Domain adaptation; Sim-to-real; Physics-guided machine learning; Ground-penetrating radar; Material property estimation; Soil moisture

\section{Introduction} \label{Introduction}
In nature, soil is often overlaid by a top layer of vegetation or organic matter, known as biomass. Determining soil moisture is critical in precision agriculture \cite{Liu_2016}, water resource management \cite{Srivastava_2017}, identification of wildfire hotspots \cite{Ambadan_2020}, and assessing the likelihood and size of wildfires \cite{Jensen_2018, Krueger_2015}. The top biomass layer directly acts as wildfire fuel, exerting direct control over ignitability and spread \cite{Rao_2020}. Moreover, the biomass amount, characterized by depth or volume, is necessary for simulating wildland fires \cite{Alipour_2023}. Hence, estimating properties of multiple layers formed by soil and the overlaying biomass layer carries significant importance. Traditional methods of estimating soil and biomass moisture include labor-intensive gravimetric tests or in-situ sensors \cite{Fragkos_2024,SU_2014,Bourgeau‐Chavez_2010}, which are time-consuming and provide only point measurements. Moreover, in the presence of a top vegetation layer or biomass, the deployment of in-situ sensors poses additional challenges. Satellite or passive remote sensing overcomes some of these challenges by providing remote measurement, but suffers from low resolution \cite{Li_2021_soil,Kim_2019}.
 
In contrast, radar-based sensing provides high-resolution measurements, and is used across a wide range of applications, including soil and biomass moisture estimation for agriculture and wildfire assessment \cite{Pathirana_2024,Aziz_2024,Wu_2022,Wu_2019,Sinchi_2023}, permafrost characterization \cite{Kang_2025}, nondestructive testing of structures \cite{Hong_2024}, building material evaluation \cite{Alam_2024,Kaplanvural_2023}, and pavement monitoring \cite{abufares_2025asphalt}. In these applications, ground-penetrating radar (GPR) scans are often used to image the subsurface or estimate the dielectric properties of subsurface materials, such as dielectric permittivity and electrical conductivity \cite{Wu_2019,Aziz_2024_RSE,Kaplanvural_2023}. 

Automated estimation of these dielectric properties from radar scans is often conducted using waveform inversion or data-driven methods \cite{Sinchi_2023,Feng_2019,Haruzi_2022,Feng_2023,Patsia_2023,Liu_2024,Xue_2024,Zhang_2025,Aziz_2024_RSE}. Waveform inversion is the process of estimating the subsurface material properties from recorded radar signals. It involves global optimization or model updating to iteratively update and optimize material properties so that the simulated response matches the recorded radar response in the real world. This often requires hundreds or thousands of forward simulations, which is computationally expensive, even for estimating multiple material properties for a single scenario. On the other hand, data-driven methods, such as deep learning, overcome this limitation by yielding fast inference and retrieval of material properties. However, deep learning methods require a large amount of labeled real-world data for adequate training and accurate prediction, which is often costly and impractical to obtain. In radar-based subsurface sensing, obtaining ground truth data typically requires collecting soil samples for gravimetric analysis and installing through-depth sensors within the soil, which is labor-intensive and time-consuming. Synthetic data generated by physics-based simulations offers the advantages of being easily obtainable and capable of representing the physics of the phenomenon. When the training data is derived from a physics-based model, the resulting framework can be regarded as a physics-guided machine learning paradigm \cite{Faroughi_2024}. However, simulated data cannot fully capture the complex factors that influence the actual system response, resulting in a domain discrepancy between the simulated data (source domain) and real-world data (target domain). Consequently, if the data-driven model is trained on the simulated data, it may not perform well on the real-world data due to the domain shift problem. 

To mitigate the domain shift inherent in sim-to-real transfer problems, numerous techniques have been proposed to minimize the discrepancy between simulated and real-world domains. Model updating is a widely adopted approach that calibrates the parameters of a simulation model so that its dynamics closely match those observed in real systems \cite{Li_2025}. However, model updating alone cannot fully eliminate the domain gap and often serves as a crucial pre-processing step before applying domain adaptation or domain randomization techniques for sim-to-real transfer \cite{Wiberg_2024}. Domain randomization, as another sim-to-real transfer technique, enhances the diversity of simulated data by randomizing key simulation parameters, thereby enabling the trained model to generalize more effectively to real-world variations \cite{Tobin_2017}. Nevertheless, this approach comes at the cost of an extensive number of training samples to learn a task, and excessive randomization across multiple parameters can significantly increase task complexity, making it challenging for the model to converge to optimal solutions \cite{Josifovski_2022}. 

In contrast, rather than relying solely on model calibration or extensive parameter randomization, domain adaptation, a transfer learning technique, leverages shared representations between domains to transfer knowledge learned from a source domain for performing inference tasks in a different but related target domain, thereby providing a promising solution to the sim-to-real problem. It has been widely applied across various sim-to-real tasks, including structural health monitoring \cite{Gardner_2020,Wang_2025_JEM}, machinery fault diagnosis \cite{Lou_2022}, image-based condition assessment \cite{alipour2020big}, semantic segmentation \cite{Benigmim_2023}, object detection \cite{DeBortoli_2021}, robotic manipulation \cite{Bousmalis_2018}, autonomous navigation \cite{Hu_2022}, deformation detection \cite{Sol_2024}, and photoacoustic tomography \cite{Wang_2025}, among others.

Depending on how shared representations between domains are established, domain adaptation techniques can be categorized into three main approaches: feature-based, instance-based, and parameter-based approaches. Feature-based approaches aim to learn a common feature representation across domains such that the feature distributions of the source and target domains become closely aligned. For example, transfer component analysis (TCA) learns transfer components across domains in a reproducing kernel Hilbert space, where data properties are preserved and data distributions from different domains are brought closer together \cite{Pan_2011}. The domain adversarial neural network (DANN) employs a feature extractor and a domain discriminator trained in an adversarial manner to learn domain-invariant feature representations \cite{Ganin_2016}. The maximum mean discrepancy (MMD) has been widely used as a measure of distributional discrepancy between source and target domains, and an invariant feature representation across domains can be obtained by minimizing the MMD \cite{Yan_2017}. Instance-based approaches aim to identify source-domain samples that are most relevant to the target domain and enhance their contribution by assigning them higher importance weights during domain adaptation \cite{Zhang_2018,Cao_2018}. Parameter-based approaches aim to adapt the parameters of a deep neural network trained on substantial source-domain data by fine-tuning the weights of the upper layers with a limited amount of target-domain data \cite{Yosinski_2014,Chen_2020}. Depending on the availability of labeled target-domain data, domain adaptation techniques can be categorized into unsupervised domain adaptation, which does not require labeled target-domain data \cite{Ganin_2016}, and supervised domain adaptation, which utilizes a limited amount of labeled target-domain data \cite{Motiian_2017,Singh_2021}. Overall, feature-based unsupervised domain adaptation is a particularly attractive technique, as it eliminates the need for labeled target-domain data while enabling the learning of domain-invariant representations that improve generalization to the target domain.

Recently, several studies have employed parameter-based domain adaptation to reduce the reliance on large labeled datasets for training deep learning models in GPR-related applications. For instance, Bralich \textit{et al.} \cite{Bralich_2017} and Reichman \textit{et al.} \cite{Reichman_2017} pre-trained convolutional neural network (CNN) models for buried target detection using the Canadian Institute For Advanced Research (CIFAR-10) dataset and subsequently fine-tuned them with a small GPR dataset. Enver and Yüksel \cite{Aydin_2019} employed a CNN model pre-trained on the ImageNet Large Scale Visual Recognition Challenge (ILSVRC 2012) dataset and transferred its learned weights to adapt the model for buried wire detection with GPR data. However, transfer learning during the pre-training phase offers limited benefit, as the fine-tuning phase remains prone to overfitting due to the scarcity of training samples \cite{Tong_2020}. Moreover, a few studies have applied feature-based domain adaptation to align GPR features between the source and target domains, but these efforts have been limited to single material parameter estimation \cite{Imai_2025}, few-shot learning \cite{Li_2026}, and classification tasks \cite{Huang_2023,Oturak_2022}. Therefore, limited investigation has been conducted on subsurface material parameter retrieval using unsupervised domain adaptation and GPR signals. While DANN has been primarily applied to classification tasks \cite{Ganin_2016}, its effectiveness in regression problems, which are inherently more challenging, remains to be further investigated.

In material parameter estimation using radar signals, the finite-difference time-domain (FDTD) method can be utilized to generate synthetic radar scans by solving Maxwell's equations numerically. A data-driven model can then be trained in a supervised manner using the simulated data. However, even after calibration of the FDTD model, the simulated radar signals (source domain) may still differ from the measured radar signals (target domain). The main factors contributing to this domain gap are summarized as follows, and the proposed physics-guided hierarchical domain adaptation approach is expected to bridge these discrepancies between the two domains:

\begin{itemize}
\item \textbf{Model assumptions}: The FDTD model assumes a waveform type (e.g., Gaussian) that is to be transmitted by the radar, whereas the true waveform of a real GPR is unknown and often not sufficiently specified by manufacturers, and thus may deviate from the assumptions.
\item \textbf{Model simplification}: The 2D version of the FDTD model is adopted to reduce computational costs; however, the real-world experimental setup is inherently three-dimensional. Numerical discretization errors and convergence limitations in the FDTD method can also introduce inaccuracies in the simulated radar scans.
\item \textbf{Boundary conditions}: The 2D FDTD model cannot capture reflections from the side barriers of the material container, which are present in the experimental setup, but may not be adequately represented in simulations.
\item \textbf{Material uncertainties}: In the real world, material layers are typically not perfectly homogeneous, unlike the FDTD model, which assumes homogeneity. Furthermore, the FDTD model cannot account for irregular or non-smooth material interfaces, which are present in actual soil and vegetation layers.
\end{itemize}

This study presents a novel sim-to-real domain adaptation framework that leverages physics-guided machine learning, hierarchical domain adaptation, and deep adversarial learning to achieve robust estimation of subsurface material properties, including permittivity, conductivity, and depth, from GPR signals. To this end, five innovative approaches are proposed: hierarchical DANN (HierDANN), physics-guided DANN-1 (PhyDANN-1), hierarchical physics-guided DANN-1 (HierPhyDANN-1), physics-guided DANN-2 (PhyDANN-2), and hierarchical physics-guided DANN-2 (HierPhyDANN-2). To incorporate physical constraints into the adversarial learning process, both source and target signals are reconstructed using the latent features generated by the feature extractor. This signal reconstruction helps preserve the essential physics-based features embedded in the radar signals that are critical for accurate material property estimation. Instead of performing an all-in-one material parameter retrieval, the hierarchical domain adaptation estimates material parameters sequentially by leveraging previously estimated parameters to predict subsequent ones, thereby reducing the task complexity and enhancing estimation performance. The order of parameters to be estimated in the hierarchical domain adaptation is determined using variance-based sensitivity analysis (i.e., the Sobol’s method). The proposed approaches were systematically compared with state-of-the-art methods, including the 1D CNN trained solely on simulated data and the DANN that predicts multiple parameters within a single network, under various test conditions encompassing single- and two-layer laboratory tests as well as single- and two-layer field tests. The proposed approaches outperform the baseline approaches by achieving higher $R$ values, lower Bias between the estimated and measured parameter values, and smaller standard deviations of the estimations, while accurately capturing soil moisture variations before and after rainfall events. It has been shown that approaches incorporating hierarchical domain adaptation achieve superior performance in material property estimation. This capability highlights its potential as a powerful AI technique for subsurface parameter retrieval, crucial for wildfire risk assessment and precision agriculture. Additionally, compared to the iterative model updating approaches previously used for subsurface material property retrieval \cite{Aziz_2024_RSE}, the proposed data-driven method offers significantly faster inference.

The remainder of this paper is organized as follows. Section~\ref{Methodology} provides an overview of the proposed physics-guided hierarchical domain adaptation framework. Section~\ref{illustrative_example_setups} presents the setups of the illustrative examples conducted in this study, including the single-layer material (soil) and the two-layer material (wood shavings over soil) in laboratory tests, as well as the single-layer material (soil) and the two-layer materials (leaves or wood chips over soil) in field tests. Section~\ref{Results_and_Discussion} presents and discusses the results obtained from the illustrative examples. The Pearson correlation coefficient $R$, Bias, root mean squared error (RMSE), and unbiased RMSE (ubRMSE) between the estimated and measured parameter values, the standard deviation of the estimations, as well as the inference time, are compared and evaluated across the proposed approaches and the baseline methods. The capabilities and limitations of the proposed approach are discussed. Concluding remarks and future work are presented in Section~\ref{Conclusions}.

\section{Methodology} \label{Methodology}
\subsection{Overview}
\begin{figure*}[ht]
\centering
\includegraphics[width=1\linewidth]{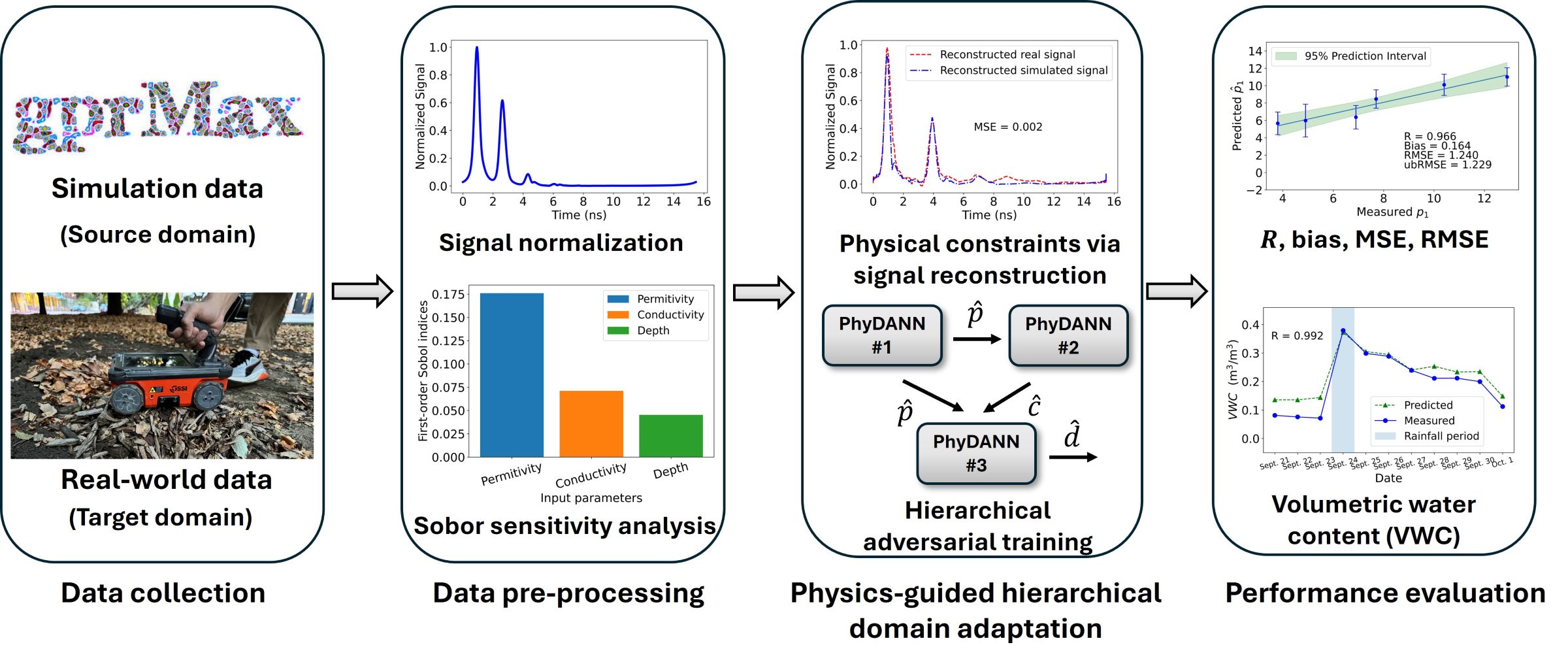}
\caption{\centering Overview of the proposed physics-guided hierarchical domain adaptation framework.}
\label{Fig_1}
\end{figure*}

Radar signals are employed to estimate subsurface material properties. To generate synthetic data, the FDTD method was employed to simulate the propagation of electromagnetic waves in multi-layered material media. In addition, a smaller set of real-world signals (target domain) is collected using a handheld GPR. To enhance consistency and improve model performance, both simulated and real-world signals, along with the simulated data labels, are normalized to the range [0, 1]. The Sobol sensitivity analysis is then applied to the simulated radar signals to determine the relative importance of material parameters, including permittivity, conductivity, and depth. Guided by this importance ranking, multiple DANNs or PhyDANNs are trained in a hierarchical manner, where the parameter values predicted by previous DANNs or PhyDANNs are used to train subsequent ones. Physical constraints are integrated into the adversarial learning framework by reconstructing radar signals from the domain-invariant features produced by the feature extractor, thereby preserving the essential physics-based features embedded in the signals that are required for accurate material property estimation. The effectiveness of the proposed approaches is evaluated using various metrics, including the correlation coefficient $R$, Bias, RMSE, and ubRMSE between the estimated and measured parameter values, and compared with those obtained by the state-of-the-art methods. Field tests are further conducted to validate the approach by tracking variations in soil moisture (volumetric water content) induced by rainfall. The overview of the proposed physics-guided hierarchical domain adaptation framework is shown in Figure \ref{Fig_1}.

\subsection{Numerical Modeling and Model Calibration}
A large dataset of simulated signals with varying parameter values (source domain) is generated using gprMax  \cite{Warren_2016}, which models electromagnetic wave propagation by solving Maxwell’s equations using the FDTD method. Two types of parameters need to be defined in the FDTD simulation: material parameters and radar parameters. Internal radar parameters that are intrinsic to the sensing system include the distance between the transmitting and receiving radar antennas (i.e., bistatic separation), the shape of the transmitted waveform, and the frequency of the waveform, among others.  Material parameters refer to extrinsic properties of the subsurface materials under investigation, such as permittivity, conductivity, and depth. Training data were generated by performing simulations with varying values of the extrinsic material parameters. However, the authors have previously shown that prior to conducting these simulations, the intrinsic parameters needed to be calibrated to ensure the FDTD simulations follow and replicate the experimental GPR signals generated by the specific radar being used \cite{Aziz_2025}. We performed systematic calibration to determine the optimal values of the intrinsic radar parameters corresponding to the real GPR (GSSI StructureScan miniXT) used in this study. The process, described as intrinsic radar calibration, is detailed in a previous study by the authors \cite{Aziz_2025}. All subsequent simulations for training data generation were conducted using these optimal or calibrated radar parameters.

\subsection{Data Pre-processing}
\begin{figure*}[ht]
     \centering
     \begin{subfigure}[t]{0.34\textwidth}
         \centering
         \includegraphics[width=1\linewidth]{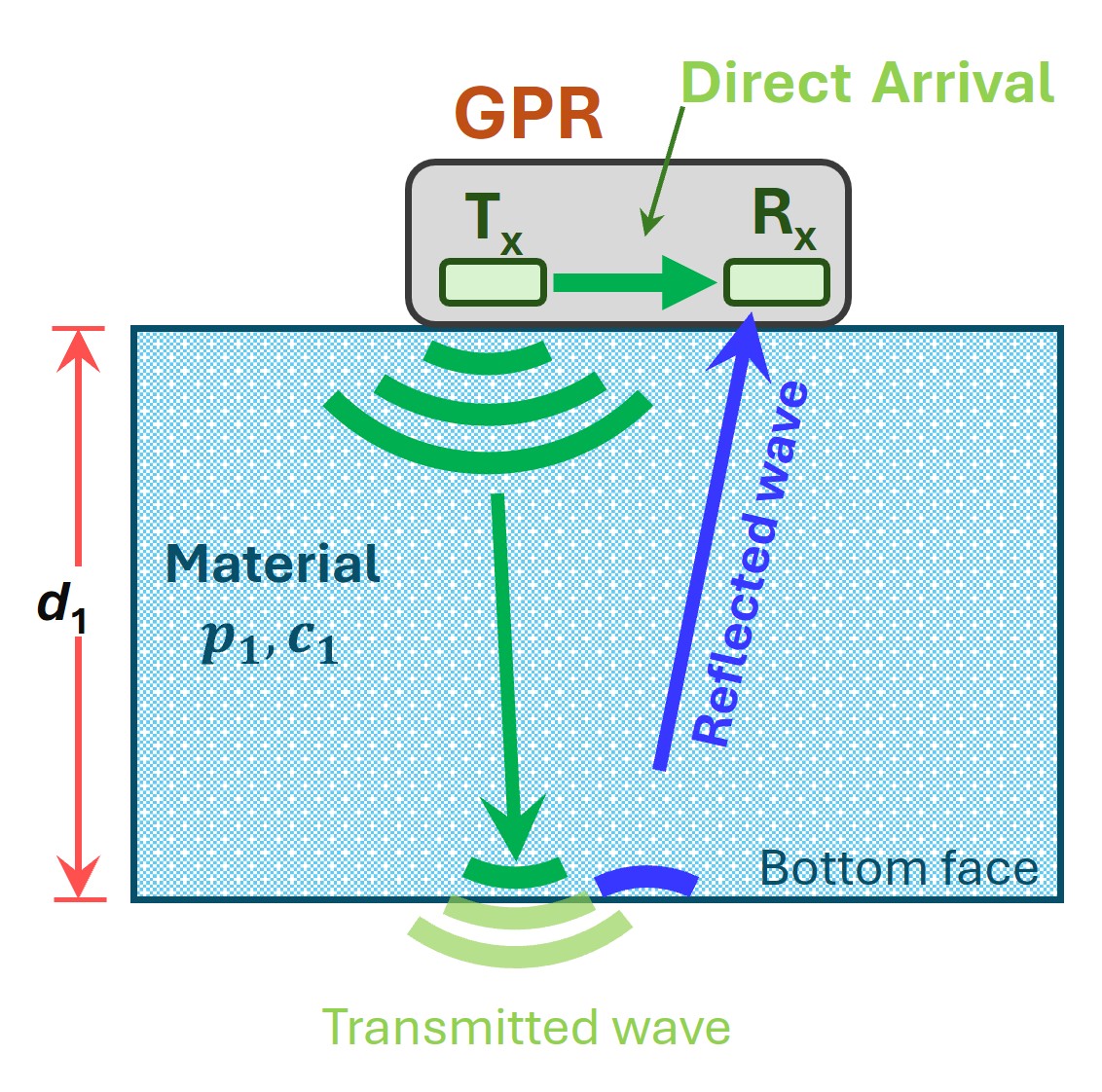}
         \captionsetup{justification=centering}
         \caption{GPR scan on a single-layer material.}
         \label{Fig_21a}
     \end{subfigure}
     \begin{subfigure}[t]{0.28\textwidth}
         \centering
         \includegraphics[width=1\linewidth]{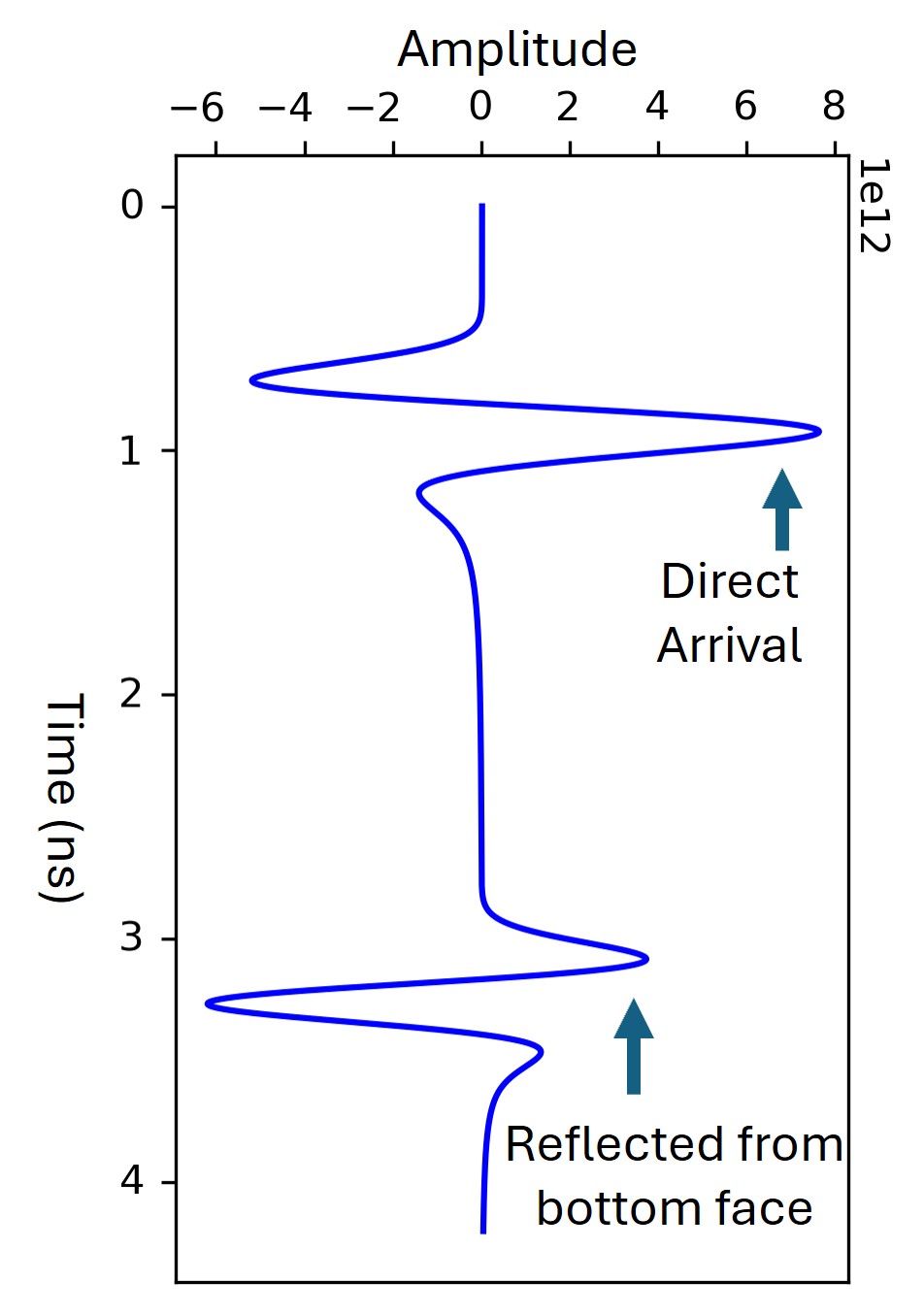}
         \captionsetup{justification=centering}
         \caption{Received A-scan.}
         \label{Fig_21b}
     \end{subfigure}
     \begin{subfigure}[t]{0.23\textwidth}
         \centering
         \includegraphics[width=1\linewidth]{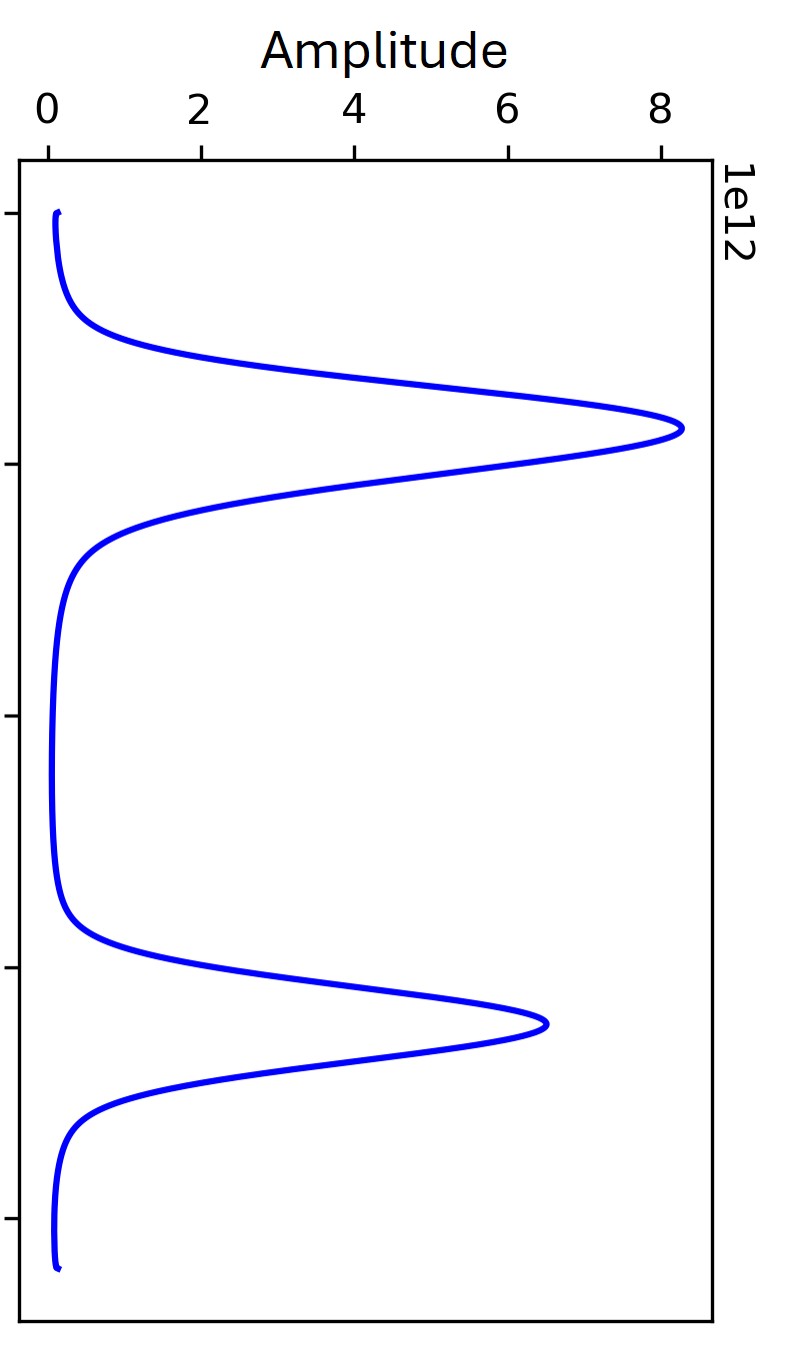}
         \captionsetup{justification=centering}
         \caption{Amplitude envelope.}
         \label{Fig_21c}
     \end{subfigure}
     \caption{\centering GPR A-scan and amplitude envelope on a single-layer material.}
     \label{Fig_21}
\end{figure*}

A GPR antenna transmits an electromagnetic wave via a transmitter and receives the reflected signal through a receiver (Figure \ref{Fig_21a}). The received waveform is a one-dimensional signal referred to as an A-scan (Figure \ref{Fig_21b}). The amplitude envelope (Figure \ref{Fig_21c}) of the received A-scan can be obtained from the magnitude of its analytic signal, as expressed in Equation (\ref{Eq_1}):

\begin{equation} \label{Eq_1}
\textbf{\textit{x}}_{env} = |F^{-1}\left(F(\textbf{\textit{x}}_A\right)2U|,
\end{equation}

\noindent where $\textbf{\textit{x}}_{env}$ is the amplitude envelope; $\textbf{\textit{x}}_A$ is the received A-scan; $F$ is the Fourier transform; and $U$ is the unit step function. $\textbf{\textit{x}}_{env}$ spans a wide range of values. However, excessively large or small input values to the neural network may lead to features with larger scales dominating the model and slower training convergence \cite{Kim_2025, Singh_2020}. Therefore, to improve model performance and enhance training stability, $\textbf{\textit{x}}^{\prime}_{env}$ is normalized to the range [0, 1] using Equation (\ref{Eq_2}), and the normalized data are used as the model input:

\begin{equation} \label{Eq_2}
\textbf{\textit{x}}^{\prime}_{env} = \frac{\textbf{\textit{x}}_{env}}{\text{max}\left(\textbf{\textit{x}}_{env}\right)}.
\end{equation}

The labels corresponding to each radar signal include the permittivity, conductivity, and depth of the subsurface material, with each parameter exhibiting a distinct range of values. The model may allocate more effort to learning targets with larger magnitudes. To ensure that all targets are treated with equal importance, the labels for each material parameter are also normalized to the range [0, 1] using min-max normalization as follows:

\begin{equation} \label{Eq_3}
\textbf{\textit{y}}^{\prime}_{i} = \frac{\textbf{\textit{y}}_{i} - \text{min}\left(\textbf{\textit{y}}_{i}\right)}{\text{max}\left(\textbf{\textit{y}}_{i}\right) - \text{min}\left(\textbf{\textit{y}}_{i}\right)},
\end{equation}

\noindent where $\textbf{\textit{y}}_{i}$ represents the labels associated with one of the material parameters.

\subsection{1D Convolutional Neural Network} \label{1DCNN}
\begin{figure*}[ht]
\centering
\includegraphics[width=0.85\linewidth]{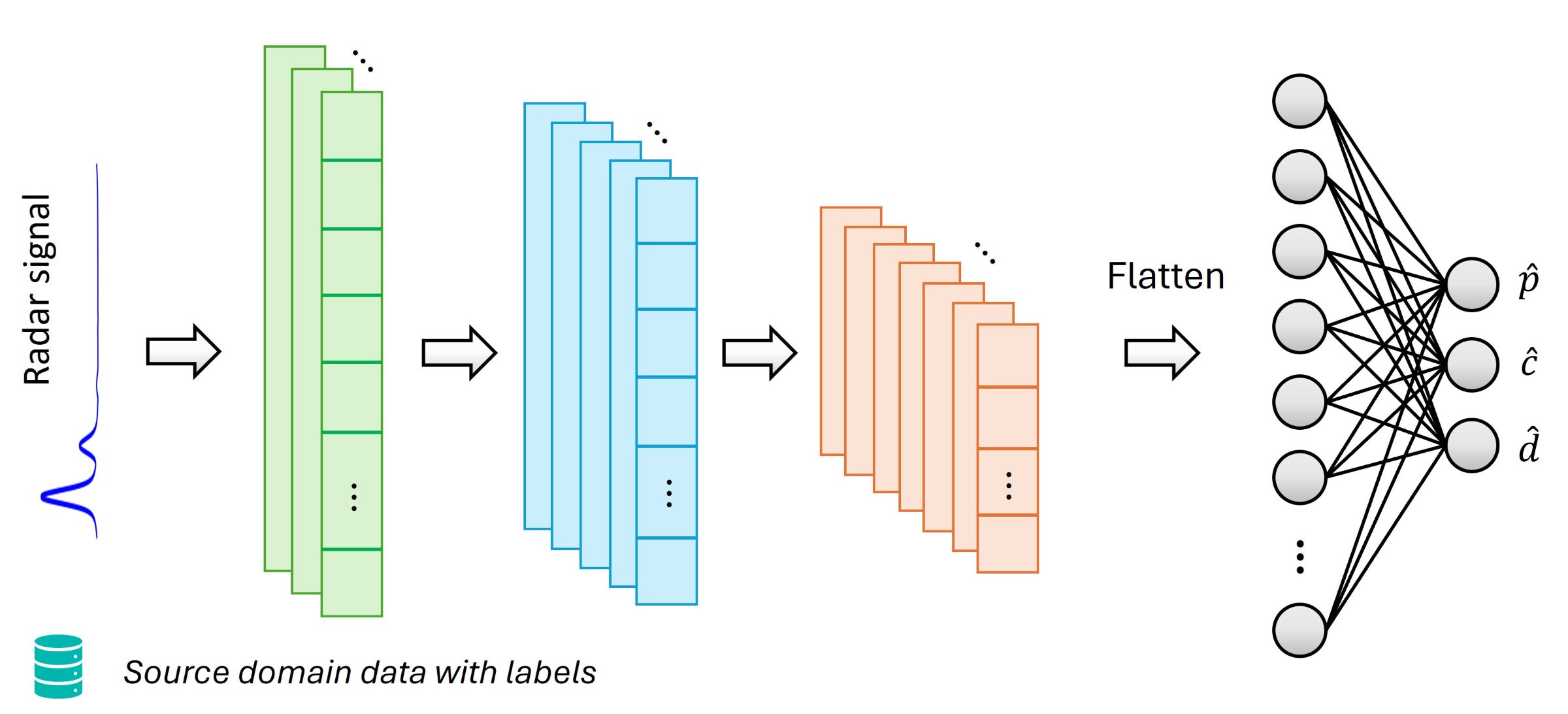}
\caption{\centering A schematic diagram of the 1D CNN.}
\label{Fig_2}
\end{figure*}

Supervised learning aims to derive a function that maps inputs to desired outputs by learning from labeled data. CNNs, a widely used deep learning architecture, are particularly effective for processing data with a known grid-like topology. The 2D CNN has been widely used in computer vision, as it is designed to process two-dimensional input data such as images and video frames \cite{Alzubaidi_2021}. The 1D CNN, on the other hand, specializes in processing 1D signals \cite{Kiranyaz_2021}. Previous studies have applied 1D CNNs for soil moisture estimation using GPR signals \cite{Vahidi_2025}. In this study, a supervised 1D CNN is employed as a baseline approach for comparison with the proposed method. The forward propagation of each layer in a 1D CNN can be expressed as follows:

\begin{equation} \label{Eq_4}
y^{l}_{k}=f\left(b^{l}_{k}+\sum^{N_{l-1}}_{i=1}w^{l-1}_{ik} \star s^{l-1}_{i} \right),
\end{equation}

\noindent where $\star$ denotes the valid cross-correlation operator. $y^{l}_{k}$ is the  intermediate output of the $k^{th}$ channel at the $l^{th}$ layer. $b^{l}_{k}$ is the bias of the $k^{th}$ channel at the $l^{th}$ layer. $w^{l-1}_{ik}$ is the kernel from the $i^{th}$ channel at the $(l-1)^{th}$ layer to the $k^{th}$ channel at the $l^{th}$ layer. $s^{l-1}_{i}$ is output of the $i^{th}$ channel at the $(l-1)^{th}$ layer. $N_{l-1}$ is the number of channels at the $(l-1)^{th}$ layer. $f(\cdot)$ is the activation function. The length of the output from a 1D convolutional layer, referred to as the feature map, is determined by the sizes of the input, kernel, padding, and stride, as given by Equation (\ref{Eq_5}):

\begin{equation} \label{Eq_5}
F_s = \frac{I_s-K_s+2P_s}{S_s}+1,
\end{equation}

\noindent where $F_s$, $I_s$, $K_s$, $P_s$, $S_s$ denote the sizes of the output, input, kernel, padding, and stride, respectively. As the number of layers increases, the length of the feature map decreases. To account for this change, the kernel size and stride are reduced accordingly. In contrast, the depth of the feature map increases with the number of kernels. A fully connected layer, in which each neuron is connected to all neurons in the preceding layer, follows the convolutional layers to integrate the extracted features and perform material parameter estimation. A schematic diagram of the 1D CNN is shown in Figure \ref{Fig_2}. The detailed architecture of the 1D CNN is summarized in Table \ref{T1}.

\begin{table*}[ht]
     \small\sf\centering
     \caption{\centering Architecture of 1D CNN.\label{T1}}
     \begin{tabular}{ccccccc}
         \toprule
         Layer & \makecell{Kernel \\ number} & \makecell{Kernel \\ size} & \makecell{Stride \\ size} & \makecell{Input \\ size} & \makecell{Output \\ size} & Activation \\
         \cmidrule{1-7}
         Convolution & $32$ & $5$ & $5$ & $1 \times 6560$ & $32 \times 1312$ & LeakyReLU \\
         Convolution & $64$ & $5$ & $5$ & $32 \times 1312$ & $64 \times 262$ & LeakyReLU \\
         Convolution & $128$ & $4$ & $4$ & $64 \times 262$ & $128 \times 65$ & LeakyReLU \\
         Convolution & $256$ & $4$ & $4$ & $128 \times 65$ & $256 \times 16$ & LeakyReLU \\
         Convolution & $512$ & $3$ & $3$ & $256 \times 16$ & $512 \times 5$ & LeakyReLU \\
         Convolution & $10240$ & $3$ & $3$ & $512 \times 5$ & $10240 \times 1$ & LeakyReLU \\
         Flatten & --- & --- & --- & $10240 \times 1$ & $10240$ & ---\\
         Full connection & --- & --- & --- & $10240$ & $M$ & Linear \\
         \bottomrule
     \end{tabular}
\end{table*}

The 1D CNN is trained using labeled simulated data (i.e., source domain data) and evaluated on real-world data (i.e., target domain data). The training process is conducted in a supervised manner by minimizing the mean squared error (MSE) between the predicted and true values. The loss function is expressed as follows:

\begin{equation} \label{Eq_6}
\mathcal{L}_{1D-CNN} = \frac{1}{NM} \sum_{j=1}^{N} \sum_{i=1}^{M} \left(y_{i}^{(j)} - \hat{y}_{i}^{(j)}\right)^2,
\end{equation}

\noindent where $N$ is the number of samples, and $M$ is the number of material parameters to be estimated. $y$ is the true value, and $\hat{y}$ is the predicted value. 

\subsection{Domain Adversarial Neural Network} \label{DANN}
Domain adaptation, as a subcategory of transfer learning, allows the learned knowledge from a source domain to be transferred to a different but related target domain. Unsupervised domain adaptation learns a mapping between domains by utilizing labeled data from the source domain and unlabeled data from the target domain. The source domain $\mathcal{D}_s = \{\mathcal{X}_s,\mathcal{Y}_s\}$ is composed of the sample space $\mathcal{X}_s$ and the corresponding label space $\mathcal{Y}_s$. The marginal and joint probability distributions of the source domain are denoted as $p(X_s)$ and $p(X_s,Y_s)$, respectively, where $X_s = \left\{\textbf{\textit{x}}^{(j)}_{s}\right\}_{j=1}^{N_s} \in \mathcal{X}_s$, $Y_s = \left\{\textit{y}^{(j)}_{s}\right\}_{j=1}^{N_s} \in \mathcal{Y}_s$, with $N_s$ representing the number of labeled source samples. Similarly, the target domain $\mathcal{D}_t = \{\mathcal{X}_t,\mathcal{Y}_t\}$ is composed of the sample space $\mathcal{X}_t$ and the corresponding label space $\mathcal{Y}_t$. The marginal and joint probability distributions of the target domain are denoted as $p(X_t)$ and $p(X_t,Y_t)$, respectively, where $X_t = \left\{\textbf{\textit{x}}^{(j)}_{t}\right\}_{j=1}^{N_t} \in \mathcal{X}_t$, $Y_t = \left\{\textit{y}^{(j)}_{t}\right\}_{j=1}^{N_t} \in \mathcal{Y}_t$, with $N_t$ representing the number of unlabeled target samples. Since the source and target domains follow different data distributions due to the domain gap, their joint probabilities are not identical $p(X_s,Y_s) \neq p(X_t,Y_t)$, which implies $p(X_s)p(Y_s \mid X_s) \neq p(X_t)p(Y_t \mid X_t)$. Domain adaptation focuses on the specific case in which the source and target domains share the same feature and label spaces, $\mathcal{X}_s = \mathcal{X}_t$ and $\mathcal{Y}_s = \mathcal{Y}_t$, but differ in their marginal and conditional probability distributions, i.e., $p(X_s) \neq p(X_t)$ and $p(Y_s \mid X_s) \neq p(Y_t \mid X_t)$ \cite{Gardner_2020,Kouw_2018}. Feature-based domain adaptation aims to learn a feature extractor $G_f(\cdot;\theta_f)$ with trainable parameter $\theta_f$ that aligns the source and target feature distributions, i.e., $p(G_f(X_s)) \approx p(G_f(X_t))$ and $p(Y_s \mid G_f(X_s)) \approx p(Y_t \mid G_f(X_t))$. In addition, it seeks to learn a material property estimator $G_y(\cdot;\theta_y)$ with trainable parameter $\theta_y$ that minimizes the target risk, defined as $\mathcal{R}_{\mathcal{D}_t} = P_{(\textbf{\textit{x}},\textit{y}) \sim \mathcal{D}_t} \left(G_y(G_f(\textbf{\textit{x}})) \neq y \right)$.

\begin{figure*}[ht]
\centering
\includegraphics[width=0.85\linewidth]{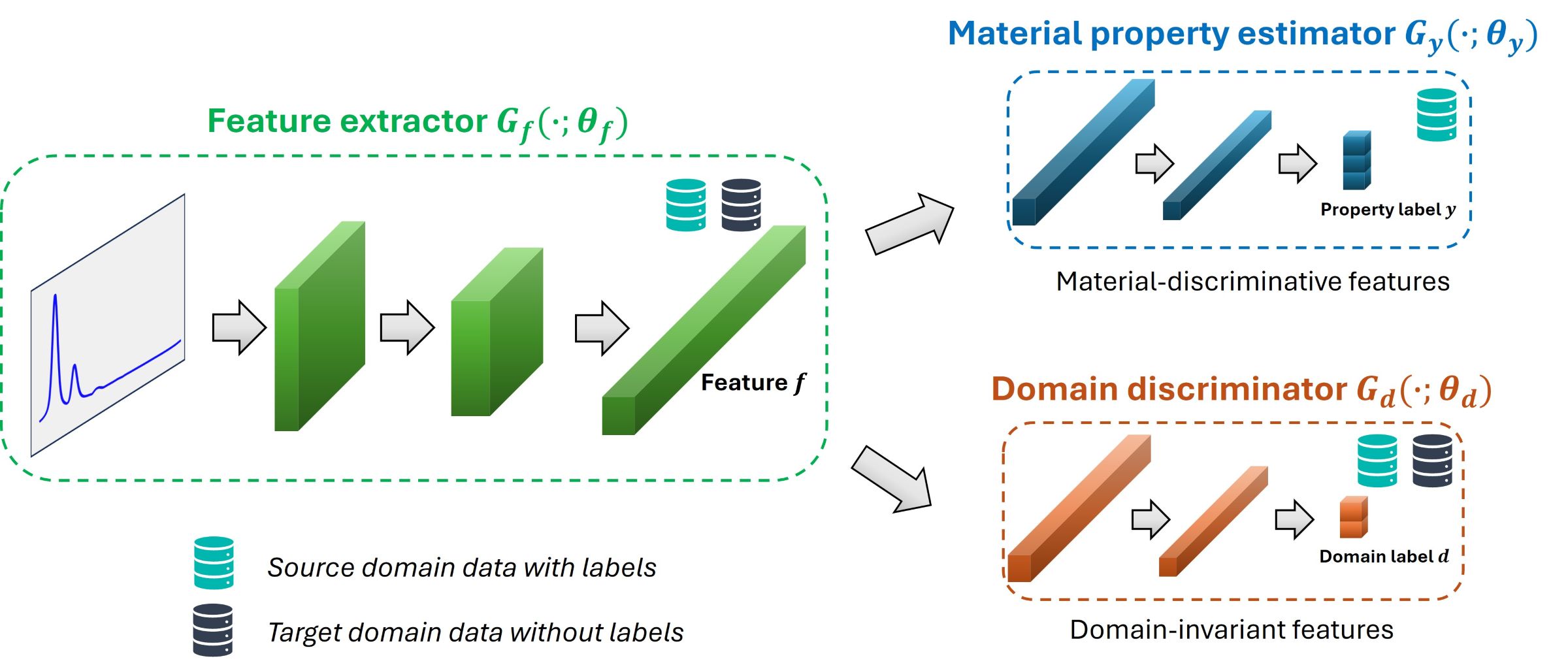}
\caption{\centering A schematic diagram of the DANN.}
\label{Fig_3}
\end{figure*}

The domain adversarial neural network (DANN) is a feature-based domain adaptation framework designed to minimize the distributional discrepancy between the source and target domains by learning domain-invariant representations through adversarial training \cite{Ganin_2016}. The feature extractor $G_f$ takes radar signals, either from the labeled source domain or the unlabeled target domain, as input and generates latent features. A domain discriminator $G_d(\cdot;\theta_d)$, parameterized by $\theta_d$, then receives these extracted features and attempts to distinguish whether they originate from the source or target domains. The feature extractor $G_f$ and the domain discriminator $G_d$ are trained in an adversarial manner. Specifically, the feature extractor $G_f$ aims to learn domain-invariant representations that can mislead $G_d$ into confusing the source and target domains. In contrast, $G_d$ is trained to correctly distinguish whether the extracted features originate from the source or target domain, resisting the deception introduced by $G_f$. In another branch of the framework, the material property estimator $G_y$ takes the features extracted from the labeled source data as input and predicts the corresponding values of the material parameters. The feature extractor $G_f$ and the material property estimator $G_y$ are trained in a supervised manner, enabling the DANN to learn material-discriminative features for accurate retrieval of material properties. A schematic diagram of the DANN is shown in Figure \ref{Fig_3}. The feature extractor $G_f$, domain discriminator $G_d$, and material property estimator $G_y$ are optimized by solving the following optimization problem:

\begin{equation} \label{Eq_7}
\min_{\theta_f,\theta_y} \left[ \mathcal{L}_{reg}\left( \theta_f,\theta_y \right) - \min_{\theta_d} \lambda \mathcal{L}_{adv} \left( \theta_f,\theta_d \right) \right],
\end{equation}

\begin{equation} \label{Eq_8}
\mathcal{L}_{reg}=\frac{1}{NM} \sum_{j=1}^{N} \sum_{i=1}^{M} \left(y_{i}^{(j)} - \hat{y}_{i}^{(j)}\right)^2,
\end{equation}

\begin{equation} \label{Eq_9}
\mathcal{L}_{adv}=-\mathop{\mathbb{E}_{\textbf{\textit{x}}_{s}\sim\mathcal{X}_s}}\left[\log\left(G_d\left( G_f(\textbf{\textit{x}}_{s})\right)\right)\right]-\mathop{\mathbb{E}_{\textbf{\textit{x}}_{t}\sim\mathcal{X}_t}}\left[\log\left(1-G_d\left( G_f(\textbf{\textit{x}}_{t})\right)\right)\right],
\end{equation}

\noindent where $\mathcal{L}_{reg}$ is the regression loss of the material property estimator $G_y$, which uses the extracted features to predict the material parameter values for the source domain. $\mathcal{L}_{adv}$ is the adversarial loss of the domain discriminator $G_d$, which distinguishes the extracted features between the source and target domains. $\lambda$ is a hyper-parameter used to tune the trade-off between domain-invariance and material-discriminativeness of the extracted features. In this study, $\lambda$ gradually varies from 0 to 1 during training, which can be defined as follows:

\begin{equation} \label{Eq_17}
\lambda = \frac{2}{1+e^{-10 \times p}}-1
\end{equation}

\noindent where $p$ linearly goes from 0 to 1 according to this formula:

\begin{equation} \label{Eq_18}
p = \frac{\msfsl{current \; iteration}}{\msfsl{total \; iterations}}.
\end{equation}

To enable adversarial training between the feature extractor $G_f$ and the domain discriminator $G_d$, a gradient reversal layer (GRL) is inserted between them. During forward propagation, the GRL acts as an identity mapping, while during backpropagation, it multiplies the gradient from the domain discriminator $G_d$ by $-\lambda$ before passing it to the feature extractor $G_f$. Consequently, during backpropagation, the trainable parameters are updated as follows:

\begin{equation} \label{Eq_10}
\theta_f \leftarrow \theta_f - \mu \left( \frac{\partial \mathcal{L}_y}{\partial \theta_f} - \lambda \frac{\partial \mathcal{L}_d}{\partial \theta_f} \right),
\end{equation}

\begin{equation} \label{Eq_11}
\theta_y \leftarrow \theta_y - \mu \frac{\partial \mathcal{L}_y}{\partial \theta_y},
\end{equation}

\begin{equation} \label{Eq_12}
\theta_d \leftarrow \theta_d - \mu \lambda \frac{\partial \mathcal{L}_d}{\partial \theta_d},
\end{equation}

\noindent where $\mu$ is the learning rate. In this study, the exponential learning rate schedule with a decay factor of 0.9 is adopted. The training of the DANN aims to find the saddle point $(\hat{\theta}_f, \hat{\theta}_y, \hat{\theta}_d)$ through Equations \ref{Eq_10}–\ref{Eq_12}, which collectively solve the min-max optimization problem formulated in Equation \ref{Eq_7}. DANN is adopted as a baseline for comparison with the proposed approach. To enable a fair comparison with the 1D CNN, the consistent network architecture is employed for the DANN, as summarized in Table \ref{T2}.

\begin{table*}[ht]
     \small\sf\centering
     \caption{\centering Architecture of the DANN.\label{T2}}
     \begin{tabular}{cccccccc}
         \toprule
         Network & Layer & \makecell{Kernel \\ number} & \makecell{Kernel \\ size} & \makecell{Stride \\ size} & \makecell{Input \\ size} & \makecell{Output \\ size} & Activation \\
         \cmidrule{1-8}
         Feature extractor & Convolution & $32$ & $5$ & $5$ & $1 \times 6560$ & $32 \times 1312$ & LeakyReLU \\
            & Convolution & $64$ & $5$ & $5$ & $32 \times 1312$ & $64 \times 262$ & LeakyReLU \\
            & Convolution & $128$ & $4$ & $4$ & $64 \times 262$ & $128 \times 65$ & LeakyReLU \\
            & Convolution & $256$ & $4$ & $4$ & $128 \times 65$ & $256 \times 16$ & LeakyReLU \\
        Material property estimator & Convolution & $512$ & $3$ & $3$ & $256 \times 16$ & $512 \times 5$ & LeakyReLU \\
            & Convolution & $10240$ & $3$ & $3$ & $512 \times 5$ & $10240 \times 1$ & LeakyReLU \\
            & Flatten & --- & --- & --- & $10240 \times 1$ & $10240$ & --- \\
            & Full connection & --- & --- & --- & $10240$ & $M$ & Linear \\
        Domain discriminator & Convolution & $512$ & $3$ & $3$ & $256 \times 16$ & $512 \times 5$ & LeakyReLU \\
            & Convolution & $1024$ & $3$ & $3$ & $512 \times 5$ & $1024 \times 1$ & LeakyReLU \\
            & Flatten & --- & --- & --- & $1024 \times 1$ & $1024$ & --- \\
            & Full connection & --- & --- & --- & $1024$ & $2$ & Linear \\
         \bottomrule
     \end{tabular}
\end{table*}

\subsection{Physics-Guided Domain Adversarial Neural Network} \label{PhyDANN}
\begin{figure*}[ht]
\centering
\includegraphics[width=0.85\linewidth]{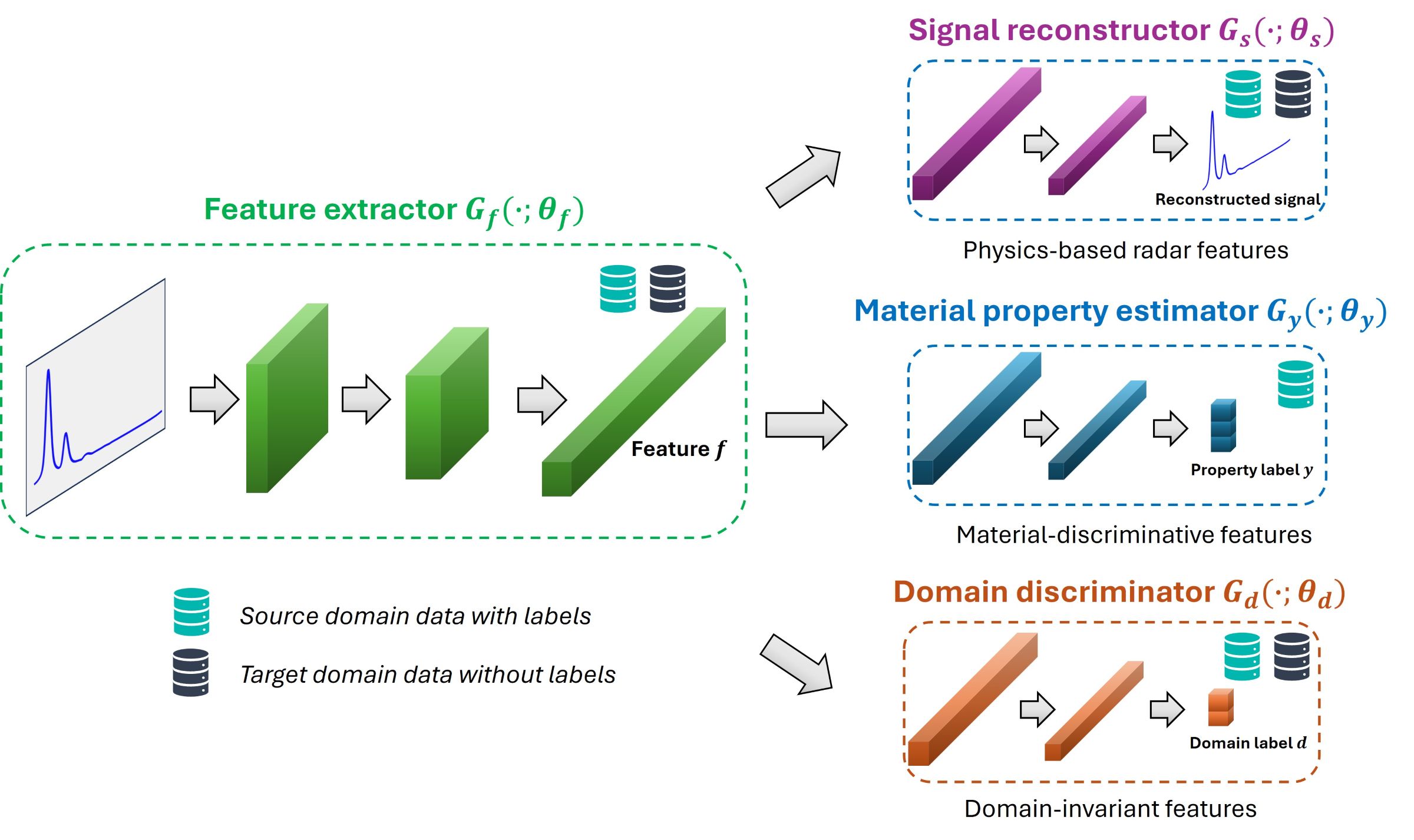}
\caption{\centering A schematic diagram of the PhyDANN-1.}
\label{Fig_19}
\end{figure*}

\begin{figure*}[ht]
\centering
\includegraphics[width=0.85\linewidth]{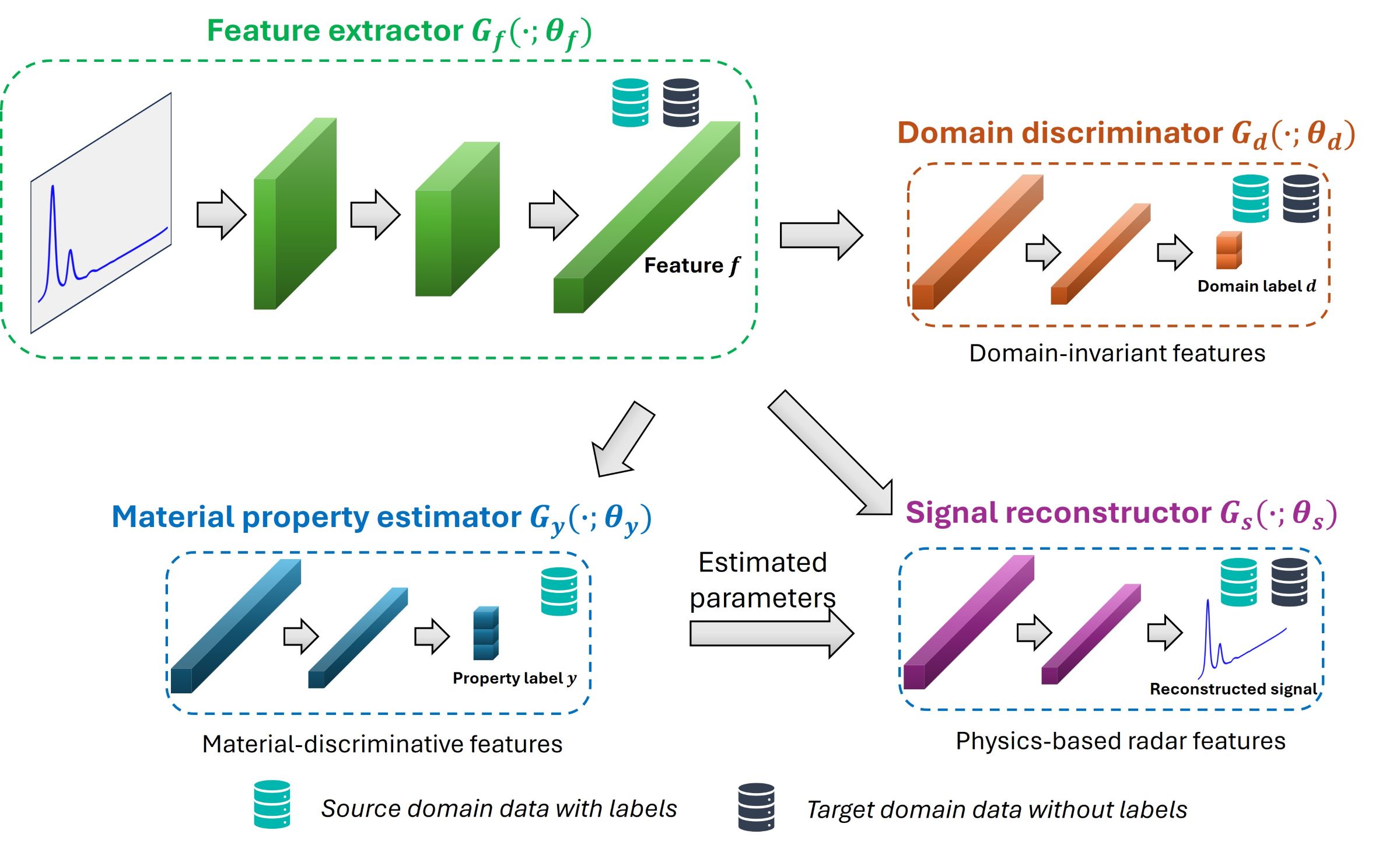}
\caption{\centering A schematic diagram of the PhyDANN-2.}
\label{Fig_20}
\end{figure*}

The adversarial learning in DANN is performed in a purely data-driven manner without incorporating any physical constraints. As a result, the learned domain-invariant features may lack sufficient discriminability with respect to different material properties. The physical features embedded in the radar signal are crucial for estimating material properties. For example, as the soil water content increases, its permittivity rises, which slows the propagation of electromagnetic waves and leads to a longer signal time delay (Figure \ref{Fig_6a}). Similarly, increasing the material depth enlarges the travel distance of the reflected wave, resulting in a longer signal time delay (Figure \ref{Fig_6c}). Moreover, higher soil water content increases soil conductivity and decreases signal amplitude, leading to smaller envelope areas and peak ratios (Figure \ref{Fig_6b}). In addition, the reduction in the centroid frequency is a physical feature in the frequency domain of the radar signal, which is caused by the attenuation of the high-frequency energy as the water content of the material increases \cite{Fang_2025}. To enable DANN to capture essential physics-based radar features during adversarial learning, we propose a physics-guided DANN (PhyDANN) that incorporates a signal reconstructor $G_s(\cdot;\theta_s)$ with trainable parameter $\theta_s$. Specifically, two types of PhyDANN have been developed. PhyDANN-1 reconstructs source and target time-domain signals using only the latent features produced by the feature extractor $G_f$. PhyDANN-2 reconstructs source and target time-domain signals using both the latent features and the material properties predicted by the material property estimator $G_y$. The schematic diagrams of the PhyDANN-1 and PhyDANN-2 are shown in Figure \ref{Fig_19} and \ref{Fig_20}, respectively. The loss function of PhyDANN can be expressed as:

\begin{equation} \label{Eq_19}
\min_{\theta_f,\theta_y, \theta_s} \left[ \mathcal{L}_{reg}\left( \theta_f,\theta_y \right) + \mathcal{L}_{reconst}\left( \theta_f,\theta_s \right)- \min_{\theta_d} \lambda \mathcal{L}_{adv} \left( \theta_f,\theta_d \right) \right],
\end{equation}

\begin{equation} \label{Eq_20}
\mathcal{L}_{reconst}=\frac{1}{ND} \sum_{j=1}^{N} \sum_{i=1}^{D} \left(x_{i}^{(j)} - \hat{x}_{i}^{(j)}\right)^2,
\end{equation}

\noindent where $N$ is the number of samples, and $D$ is the number of data points in the radar signal. $x$ is the original signal value, and $\hat{x}$ is the reconstructed signal value. The network architectures of the signal reconstructor $G_f$ in PhyDANN-1 and PhyDANN-2 are summarized in Tables \ref{T11} and \ref{T12}, respectively.

\begin{table*}[ht]
     \small\sf\centering
     \caption{\centering Architecture of the signal reconstructor of PhyDANN-1.\label{T11}}
     \begin{tabular}{lcccccc}
         \toprule
         Layer & Kernel number & Kernel size & Stride & Input size & Output size & Activation \\
         \cmidrule{1-7}
        Upsample & \multicolumn{3}{c}{\textit{Linear interpolation}} & $256 \times 10$ & $256 \times 41$ & --- \\
        Convolution & $128$ & $3$ & $1$ & $256 \times 41$ & $128 \times 41$ & LeakyReLU \\
        Upsample & \multicolumn{3}{c}{\textit{Linear interpolation}} & $128 \times 41$ & $128 \times 164$ & --- \\
        Convolution & $64$ & $3$ & $1$ & $128 \times 164$ & $64 \times 164$ & LeakyReLU\\
        Upsample & \multicolumn{3}{c}{\textit{Linear interpolation}} & $64 \times 164$ & $64 \times 821$ & --- \\
        Convolution & $32$ & $3$ & $1$ & $64 \times 821$ & $32 \times 821$ & LeakyReLU\\
        Upsample & \multicolumn{3}{c}{\textit{Linear interpolation}} & $32 \times 821$ & $32 \times 6560$ & --- \\
        Convolution & $1$ & $3$ & $1$ & $32 \times 6560$ & $1 \times 6560$ & Linear\\
         \bottomrule
     \end{tabular}
\end{table*}

\begin{table*}[ht]
     \small\sf\centering
     \caption{\centering Architecture of the signal reconstructor of PhyDANN-2.\label{T12}}
     \begin{tabular}{lcccccc}
         \toprule
         Layer & Kernel number & Kernel size & Stride & Input size & Output size & Activation \\
         \cmidrule{1-7}
        Full connection & \multicolumn{3}{c}{\textit{Parameter embedding}} & $M$ & $64$ & LeakyReLU \\
        Full connection & \multicolumn{3}{c}{\textit{Parameter embedding}} & $64$ & $256$ & Linear \\
        Summation & \multicolumn{3}{c}{\textit{Parameter and feature fusion}} & $256 \times 10$ & $256 \times 10$ & --- \\
        Upsample & \multicolumn{3}{c}{\textit{Linear interpolation}} & $256 \times 10$ & $256 \times 41$ & --- \\
        Convolution & $128$ & $3$ & $1$ & $256 \times 41$ & $128 \times 41$ & LeakyReLU \\
        Upsample & \multicolumn{3}{c}{\textit{Linear interpolation}} & $128 \times 41$ & $128 \times 164$ & --- \\
        Convolution & $64$ & $3$ & $1$ & $128 \times 164$ & $64 \times 164$ & LeakyReLU\\
        Upsample & \multicolumn{3}{c}{\textit{Linear interpolation}} & $64 \times 164$ & $64 \times 821$ & --- \\
        Convolution & $32$ & $3$ & $1$ & $64 \times 821$ & $32 \times 821$ & LeakyReLU\\
        Upsample & \multicolumn{3}{c}{\textit{Linear interpolation}} & $32 \times 821$ & $32 \times 6560$ & --- \\
        Convolution & $1$ & $3$ & $1$ & $32 \times 6560$ & $1 \times 6560$ & Linear\\
         \bottomrule
     \end{tabular}
\end{table*}

\subsection{Hierarchical Domain Adversarial Neural Network} \label{HierDANN}
\begin{figure*}[ht]
\centering
\includegraphics[width=0.9\linewidth]{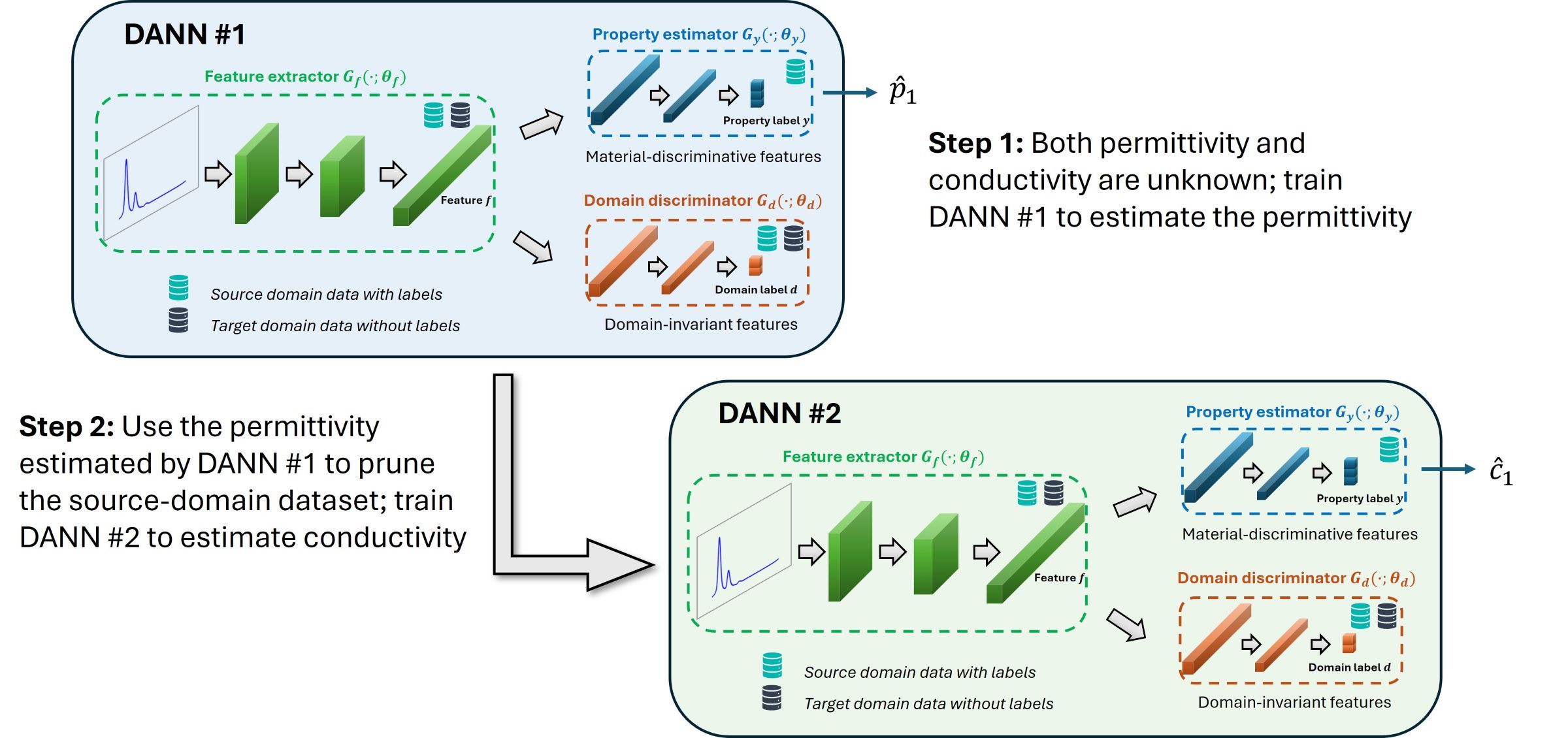}
\caption{\centering Hierarchical domain adaptation in the HierDANN.}
\label{Fig_4}
\end{figure*}

The 1D CNN, DANN, and PhyDANN introduced in Sections \ref{1DCNN}, \ref{DANN}, and \ref{PhyDANN} are designed to estimate multiple material properties with $M$ parameters (e.g., permittivity, conductivity, and depth of each layer) simultaneously. However, compared to single-output regression, multi-output regression is more challenging because it should model and interpret the interdependencies among multiple targets \cite{Borchani_2015}. Moreover, domain adaptation for multi-output regression is trained on a source-domain dataset that captures a wide range of possible combinations of target variables, thereby increasing the training complexity. Additionally, when the source label space is larger than the target label space, this partial domain adaptation setting may further exacerbate the well-known negative transfer bottleneck \cite{Cao_2018}. To improve the performance of domain adaptation in estimating multiple material properties, a hierarchical DANN (HierDANN) is developed, where the material parameters are estimated sequentially, and each estimated parameter is treated as known for subsequent estimations. Accordingly, at each subsequent step, the source-domain dataset is pruned by retaining only the radar signals corresponding to the known material parameter value, thereby enabling more stable and effective adversarial training of the DANN. As an example of two-parameter estimation (e.g., permittivity and conductivity), the hierarchical domain adaptation process of the proposed HierDANN is illustrated in Figure \ref{Fig_4}. It is worth noting that, unlike single-output regression, which ignores the interdependence among target variables, the proposed HierDANN leverages these interdependencies by treating each estimated parameter as known for the subsequent estimation.

The order of the material parameters to be estimated can be determined through variance-based sensitivity analysis, which quantifies the proportion of the model output variance attributed to each input parameter and assesses their relative importance in influencing the model output. The Sobol’s method \cite{Sobol_2001}, as a global sensitivity analysis method, has been utilized to measure the importance of the input parameters in various models \cite{Homma_1996,Saltelli_2010,Ojha_2022}. Specifically, the first-order Sobol indices, which calculate the contribution of individual input parameters to the variance of the model output, are adopted to determine the sequence of the material parameters to be estimated in the HierDANN. Given a model of the form $Y=f(X_1,X_2,...,X_M)$, the first-order Sobol index $S_i$ for a generic factor $X_i$ can be written as:

\begin{equation} \label{Eq_13}
S_i = \frac{V_{X_i} \left( E_{\mathbf{X}_{\sim i}} (Y \mid X_i) \right)}{V(Y)},
\end{equation}

\noindent where $\mathbf{X}_{\sim i}$ denotes the vector containing all input factors except $X_i$. The expectation $E_{\mathbf{X}_{\sim i}}(Y \mid X_i)$ represents the mean of $Y$ over all possible realizations of $\mathbf{X}_{\sim i}$ while keeping $X_i$ fixed. The variance $V_{X_i}\left(E_{\mathbf{X}_{\sim i}}(Y \mid X_i)\right)$ is then computed over all possible values of $X_i$, ranging from zero to the total output variance $V(Y)$. The portion of the model output variance attributed to the factor $X_i$ can be estimated as follows:

\begin{equation} \label{Eq_14}
V_{X_i} \left( E_{\mathbf{X}_{\sim i}} (Y \mid X_i) \right) = 
\frac{1}{N} \sum_{j=1}^{N} f(\mathbf{A})_j 
\left( f\left( \mathbf{B}_A^{(i)} \right)_j - f(\mathbf{B})_j \right),
\end{equation}

\noindent where $\mathbf{A}$ and $\mathbf{B}$ are sample matrices of input factors with size $N \times M$. $\mathbf{B}_A^{(i)}$ denotes a matrix where column $i$ comes from matrix $\mathbf{B}$ and all other $k-1$ columns come from matrix $\mathbf{A}$. 

\begin{figure*}[ht]
     \centering
     \begin{subfigure}[t]{0.4\textwidth}
         \centering
         \includegraphics[width=1\linewidth]{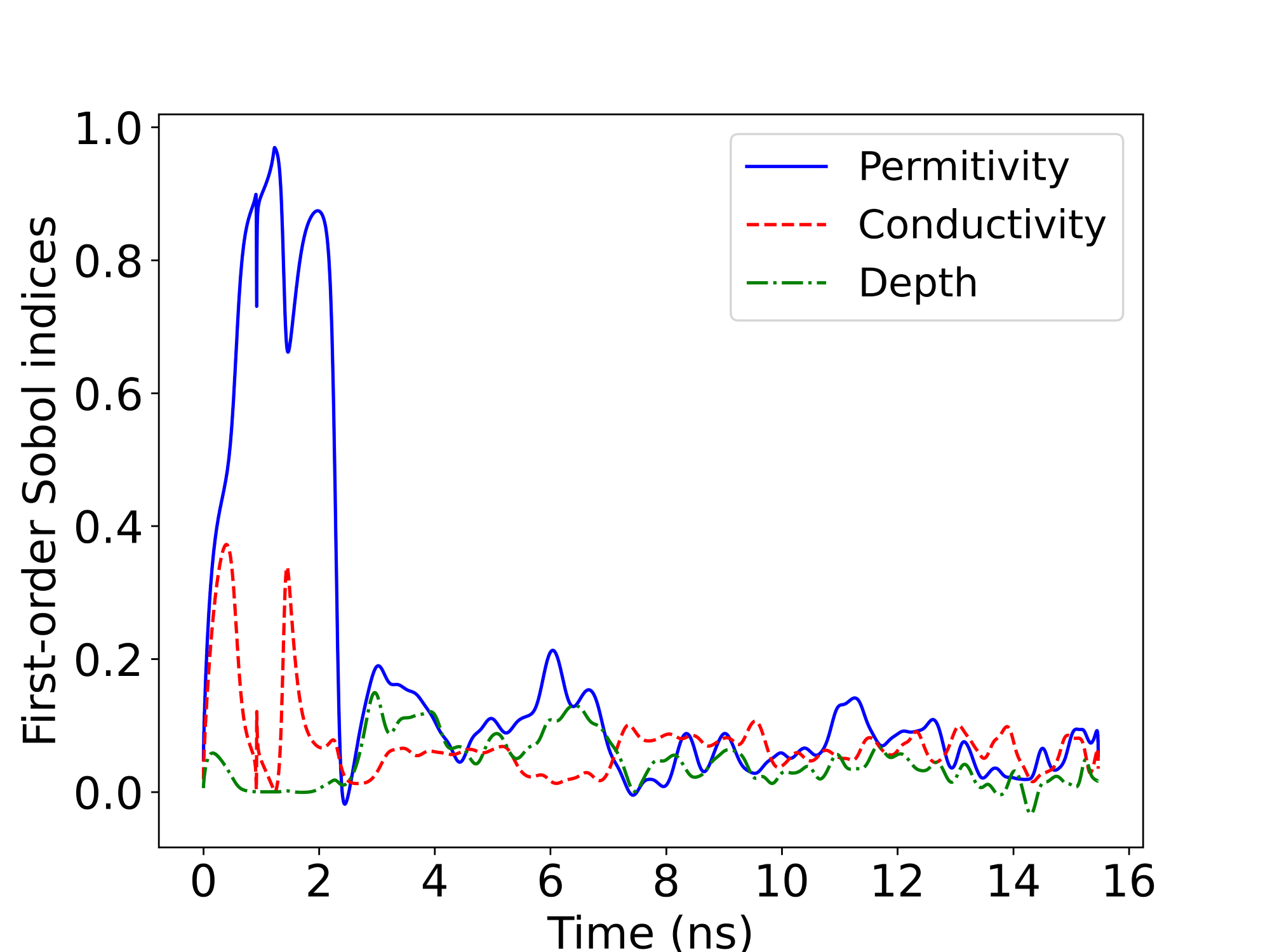}
         \captionsetup{justification=centering}
         \caption{Time history of first-order Sobol indices.}
         \label{Fig_5a}
     \end{subfigure}
     \begin{subfigure}[t]{0.4\textwidth}
         \centering
         \includegraphics[width=0.9\linewidth]{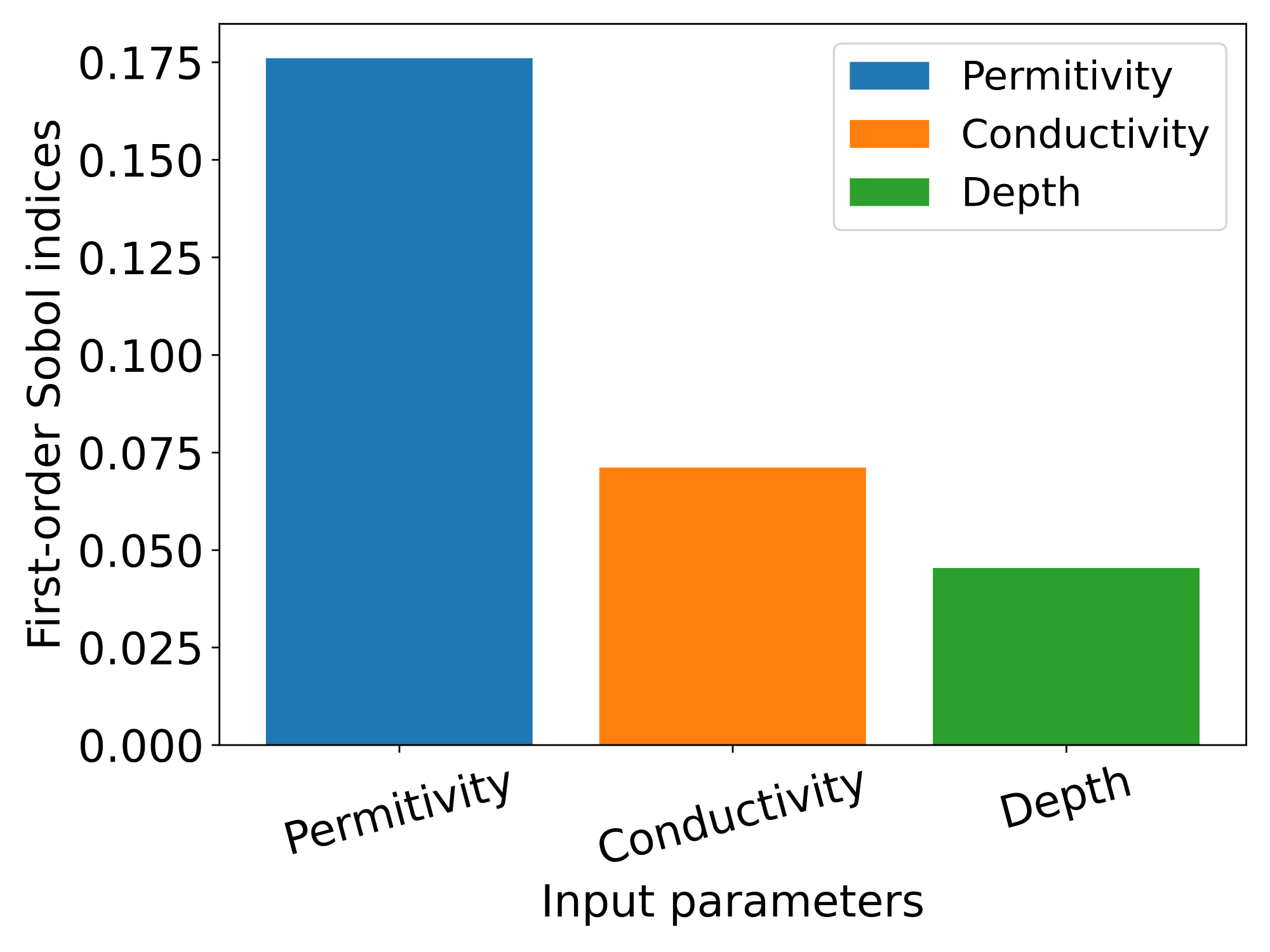}
         \captionsetup{justification=centering}
         \caption{Mean of first-order Sobol indices.}
         \label{Fig_5b}
     \end{subfigure}
     \caption{\centering Variance-based sensitivity analysis using the Sobol's method.}
     \label{Fig_5}
\end{figure*}

\begin{figure*}[ht]
     \centering
     \begin{subfigure}[t]{0.32\textwidth}
         \centering
         \includegraphics[width=1\linewidth]{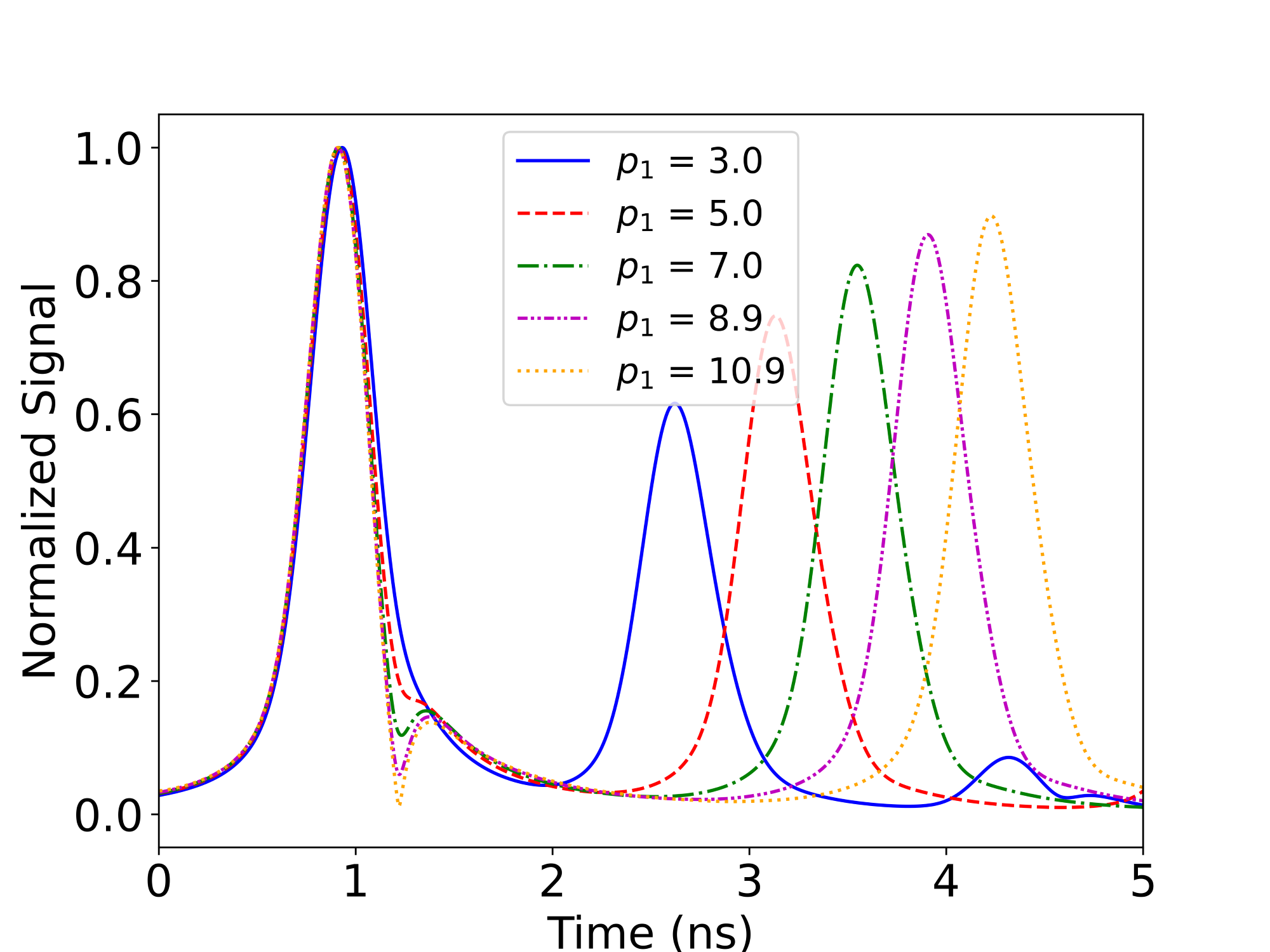}
         \captionsetup{justification=centering}
         \caption{Permittivity ($p_1$).}
         \label{Fig_6a}
     \end{subfigure}
     \begin{subfigure}[t]{0.32\textwidth}
         \centering
         \includegraphics[width=1\linewidth]{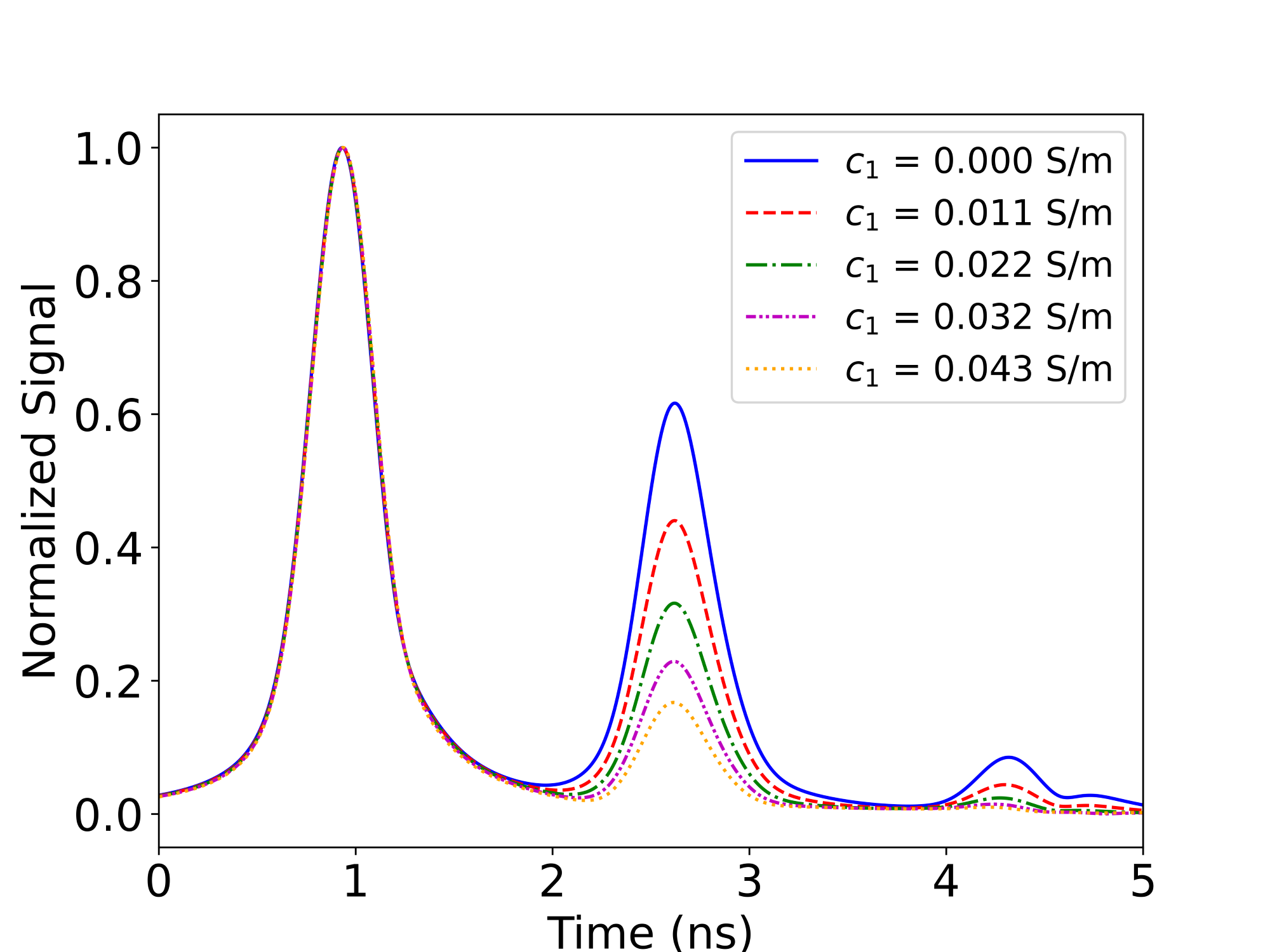}
         \captionsetup{justification=centering}
         \caption{Conductivity ($c_1$).}
         \label{Fig_6b}
     \end{subfigure}
     \begin{subfigure}[t]{0.32\textwidth}
         \centering
         \includegraphics[width=1\linewidth]{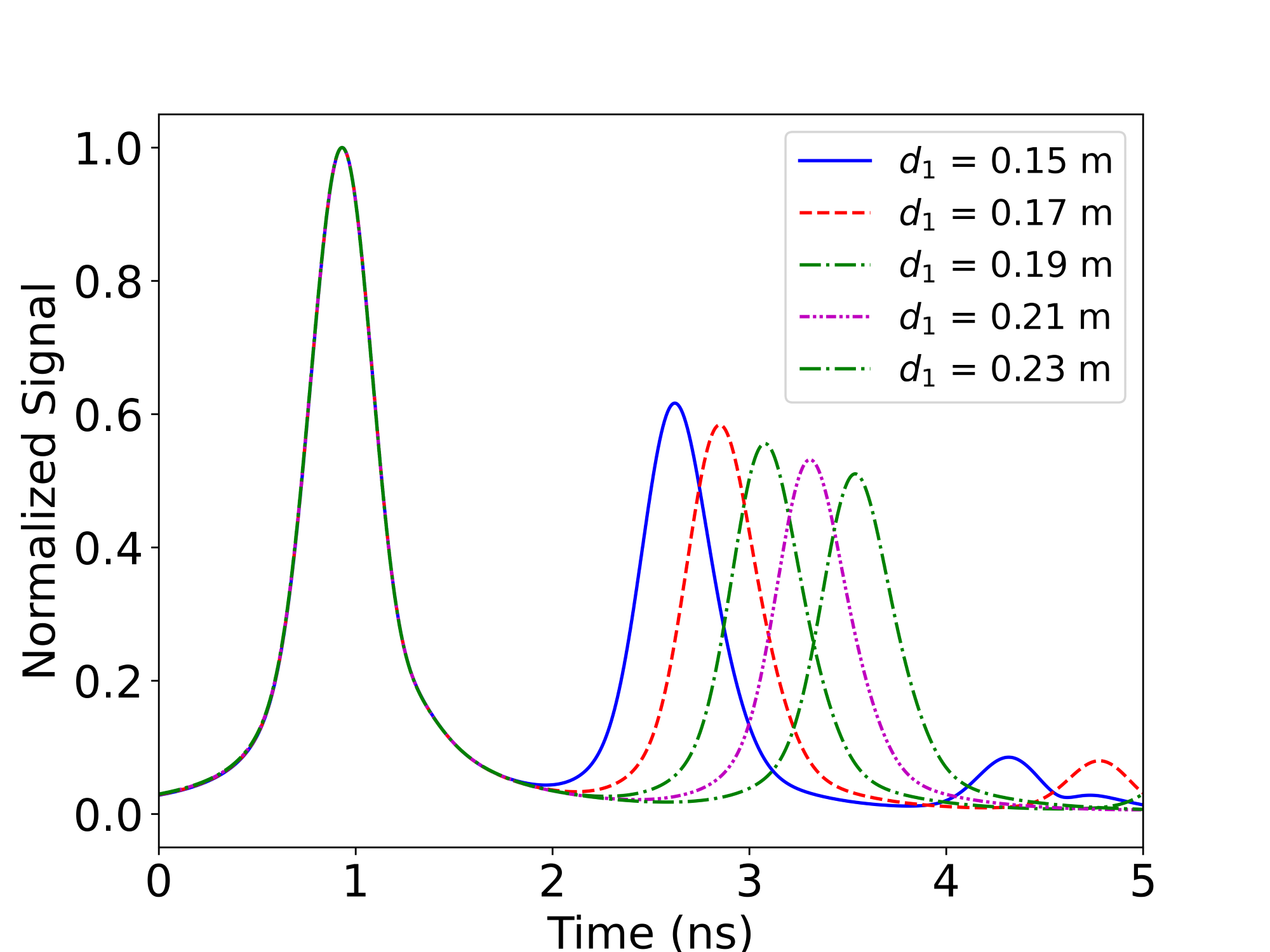}
         \captionsetup{justification=centering}
         \caption{Depth ($d_1$).}
         \label{Fig_6c}
     \end{subfigure}
     \caption{\centering Effect of material property variations on radar scans of a single-layer material.}
     \label{Fig_6}
\end{figure*}

In this study, the model $Y=f(X_1,X_2,...,X_M)$ is implemented using the FDTD simulation, which produces the radar A-scan (i.e., the output) based on material parameters including permittivity, conductivity, and depth (i.e., the inputs). The Sobol sensitivity analysis is performed on a dataset consisting of 9,261 scans, generated through simulations using 21 uniformly distributed values for each of the three material parameters: permittivity ($p_1 = 3\sim12.9$, step size 0.495), conductivity ($c_1 = 0\sim0.0054$ S/m, step size 0.0027 S/m), and depth ($d_1 = 0.15\sim0.25$ m, step size 0.005 m). According to the first-order Sobol indices shown in Figure \ref{Fig_5}, the material parameters influencing the radar scan can be ranked in order of importance as permittivity, conductivity, and depth. Figure \ref{Fig_5a} illustrates the variation of the first-order Sobol indices over time, while Figure \ref{Fig_5b} presents their averaged values over the entire time series. Figure \ref{Fig_6} illustrates the changes in the radar scan of a single-layer material due to variations in material properties, calculated using FDTD simulations. This figure shows good agreement with the first-order Sobol indices. Therefore, the sequence of the material parameters to be estimated in the HierDANN is permittivity, conductivity, and depth. To enable a fair comparison with the 1D CNN and DANN, the HierDANN adopts the same network architecture as the DANN, as summarized in Table \ref{T2}, except that its output layer consists of a single neuron, as each HierDANN network predicts only one material parameter.

Note that the proposed PhyDANN and HierDANN frameworks can be integrated, resulting in five novel sim-to-real domain adaptation approaches for subsurface material property estimation: HierDANN, PhyDANN-1, HierPhyDANN-1, PhyDANN-2, and HierPhyDANN-2. These five approaches are compared with two baseline methods, 1D-CNN and DANN, as discussed in Section \ref{Results_and_Discussion}.

\subsection{Performance Evaluation}
The permittivity and conductivity of the materials were measured with an external in-situ capacitance sensor Teros-12 \cite{Fragkos_2024}. The measured values were compared with the values predicted by our proposed methods. Four evaluation metrics, namely the Pearson correlation coefficient $R$, Bias, root mean squared error (RMSE), and unbiased RMSE (ubRMSE), were employed to assess the performance of material parameter estimation across different approaches. $R$, Bias, RMSE, and ubRMSE can be calculated as follows:

\begin{equation} \label{Eq_15}
R = \frac{\sum_{j=1}^{K} (y_j - \bar{y})(\hat{y}_j - \bar{\hat{y}})}{\sqrt{\sum_{j=1}^{K} (y_j - \bar{y})^2\sum_{j=1}^{K} (\hat{y}_j - \bar{\hat{y}})^2}},
\end{equation}

\begin{equation} \label{Eq_16}
\text{Bias} = \frac{1}{K}{\sum_{j=1}^{K} (\hat{y}_j - y_j)},
\end{equation}

\begin{equation} \label{Eq_21}
\text{RMSE} =\sqrt{\frac{1}{K} \sum_{j=1}^{K} \left(y_j - \hat{y}_j\right)^2},
\end{equation}

\begin{equation} \label{Eq_22}
\text{ubRMSE} = \sqrt{\text{RMSE}^2 - \text{Bias}^2},
\end{equation}

\noindent where $y_1$, $y_2$, ...,  $y_K$ are the measured parameter values; $\hat{y}_j$ is the predicted parameter value, which is obtained by taking the mean of the estimations from $L$ scans; $\bar{y}$ is the mean of the measured parameter values $y_1$, $y_2$, ..., $y_K$; and $\bar{\hat{y}}$ is the mean of the predicted parameter values $\hat{y}_1$, $\hat{y}_2$, ..., $\hat{y}_K$.

In evaluating the performance of the proposed methods, we prioritize the Pearson correlation coefficient $R$ as the primary criterion for selection. A high $R$ serves as the strongest indicator that the algorithm is correctly capturing the physical sensitivity of the underlying soil moisture signal. Although Bias and RMSE are reported for completeness, these metrics are often dominated by systematic offsets attributable to spatial scale mismatches or lack of site-specific calibration. To isolate the random error component from these systematic shifts, we also report the ubRMSE, which is widely recognized in soil moisture validation for its ability to decouple systematic biases (e.g., representativeness errors) from intrinsic random errors (\cite{Entekhabi_2010}). This evaluation strategy allows us to identify methods that possess high intrinsic sensitivity (i.e., high $R$), as their systematic errors (i.e., Bias) can be removed via linear calibration, whereas methods with low correlation suffer from irreducible random error that cannot be corrected.

The in-situ sensor was calibrated in the laboratory using the same soil as that used in the field study to establish a soil-specific relation between the sensor voltage values and the true volumetric water content (VWC). Here, the true VWCs were estimated using gravimetric oven drying tests of soil samples. Another independent calibration was also conducted to establish a relation between permittivity and VWC for the same soil. Using these two calibrated relations separately, the true VWCs from sensor measurements and the VWCs from predicted permittivity were obtained and compared to each other. 

\section{Illustrative Example Setups} \label{illustrative_example_setups}
The proposed physics-guided hierarchical domain adaptation approaches, HierDANN, PhyDANN-1, HierPhyDANN-1, PhyDANN-2, and HierPhyDANN-2, have been systematically investigated using various setups: the laboratory tests for single- and two-layer materials, and the field tests for single- and two-layer materials. The real-world signals (target domain) were collected using the GSSI StructureScan MiniXT with an antenna frequency of 2,700 MHz. The details of these setups are provided in this section.

\subsection{Laboratory Test for Single-Layer Material}
A single-layer material (soil) with a depth of 0.2 m was constructed in the laboratory, where various soil permittivities and conductivities were achieved by blending different amounts of water into the soil. Radar scans were collected from six specimens, each comprising 25 individual scans. A summary of the experimental data collected for the single-layer material is presented in Table \ref{T3}. Additionally, a simulated dataset consisting of 15,675 scans was generated using the FDTD simulation by varying the material parameters as follows: permittivity ($p_1 = 3-12.9$, step size 0.1), conductivity ($c_1 = 0-0.0054$ S/m, step size 0.001 S/m), and depth ($d_1 = 0.15-0.25$ m, step size 0.05 m). The simulation parameters were defined to span the range of realizations expected under laboratory testing conditions.

\begin{table*}[ht]
    \small\sf\centering
    \caption{\centering Experimental data collected for single-layer material (soil).}
    \label{T3}
    \begin{threeparttable}
        \begin{tabularx}{0.8\textwidth}{c *{5}{>{\centering\arraybackslash}X}}
            \toprule
            Specimen & $p_1$ & $c_1$ (S/m) & $d_1$ (m) & Number of scans \\
            \cmidrule(lr){1-5}
            1 & 3.2  & 0.0000 & 0.20 & 25 \\
            2 & 5.3  & 0.0097 & 0.20 & 25 \\
            3 & 7.3  & 0.0256 & 0.20 & 25 \\
            4 & 9.7  & 0.0263 & 0.20 & 25 \\
            5 & 10.3 & 0.0480 & 0.20 & 25 \\
            6 & 12.9 & 0.0513 & 0.20 & 25 \\
            \bottomrule
        \end{tabularx}
        \begin{tablenotes}
            \item[1] $p_1$, $c_1$, and $d_1$ denote the permittivity, conductivity, and depth of the soil, respectively.
        \end{tablenotes}
    \end{threeparttable}
\end{table*}

\subsection{Laboratory Test for Two-Layer Material}
\begin{figure*}[ht]
\centering
\includegraphics[width=0.6\linewidth]{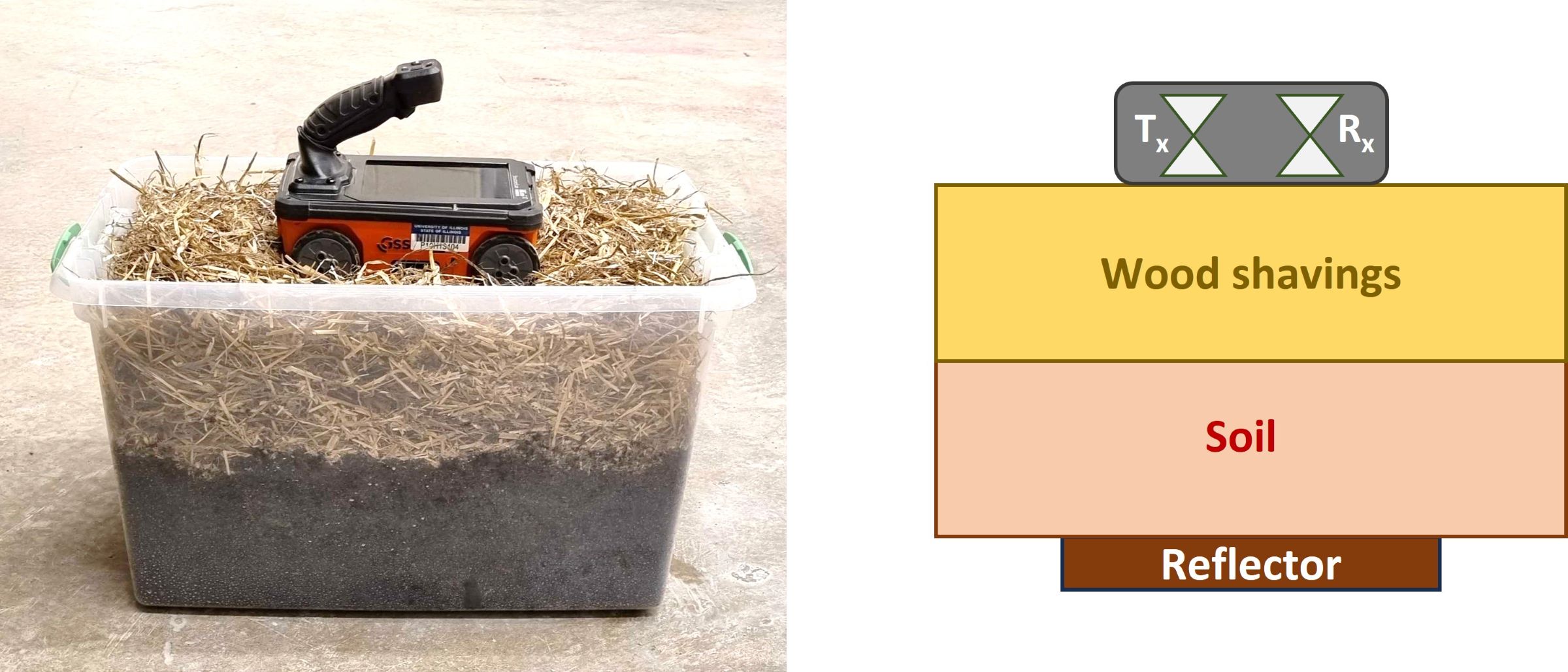}
\caption{\centering The two-layer laboratory setup.}
\label{Fig_15}
\end{figure*}

A two-layer material comprising soil and wood shavings was constructed in the laboratory, where the soil layer had a depth of 0.15 m, and the wood shavings layer was prepared with depths of 0.10 m and 0.15 m in two configurations. A metal plate is embedded at the bottom of the soil layer to enhance radar signal reflection. The two-layer laboratory setup is demonstrated in Figure \ref{Fig_15}. Moisture of both layers, and depth of the top layer were varied, resulting in a total of 18 combinations of permittivity, conductivity, and depth of the 2-layer setup (listed in Table \ref{T4}). 50 GPR scans were recorded for each of these combinations. Moreover, a simulated dataset consisting of 49,680 scans was generated using the FDTD simulation by varying the material parameters as follows: soil permittivity ($p_1 = 3.5-13$, step size 0.5), soil conductivity ($c_1 = 0-0.11$ S/m, step size 0.005 S/m), soil depth ($d_1 = 0.1-0.2$ m, step size 0.05 m), wood shavings permittivity ($p_2 = 1.5-3$, step size 0.5), wood shavings conductivity ($c_2 = 0-0.01$ S/m, step size 0.005 S/m), and wood shavings depth ($d_2 = 0.1-0.2$ m, step size 0.05 m). The simulation parameters were selected to cover the range of realizations anticipated under laboratory testing conditions.

\begin{table*}[ht]
    \small\sf\centering
    \caption{\centering Experimental data collected for two-layer material (soil-wood shavings).}
    \label{T4}
    \begin{threeparttable}
        \begin{tabularx}{0.8\textwidth}{Y *{9}{Y}}
            \toprule
            Specimen & $p_1$ & $c_1$ (S/m) & $d_1$ (m) & $p_2$ & $c_2$ (S/m) & $d_2$ (m) & Number of scans \\
            \cmidrule(lr){1-8}
            1  & 3.8  & 0.0000 & 0.15 & 1.6 & 0.0000 & 0.10 & 50 \\
            2  & 3.8  & 0.0000 & 0.15 & 1.6 & 0.0000 & 0.15 & 50 \\
            3  & 3.8  & 0.0000 & 0.15 & 2.8 & 0.0000 & 0.10 & 50 \\
            4  & 4.9  & 0.0027 & 0.15 & 1.6 & 0.0000 & 0.10 & 50 \\
            5  & 4.9  & 0.0027 & 0.15 & 1.6 & 0.0000 & 0.15 & 50 \\
            6  & 4.9  & 0.0027 & 0.15 & 2.8 & 0.0000 & 0.10 & 50 \\
            7  & 6.9  & 0.0230 & 0.15 & 1.6 & 0.0000 & 0.10 & 50 \\
            8  & 6.9  & 0.0230 & 0.15 & 1.6 & 0.0000 & 0.15 & 50 \\
            9  & 6.9  & 0.0230 & 0.15 & 2.8 & 0.0000 & 0.10 & 50 \\
            10 & 7.7  & 0.0473 & 0.15 & 1.6 & 0.0000 & 0.10 & 50 \\
            11 & 7.7  & 0.0473 & 0.15 & 1.6 & 0.0000 & 0.15 & 50 \\
            12 & 7.7  & 0.0473 & 0.15 & 2.8 & 0.0000 & 0.10 & 50 \\
            13 & 10.4 & 0.0750 & 0.15 & 1.6 & 0.0000 & 0.10 & 50 \\
            14 & 10.4 & 0.0750 & 0.15 & 1.6 & 0.0000 & 0.15 & 50 \\
            15 & 10.4 & 0.0750 & 0.15 & 2.8 & 0.0000 & 0.10 & 50 \\
            16 & 12.9 & 0.1060 & 0.15 & 1.6 & 0.0000 & 0.10 & 50 \\
            17 & 12.9 & 0.1060 & 0.15 & 1.6 & 0.0000 & 0.15 & 50 \\
            18 & 12.9 & 0.1060 & 0.15 & 2.8 & 0.0000 & 0.10 & 50 \\
            \bottomrule
        \end{tabularx}
        \begin{tablenotes}
            \item[1] $p_1$, $c_1$, and $d_1$ denote the permittivity, conductivity, and depth of the soil, respectively. 
            \item[2] $p_2$, $c_2$, and $d_2$ denote the permittivity, conductivity, and depth of the wood shavings, respectively.
        \end{tablenotes}
    \end{threeparttable}
\end{table*}

\subsection{Field Test for Single- and Two-Layer Materials}
\begin{figure*}[ht]
     \centering
     \begin{subfigure}[t]{0.32\textwidth}
         \centering
         \includegraphics[width=0.9\linewidth]{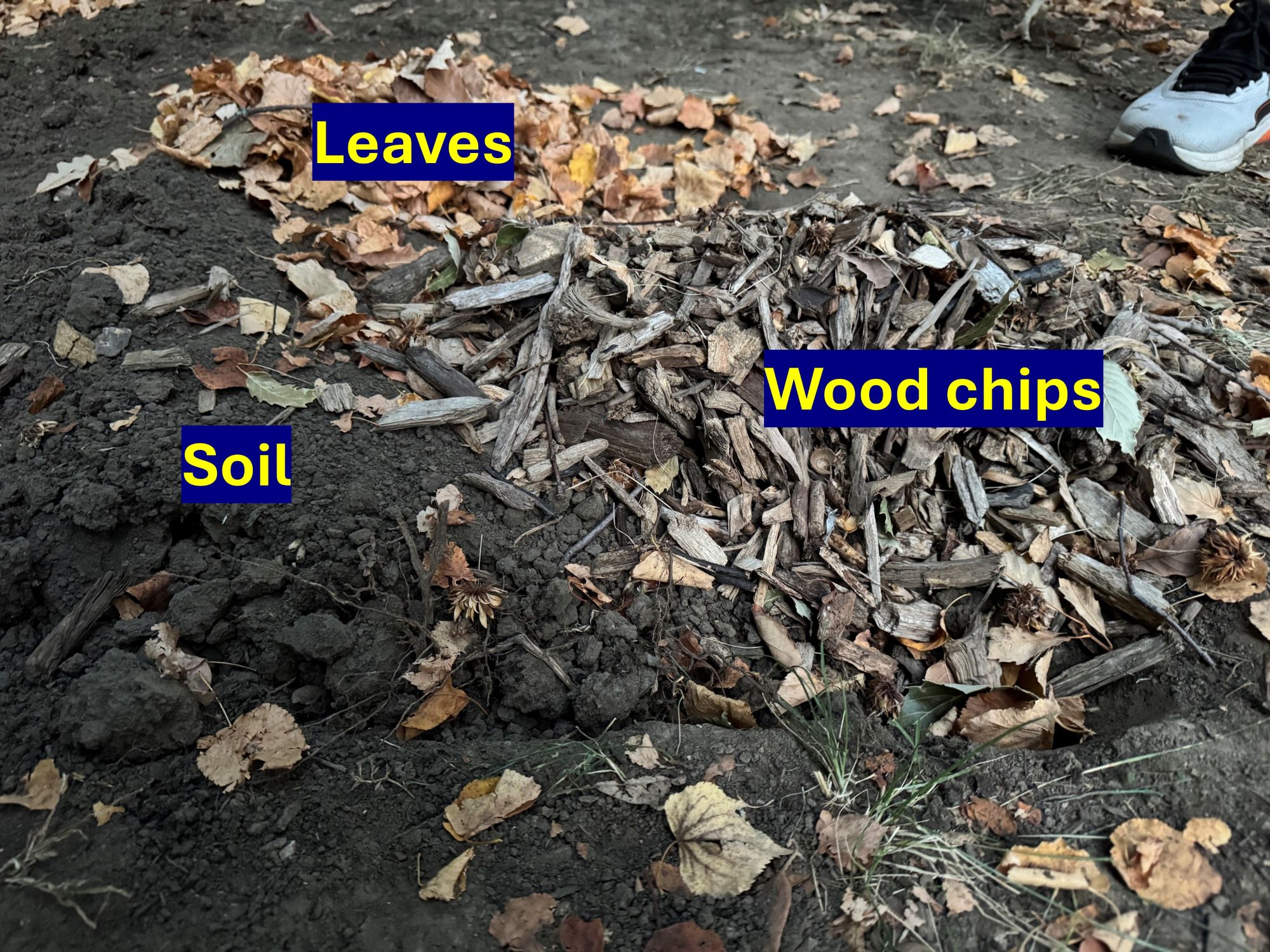}
         \captionsetup{justification=centering}
         \caption{Field test materials.}
         \label{Fig_7a}
     \end{subfigure}
     \begin{subfigure}[t]{0.32\textwidth}
         \centering
         \includegraphics[width=0.9\linewidth]{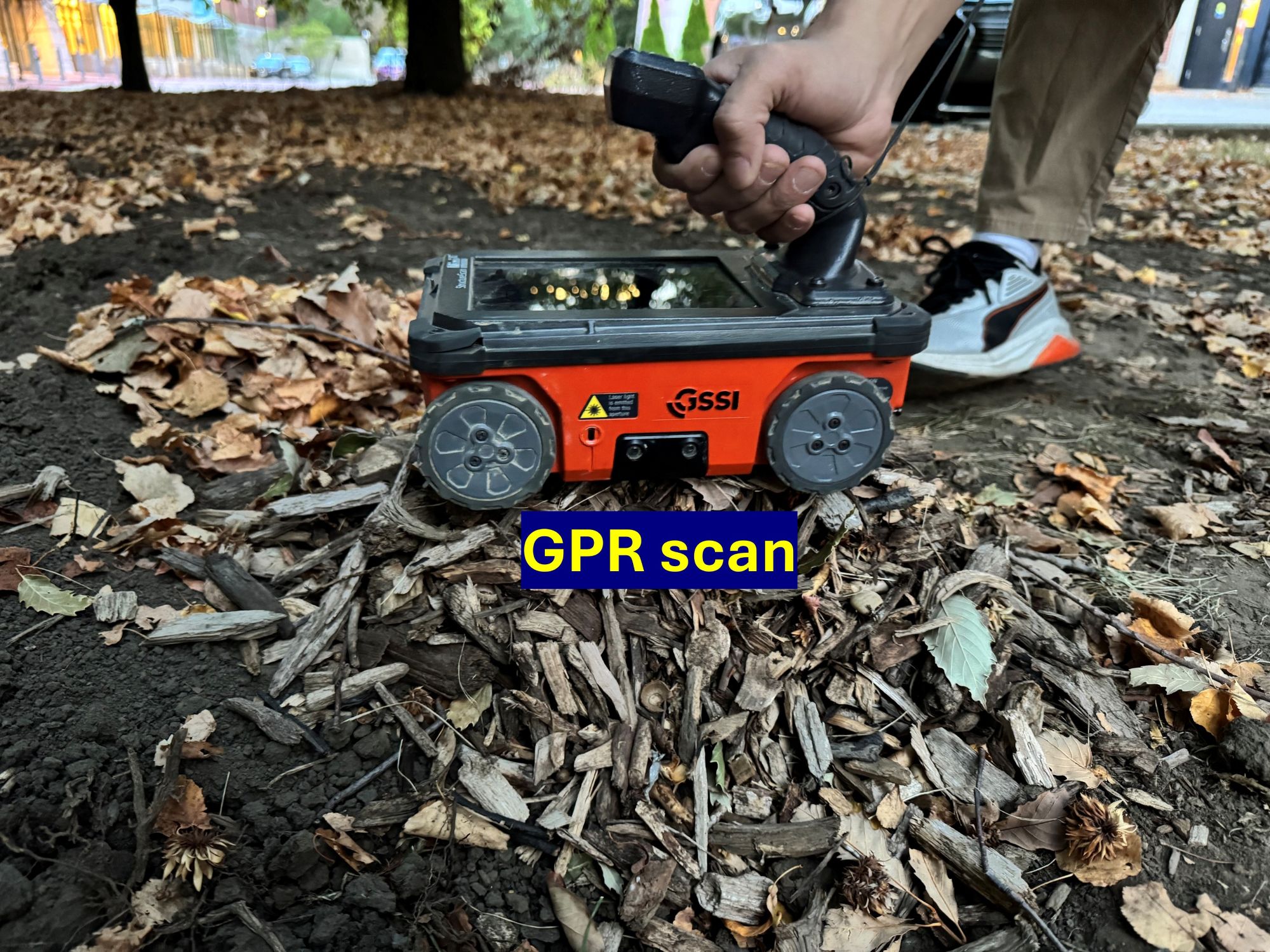}
         \captionsetup{justification=centering}
         \caption{GPR scan.}
         \label{Fig_7b}
     \end{subfigure}
     \begin{subfigure}[t]{0.18\textwidth}
         \centering
         \includegraphics[width=0.9\linewidth]{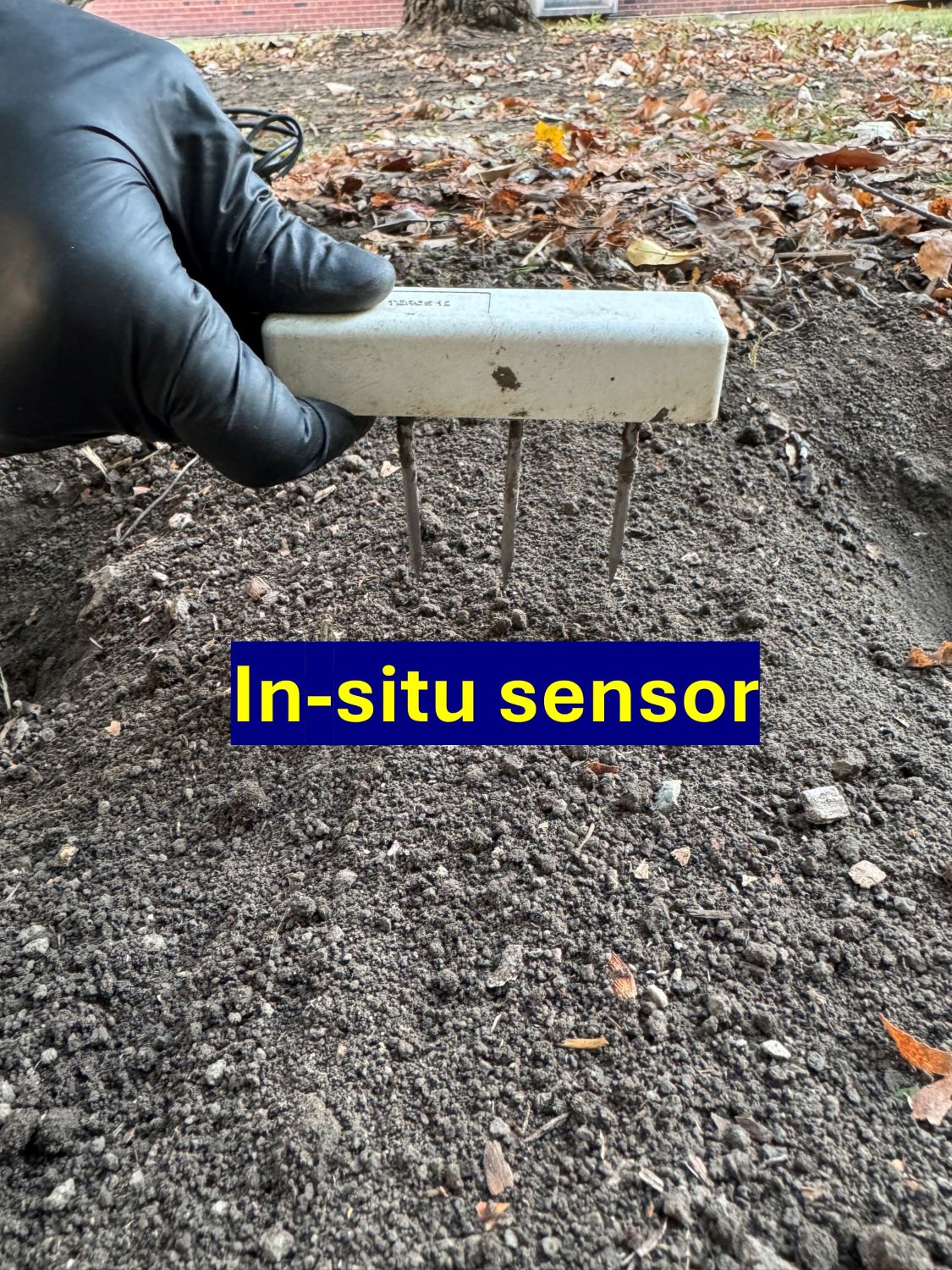}
         \captionsetup{justification=centering}
         \caption{In-situ sensor.}
         \label{Fig_7c}
     \end{subfigure}
     \caption{\centering Field test setup. Three materials were investigated: soil, leaves, and wood chips. A GPR scanner was used to collect radar signals, and an in-situ sensor was employed to measure the permittivity and conductivity of the materials.}
     \label{Fig_7}
\end{figure*}

To validate the capability of the proposed physics-guided hierarchical domain adaptation framework in estimating field soil moisture and tracking its temporal variation, a field test was conducted in a yard outside the Newmark Civil Engineering Laboratory at the University of Illinois Urbana–Champaign. A metal plate was buried approximately 0.1 m below the soil surface before data collection to ensure sufficient radar signal reflection. The single-layer material configuration consisted of bare soil, while the two-layer material configurations were created by adding a 0.1 m-thick layer of either leaves or wood chips on top of the bare soil. GPR scans were collected over the bare soil, leaves, and wood chips for 11 consecutive days, with 100 scans acquired per day for each material configuration. Figure \ref{Fig_7} demonstrates the field test setup. Table \ref{T5} presents a summary of the collected field test data for the single-layer material (soil) and two-layer materials (soil–leaves and soil–wood chips). Furthermore, for the single-layer material (soil), a simulated dataset comprising 10,092 scans was generated using the FDTD simulation by varying the material parameters as follows: soil permittivity ($p_1 = 3.5-15$, step size 0.1), soil conductivity ($c_1 = 0-0.14$ S/m, step size 0.005 S/m), soil depth ($d_1 = 0.1-0.2$ m, step size 0.05 m). For the two-layer materials, another simulated dataset comprising 77,760 scans was generated by varying the material parameters as follows: soil permittivity ($p_1 = 3.5-15$, step size 0.5), soil conductivity ($c_1 = 0-0.14$ S/m, step size 0.01 S/m), soil depth ($d_1 = 0.1-0.2$ m, step size 0.05 m), upper-layer permittivity ($p_2 = 1.0-3.5$, step size 0.5), upper-layer conductivity ($c_2 = 0-0.03$ S/m, step size 0.01 S/m), and upper-layer depth ($d_2 = 0.1-0.2$ m, step size 0.05 m). The simulation parameters were selected to span the range of realizations anticipated under field testing conditions.

\begin{table*}[ht]
    \small\sf\centering
    \caption{\centering Field test data collected for the single-layer material (soil) and two-layer materials (soil–leaves and soil–wood chips).}
    \label{T5}
    \begin{threeparttable}
        \begin{tabularx}{1\textwidth}{Y *{9}{Y}}
            \toprule
            Date & $p_1$ & $c_1$ (S/m) & $d_1$ (m) & $p_{2}^{(l)}$ & $c_{2}^{(l)}$ (S/m) & $d_{2}^{(l)}$ (m) & $p_{2}^{(w)}$ & $c_{2}^{(w)}$(S/m) & $d_{2}^{(w)}$ (m) \\
            \cmidrule(lr){1-10}
            Sept.\ 21 & 4.2  & 0.000 & 0.1 & 1.4 & 0.000 & 0.1 & 1.7 & 0.000 & 0.1 \\
            Sept.\ 22 & 4.0  & 0.000 & 0.1 & 1.3 & 0.000 & 0.1 & 1.7 & 0.000 & 0.1 \\
            Sept.\ 23 & 3.9  & 0.000 & 0.1 & 1.3 & 0.000 & 0.1 & 1.6 & 0.000 & 0.1 \\
            Sept.\ 24 & 14.3 & 0.117 & 0.1 & 2.7 & 0.027 & 0.1 & 3.0 & 0.000 & 0.1 \\
            Sept.\ 25 & 11.3 & 0.088 & 0.1 & 1.3 & 0.000 & 0.1 & 2.0 & 0.000 & 0.1 \\
            Sept.\ 26 & 10.9 & 0.073 & 0.1 & 1.4 & 0.000 & 0.1 & 2.4 & 0.000 & 0.1 \\
            Sept.\ 27 & 9.3  & 0.061 & 0.1 & 1.3 & 0.000 & 0.1 & 1.7 & 0.000 & 0.1 \\
            Sept.\ 28 & 8.3  & 0.057 & 0.1 & 1.4 & 0.000 & 0.1 & 1.6 & 0.000 & 0.1 \\
            Sept.\ 29 & 8.4  & 0.064 & 0.1 & 1.3 & 0.000 & 0.1 & 1.6 & 0.000 & 0.1 \\
            Sept.\ 30 & 8.0  & 0.056 & 0.1 & 1.4 & 0.000 & 0.1 & 1.6 & 0.000 & 0.1 \\
            Oct.\ 1   & 5.2  & 0.027 & 0.1 & 1.3 & 0.000 & 0.1 & 1.6 & 0.000 & 0.1 \\
            \bottomrule
        \end{tabularx}
        \begin{tablenotes}
            \item[1] $p_1$, $c_1$, and $d_1$ denote the permittivity, conductivity, and depth of the soil, respectively. 
            \item[2] $p_{2}^{(l)}$, $c_{2}^{(l)}$, and $d_{2}^{(l)}$ denote the permittivity, conductivity, and depth of the leaves, respectively.
            \item[3] $p_{2}^{(w)}$, $c_{2}^{(w)}$, and $d_{2}^{(w)}$ denote the permittivity, conductivity, and depth of the wood chips, respectively.
        \end{tablenotes}
    \end{threeparttable}
\end{table*}

\section{Results and Discussion} \label{Results_and_Discussion}
\subsection{Laboratory Test for Single-Layer Material}
The proposed physics-guided hierarchical domain adaptation approaches, HierDANN, PhyDANN-1, HierPhyDANN-1, PhyDANN-2, and HierPhyDANN-2, were initially investigated on a single-layer material under laboratory conditions. Its performance was benchmarked against two baseline methods, namely the 1D CNN and DANN. As illustrated in Section \ref{Methodology}, the 1D CNN was trained on labeled simulated data and tested on experimental data (Specimens \#2–\#6), whereas the DANN and the proposed DANN variants were trained on labeled simulated data together with unlabeled experimental data (Specimens \#1–\#6) to predict the labels of the experimental scans (Specimens \#2–\#6). It should be noted that the GPR scans for Specimen \#1 were collected over a significantly different duration compared to the other specimens; therefore, Specimen \#1 was excluded from the evaluation. The sim-to-real problem is explored by performing three-material property retrievals (permittivity, conductivity, and depth) on the single-layer material (soil).

\begin{figure*}[ht]
     \centering
     \begin{subfigure}[t]{0.32\textwidth}
         \centering
         \includegraphics[width=1\linewidth]{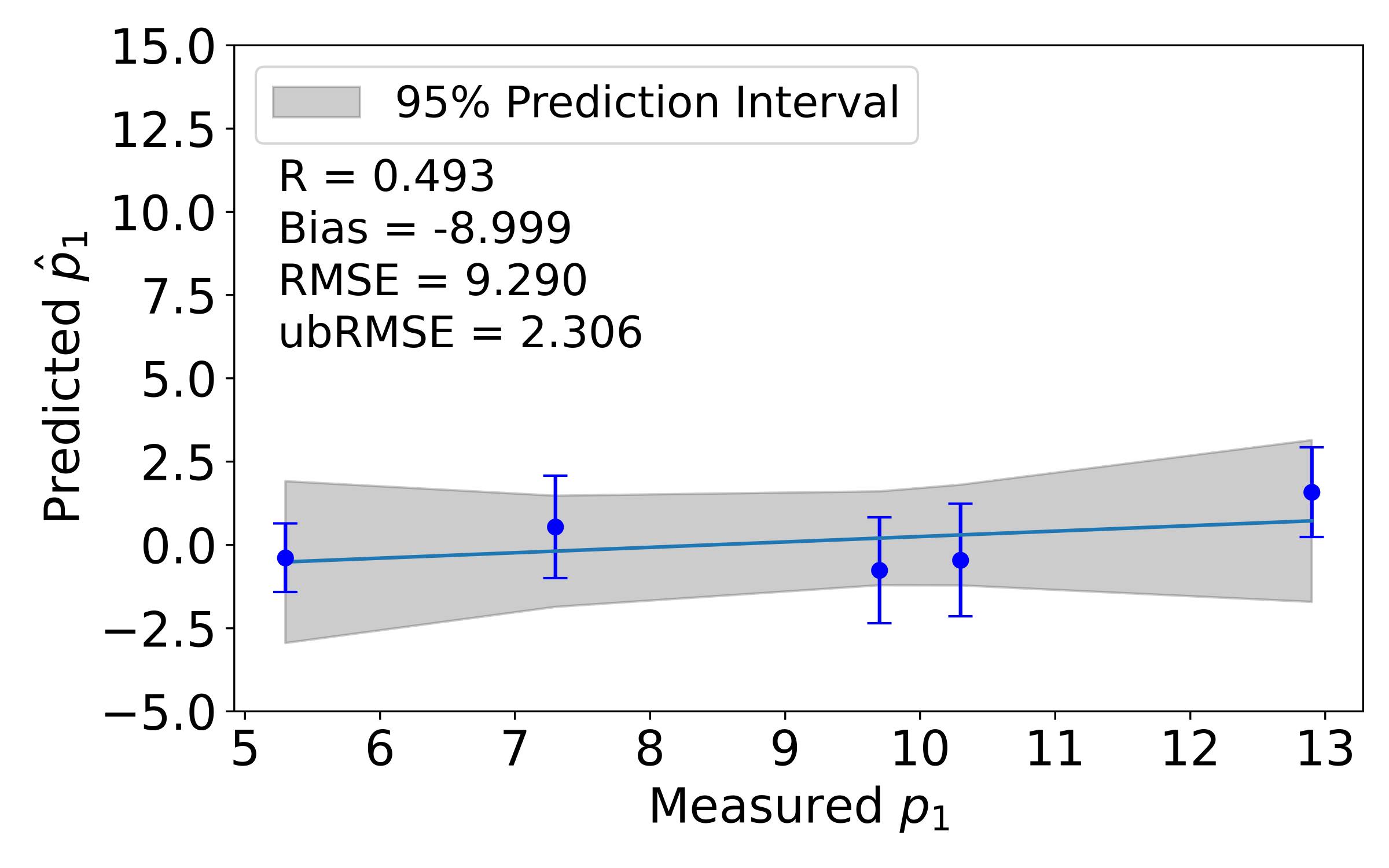}
         \captionsetup{justification=centering}
         \caption{Permittivity estimated by 1D CNN.}
         \label{Fig_10a}
     \end{subfigure}
     \begin{subfigure}[t]{0.32\textwidth}
         \centering
         \includegraphics[width=1\linewidth]{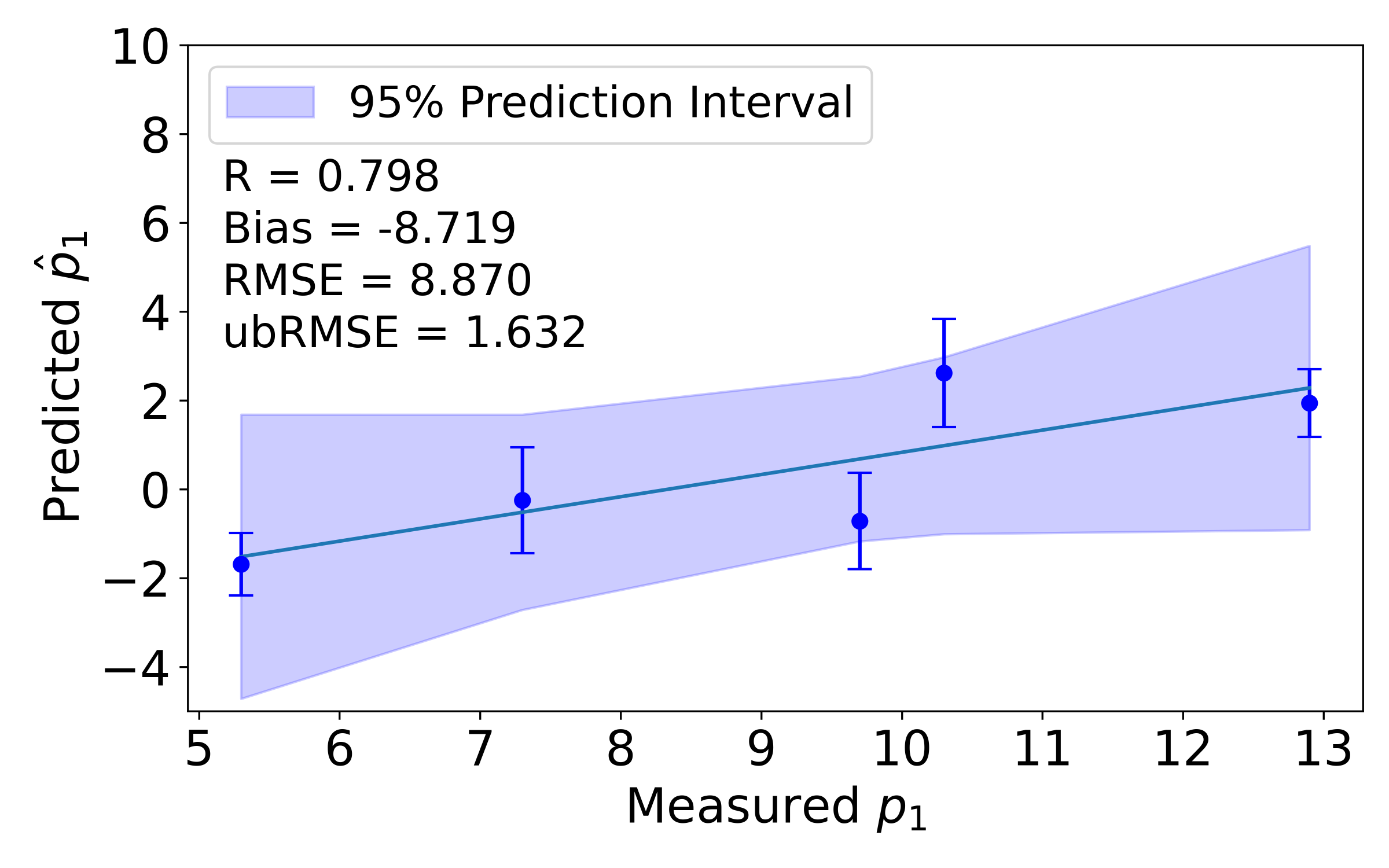}
         \captionsetup{justification=centering}
         \caption{Permittivity estimated by DANN.}
         \label{Fig_10b}
     \end{subfigure}
     \begin{subfigure}[t]{0.32\textwidth}
         \centering
         \includegraphics[width=1\linewidth]{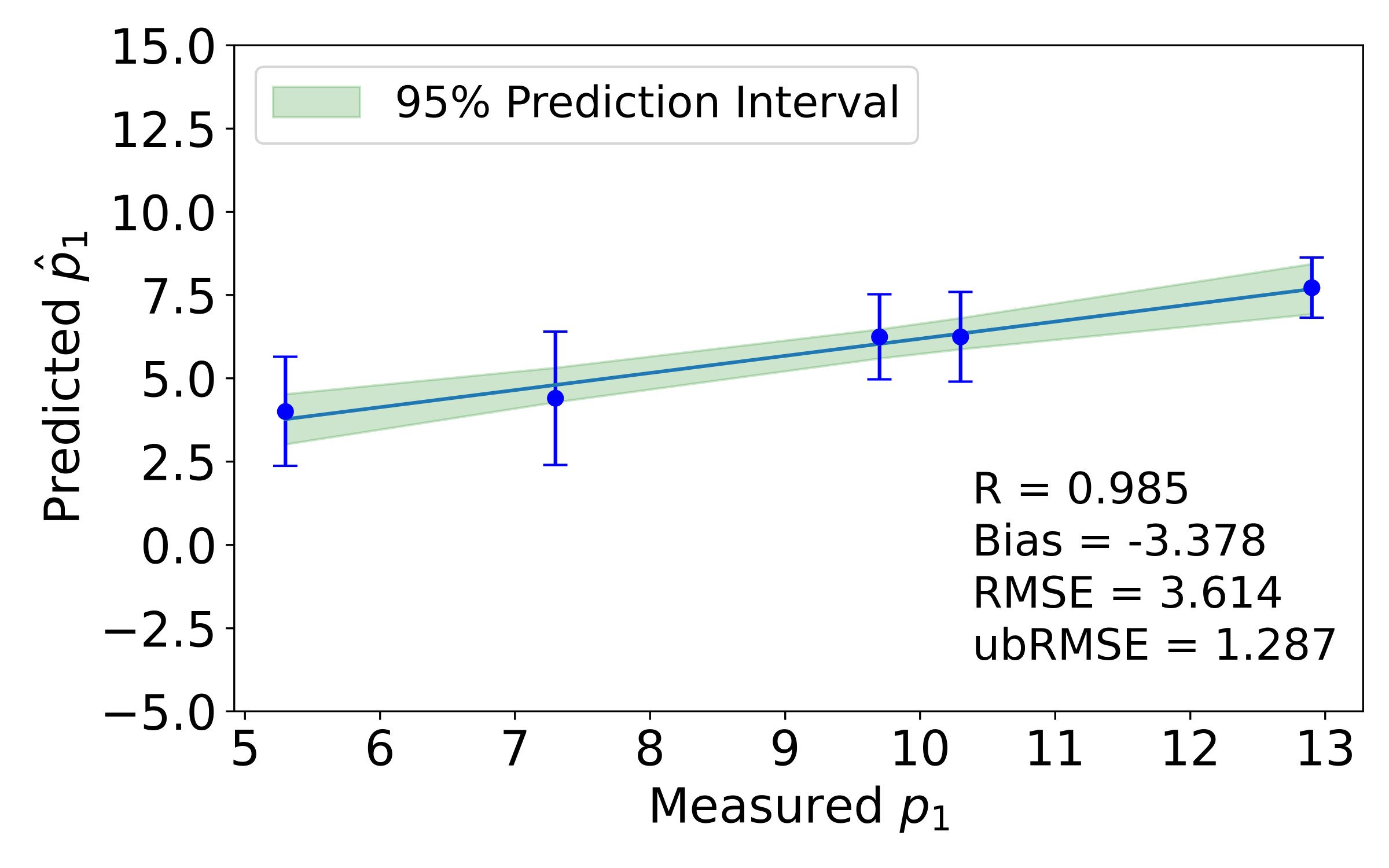}
         \captionsetup{justification=centering}
         \caption{Permittivity estimated by HierPhyDANN-2.}
         \label{Fig_10c}
     \end{subfigure}
     \begin{subfigure}[t]{0.32\textwidth}
         \centering
         \includegraphics[width=1\linewidth]{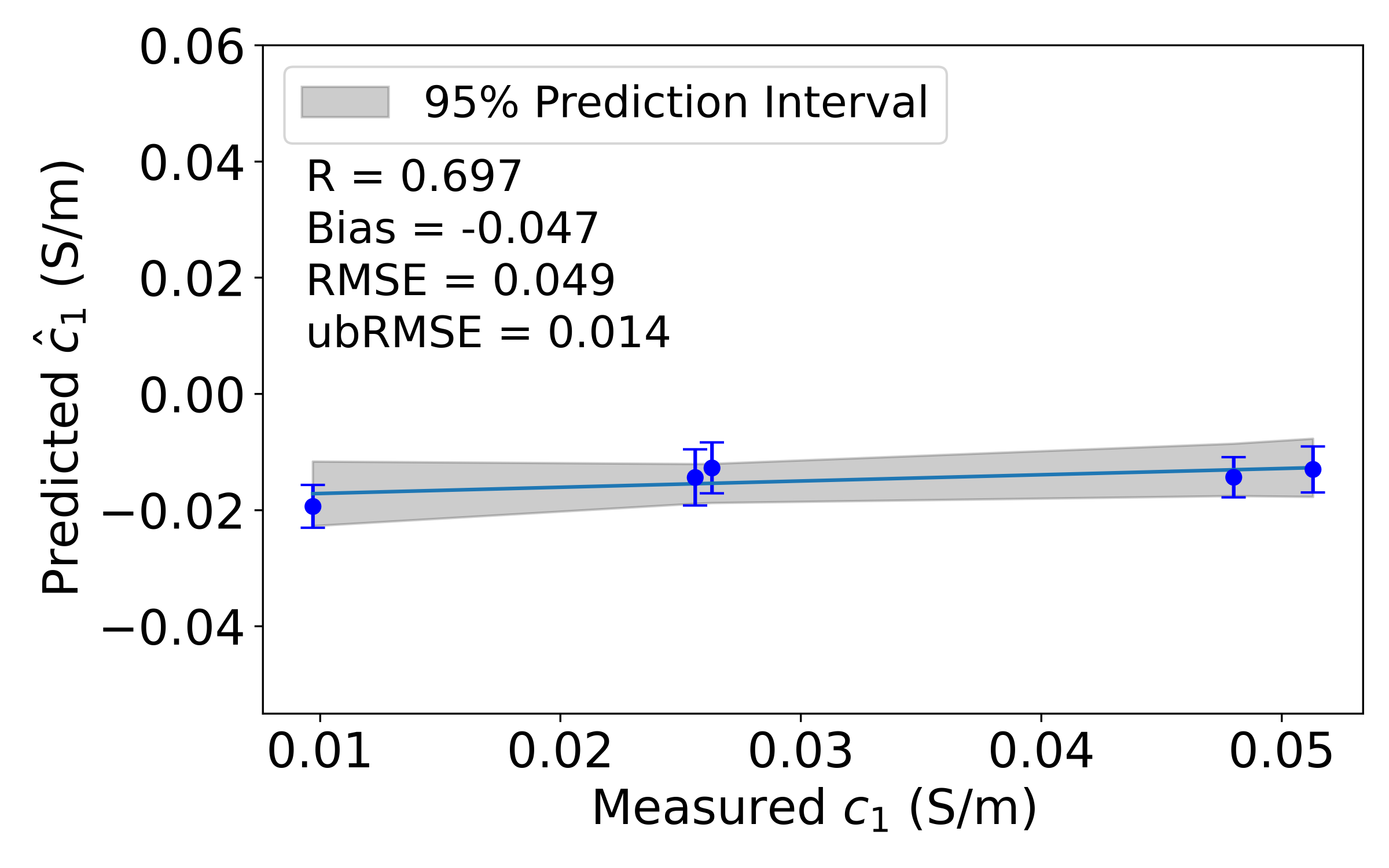}
         \captionsetup{justification=centering}
         \caption{Conductivity estimated by 1D CNN.}
         \label{Fig_10d}
     \end{subfigure}
     \begin{subfigure}[t]{0.32\textwidth}
         \centering
         \includegraphics[width=1\linewidth]{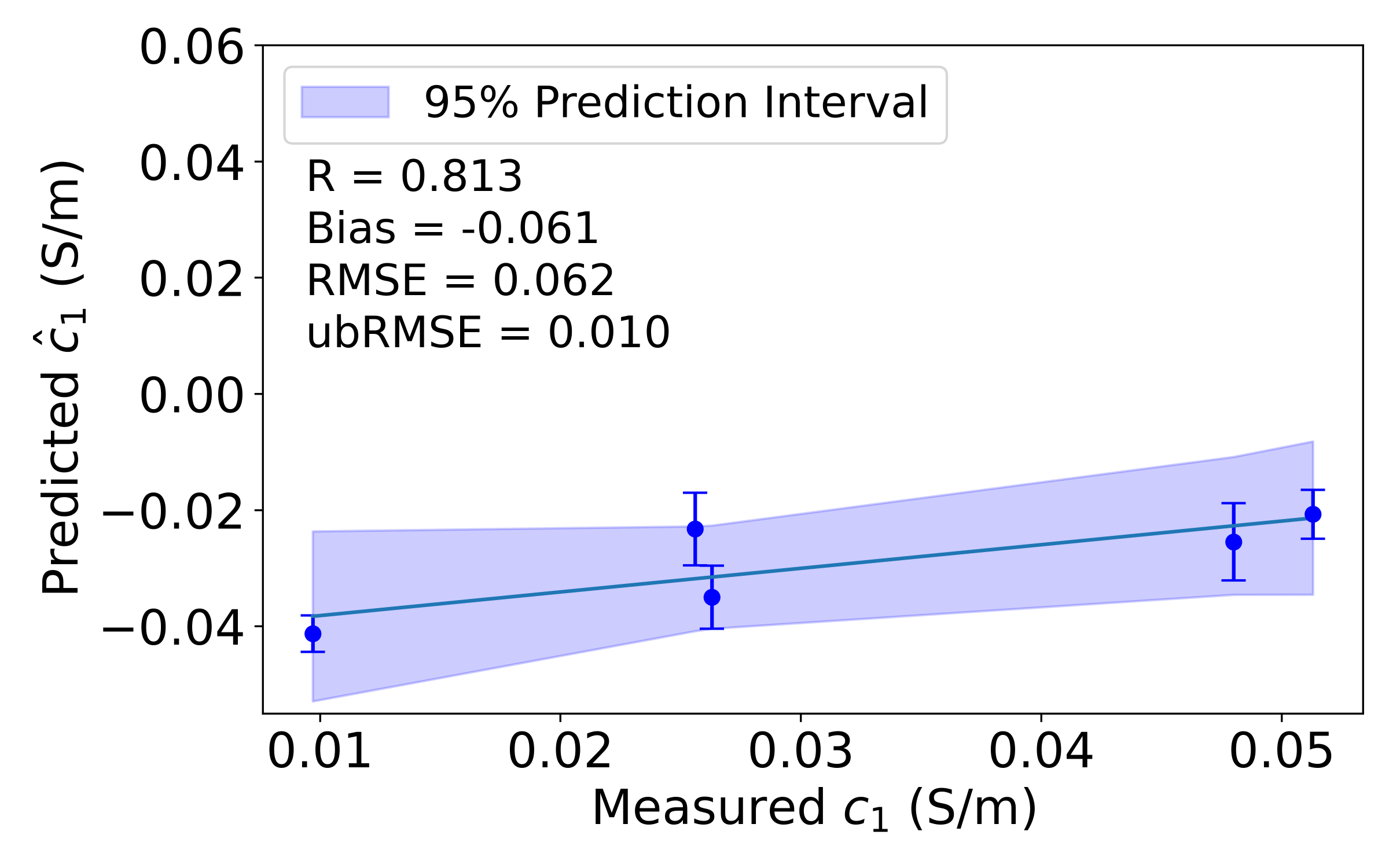}
         \captionsetup{justification=centering}
         \caption{Conductivity estimated by DANN.}
         \label{Fig_10e}
     \end{subfigure}
     \begin{subfigure}[t]{0.32\textwidth}
         \centering
         \includegraphics[width=1\linewidth]{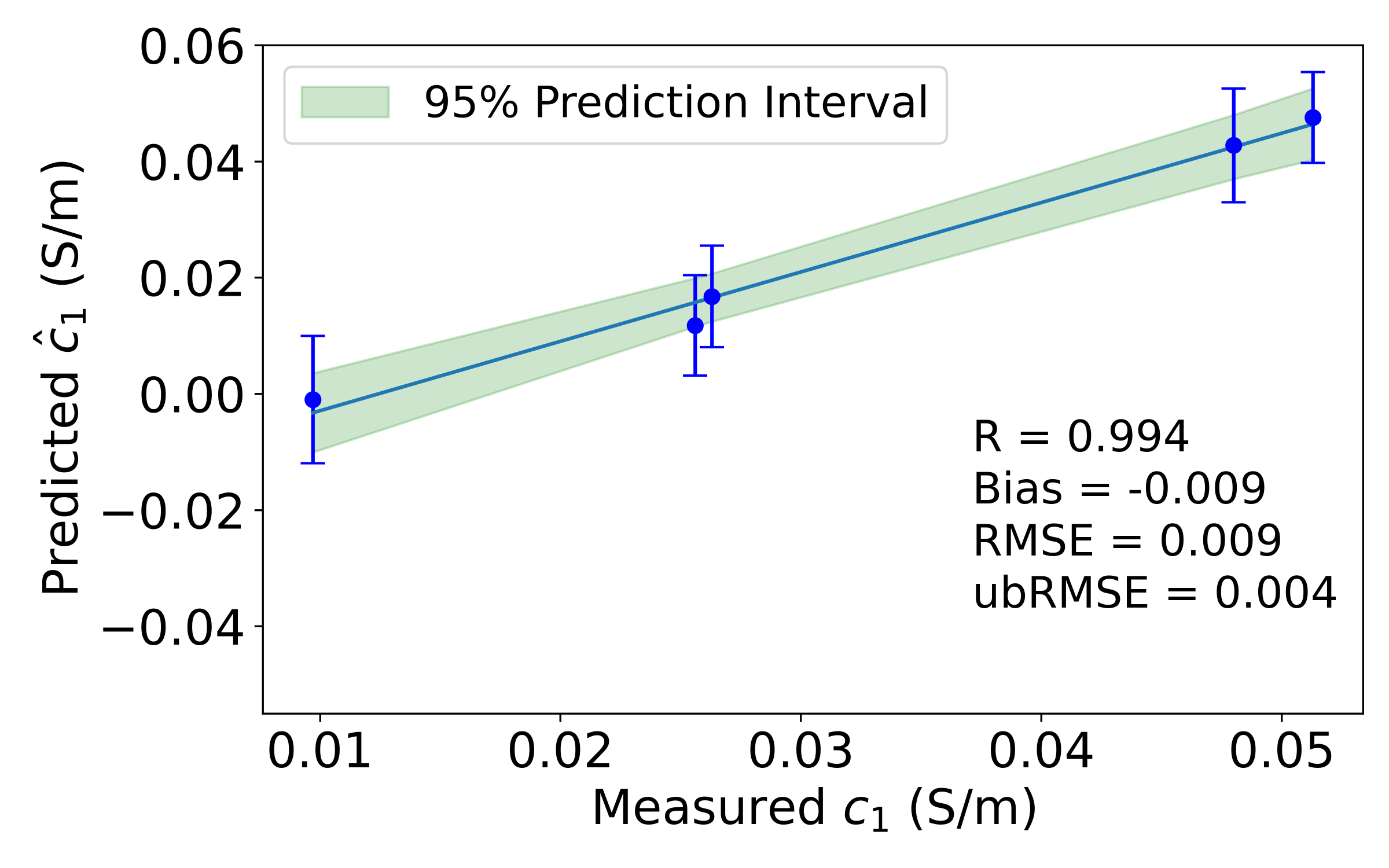}
         \captionsetup{justification=centering}
         \caption{Conductivity estimated by HierPhyDANN-2.}
         \label{Fig_10f}
     \end{subfigure}
     \begin{subfigure}[t]{0.32\textwidth}
         \centering
         \includegraphics[width=1\linewidth]{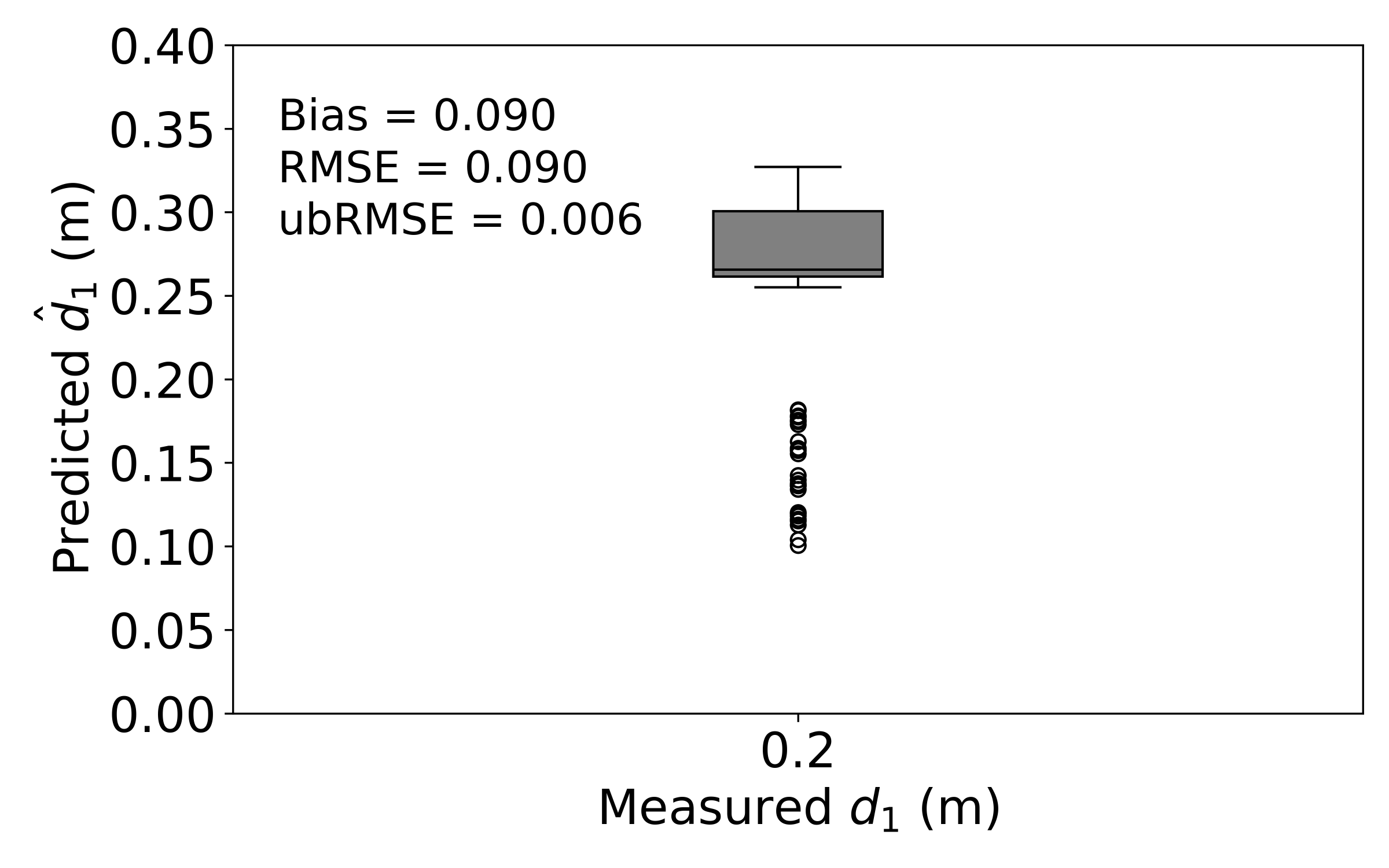}
         \captionsetup{justification=centering}
         \caption{Depth estimated by 1D CNN.}
         \label{Fig_10g}
     \end{subfigure}
     \begin{subfigure}[t]{0.32\textwidth}
         \centering
         \includegraphics[width=1\linewidth]{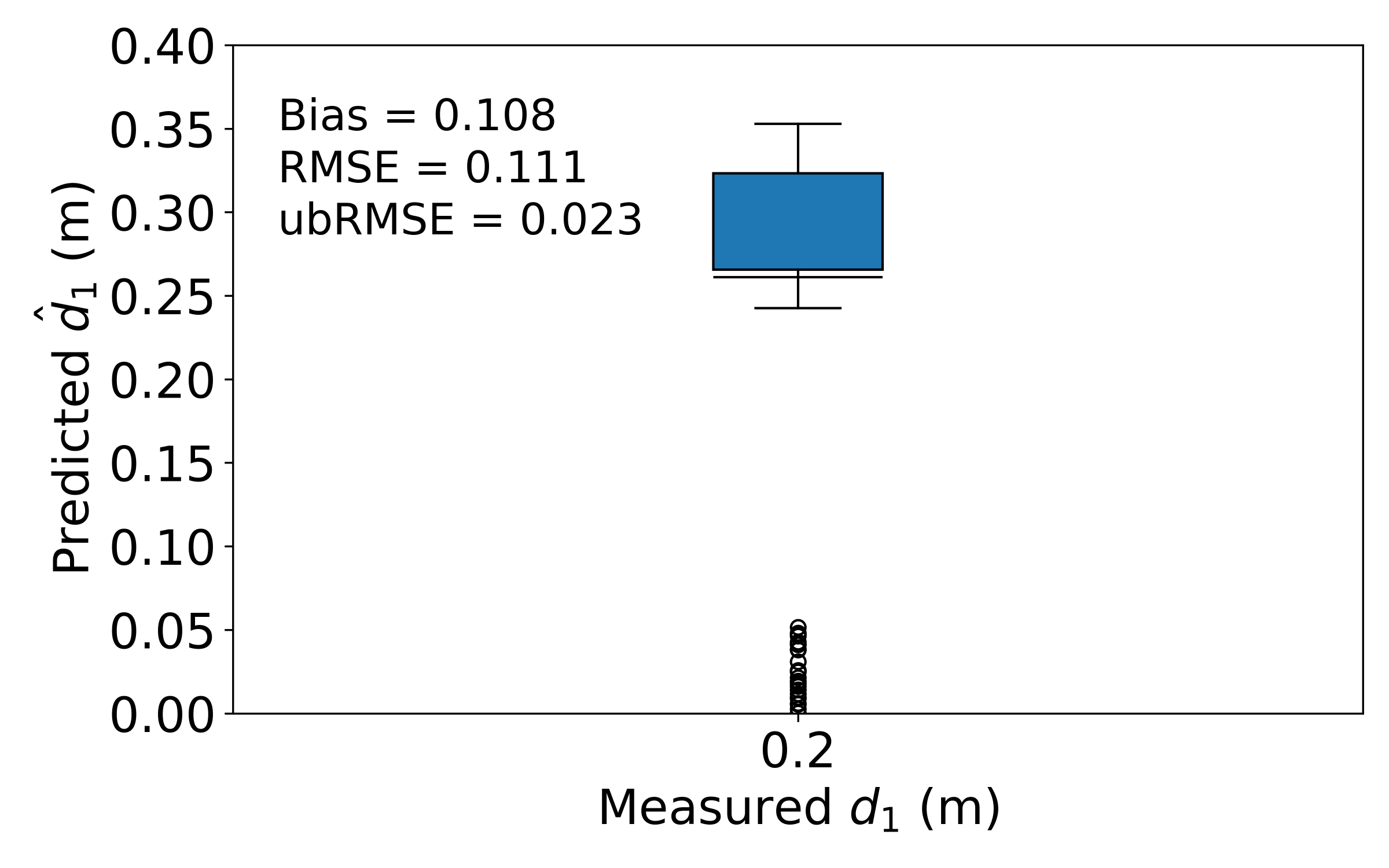}
         \captionsetup{justification=centering}
         \caption{Depth estimated by DANN.}
         \label{Fig_10h}
     \end{subfigure}
     \begin{subfigure}[t]{0.32\textwidth}
         \centering
         \includegraphics[width=1\linewidth]{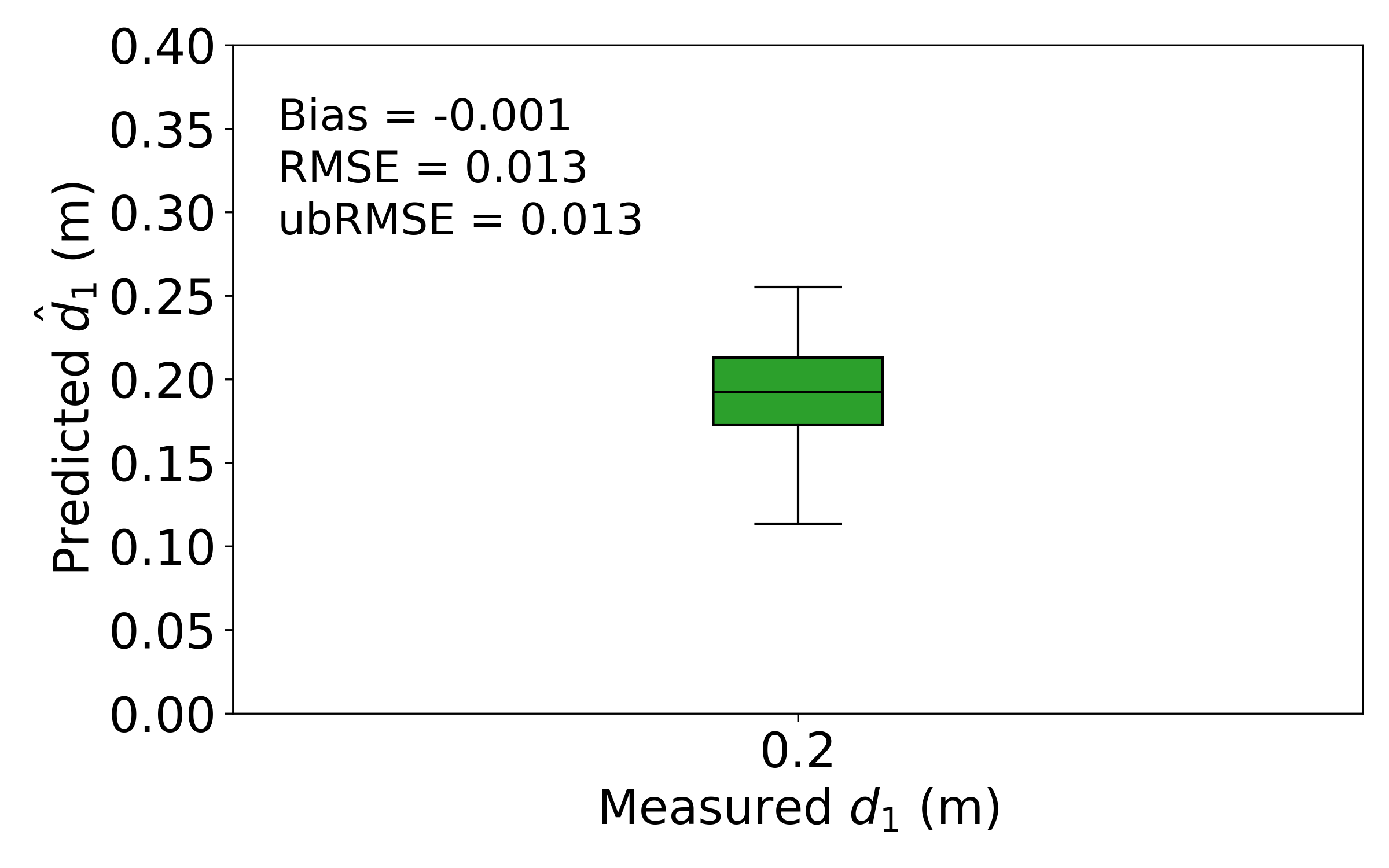}
         \captionsetup{justification=centering}
         \caption{Depth estimated by HierDANN.}
         \label{Fig_10i}
     \end{subfigure}
     \caption{\centering Single-layer three-parameter (permittivity, conductivity, and depth) estimation.}
     \label{Fig_10}
\end{figure*}

Table \ref{T8} and Figure \ref{Fig_10} present the performance comparison among the different approaches. The 1D CNN and DANN estimate all three parameters simultaneously by setting the number of neurons in the output layer to three, whereas HierDANN, HierPhyDANN-1, and HierPhyDANN-2 estimate permittivity, conductivity, and depth sequentially. For example, during conductivity estimation with HierDANN, the estimated permittivity values ($\hat{p}_1 = 1.6, 5.1, 6.3, 9.4, 11.6$) are known, and the simulated dataset is pruned to retain only the radar scans corresponding to these permittivity values. Similarly, the estimated conductivity values ($\hat{c}_1 = 0.000, 0.000, 0.003, 0.020, 0.010~\text{S/m}$) are then used to further refine the simulated dataset that has already been updated using the estimated permittivity values, for subsequent depth estimation. It should be noted that when the conductivity predicted by HierDANN, HierPhyDANN-1, and HierPhyDANN-2 is negative, a conductivity value of zero is assigned.

\begin{table*}[ht]
    \small\sf\centering
    \caption{\centering Laboratory test results of the one-layer, three-parameter estimation.}
    \label{T8}
    \resizebox{\textwidth}{!}{%
    \begin{threeparttable}
    \begin{tabular}{@{\extracolsep{\fill}}lccccccccccc}
        \toprule
        \multirow{2}{*}{Approach}
            & \multicolumn{4}{c}{$p_1$}
            & \multicolumn{4}{c}{$c_1$}
            & \multicolumn{3}{c}{$d_1$} \\
        \cmidrule(lr){2-5} \cmidrule(lr){6-9} \cmidrule(lr){10-12}
            & $R$ & Bias & RMSE & ubRMSE
            & $R$ & Bias & RMSE & ubRMSE
            & Bias & RMSE & ubRMSE \\
        \midrule
        1D CNN (baseline)        & 0.493 & -8.999 & 9.290 & 2.306 & 0.697 & -0.047 & 0.049 & 0.014 & 0.090 & 0.090 & 0.006 \\
        DANN (baseline)          & 0.798 & -8.719 & 8.870 & 1.632 & 0.813 & -0.061 & 0.062 & 0.010 & 0.108 & 0.111 & 0.023 \\
        HierDANN (ours)          & 0.973 & -2.288 & 2.540 & 1.104 & 0.938 & -0.027 & 0.028 & 0.007 & -0.001 & 0.013 & 0.013 \\
        PhyDANN-1 (ours)         & 0.966 & -9.408 & 9.434 & 0.693 & 0.863 & -0.059 & 0.059 & 0.008 & 0.133 & 0.139 & 0.040 \\
        HierPhyDANN-1 (ours)     & 0.961 & -4.473 & 4.546 & 0.814 & 0.985 & -0.066 & 0.066 & 0.003 & 0.018 & 0.037 & 0.032 \\
        PhyDANN-2 (ours)         & 0.868 & -2.511 & 2.844 & 1.334 & 0.304 & -0.043 & 0.046 & 0.016 & 0.078 & 0.081 & 0.019 \\
        HierPhyDANN-2 (ours)     & \textit{\textbf{\underline{0.985}}} & -3.378 & 3.614 & 1.287
                                  & \textit{\textbf{\underline{0.994}}} & -0.009 & 0.009 & 0.004
                                  & 0.010 & 0.030 & 0.029 \\
        \bottomrule
    \end{tabular}
    \begin{tablenotes}
        \item[1] The bold values indicate the best correlation coefficient $R$ among the compared approaches.
    \end{tablenotes}
    \end{threeparttable}
    }
\end{table*}

The results show that, in the absence of domain adaptation, the 1D CNN exhibits a weaker correlation between the predicted and measured values. Bridging the domain gap with DANN improves the correlation coefficient $R$ for both permittivity and conductivity. The proposed approaches, HierDANN, PhyDANN-1, HierPhyDANN-1, PhyDANN-2, and HierPhyDANN-2, outperform the baseline methods, 1D CNN and DANN. Specifically, by incorporating physical constraints into adversarial learning via signal reconstruction, PhyDANN-1 and PhyDANN-2 further improve the $R$ values compared with the baseline 1D CNN and DANN. By hierarchically performing domain adaptation, HierDANN, HierPhyDANN-1, and HierPhyDANN-2 generally further improve material property estimation relative to their non-hierarchical counterparts, achieving higher $R$ values and lower Bias, RMSE, and ubRMSE in most cases across all three parameters. In particular, HierPhyDANN-2 achieves the best correlation $R$ for both permittivity and conductivity estimation. As shown in Table \ref{T3}, the depth of the single-layer material is fixed at 0.2 m. The box plots of the depth estimates from all scans obtained by the 1D CNN, DANN, and HierDANN are presented in Figures \ref{Fig_10g}, \ref{Fig_10h}, and \ref{Fig_10i}, respectively. Leveraging the permittivity and conductivity estimates obtained in the earlier stages, HierDANN yields the most accurate depth estimation, exhibiting the lowest Bias and standard deviation among the compared approaches.

\begin{figure*}[ht]
     \centering
     \begin{subfigure}[t]{0.4\textwidth}
         \centering
         \includegraphics[width=1\linewidth]{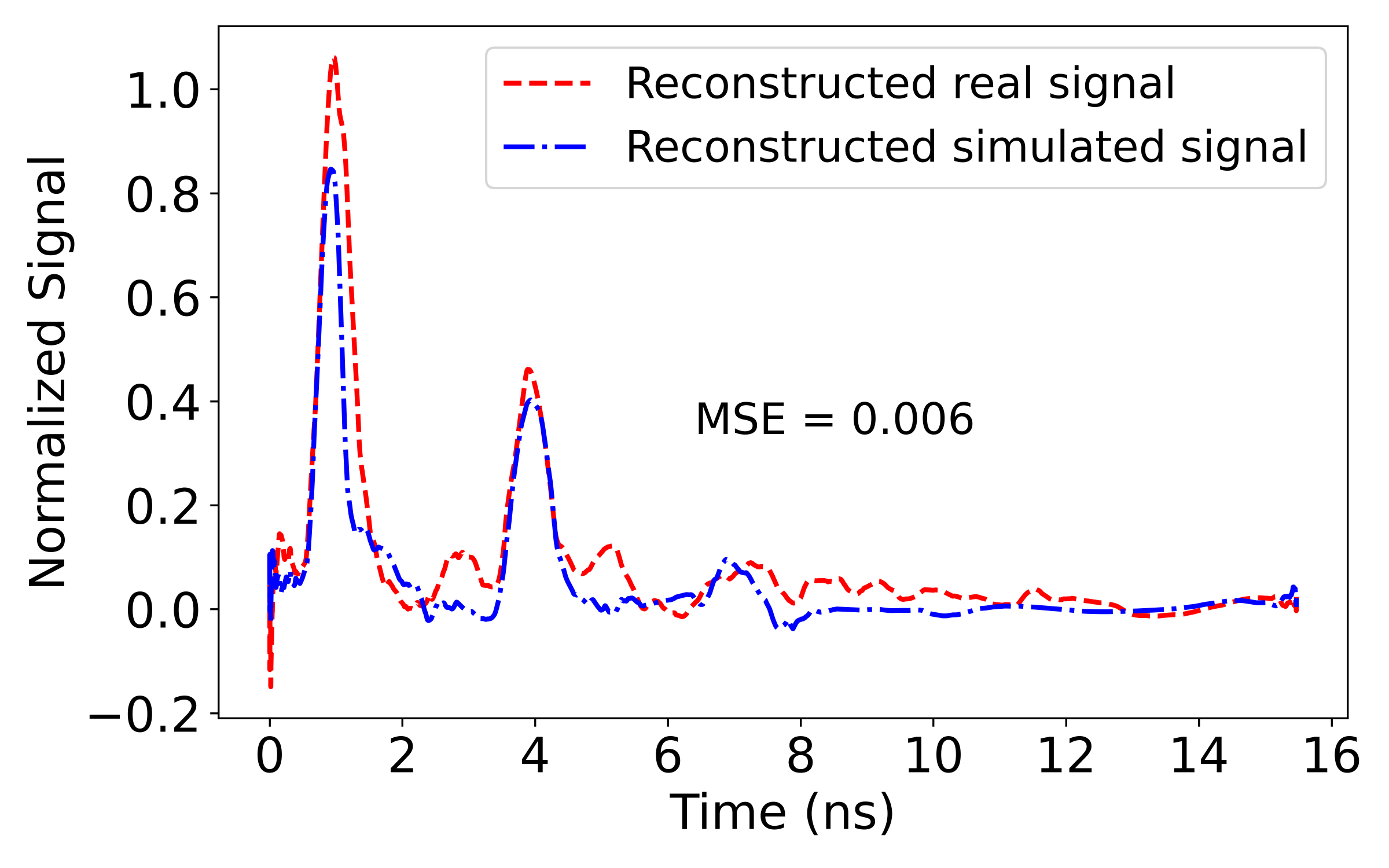}
         \captionsetup{justification=centering}
         \caption{Without domain adaptation.}
         \label{Fig_18a}
     \end{subfigure}
     \begin{subfigure}[t]{0.4\textwidth}
         \centering
         \includegraphics[width=1\linewidth]{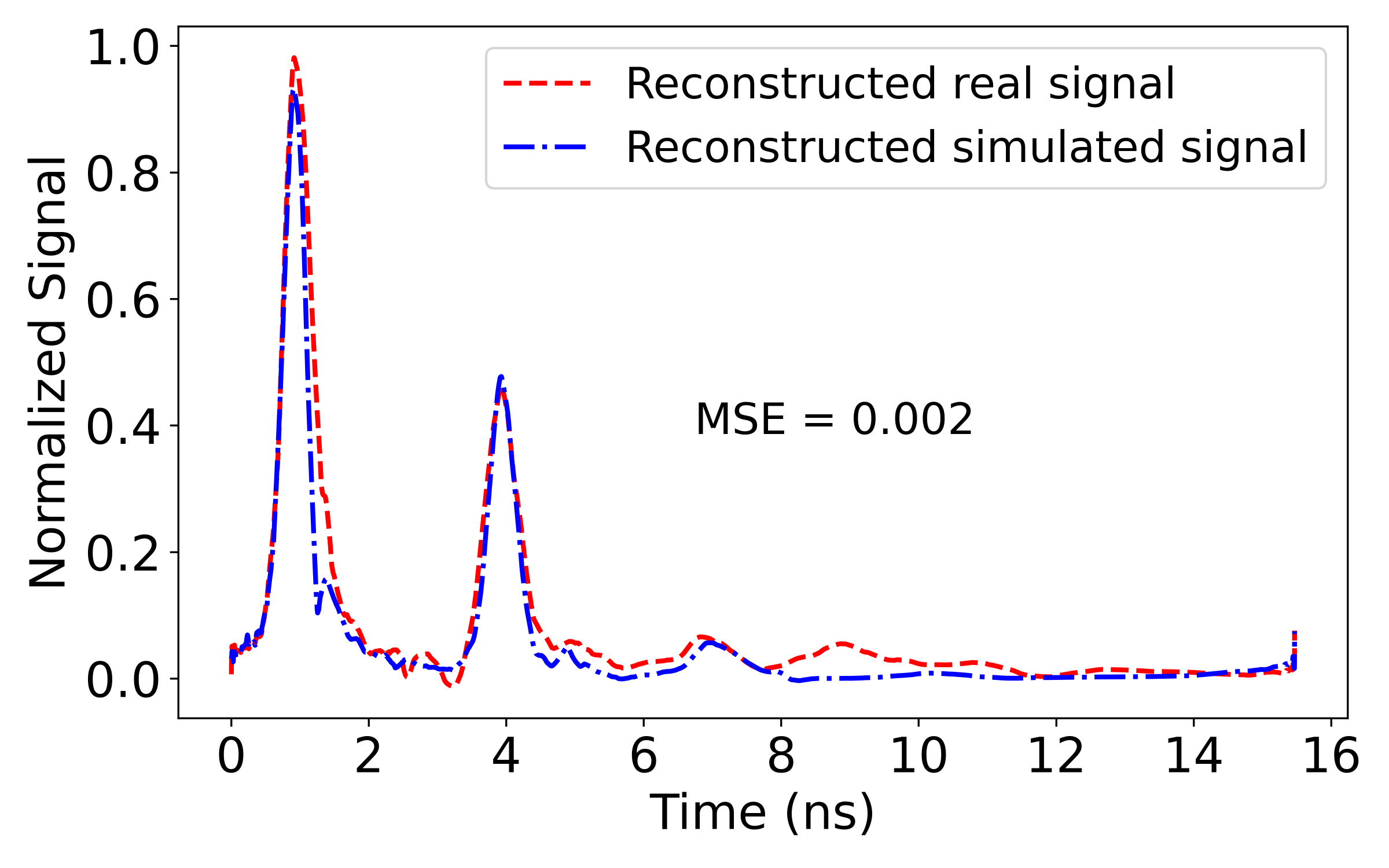}
         \captionsetup{justification=centering}
         \caption{HierPhyDANN-2.}
         \label{Fig_18b}
     \end{subfigure}
     \begin{subfigure}[t]{0.4\textwidth}
         \centering
         \includegraphics[width=1\linewidth]{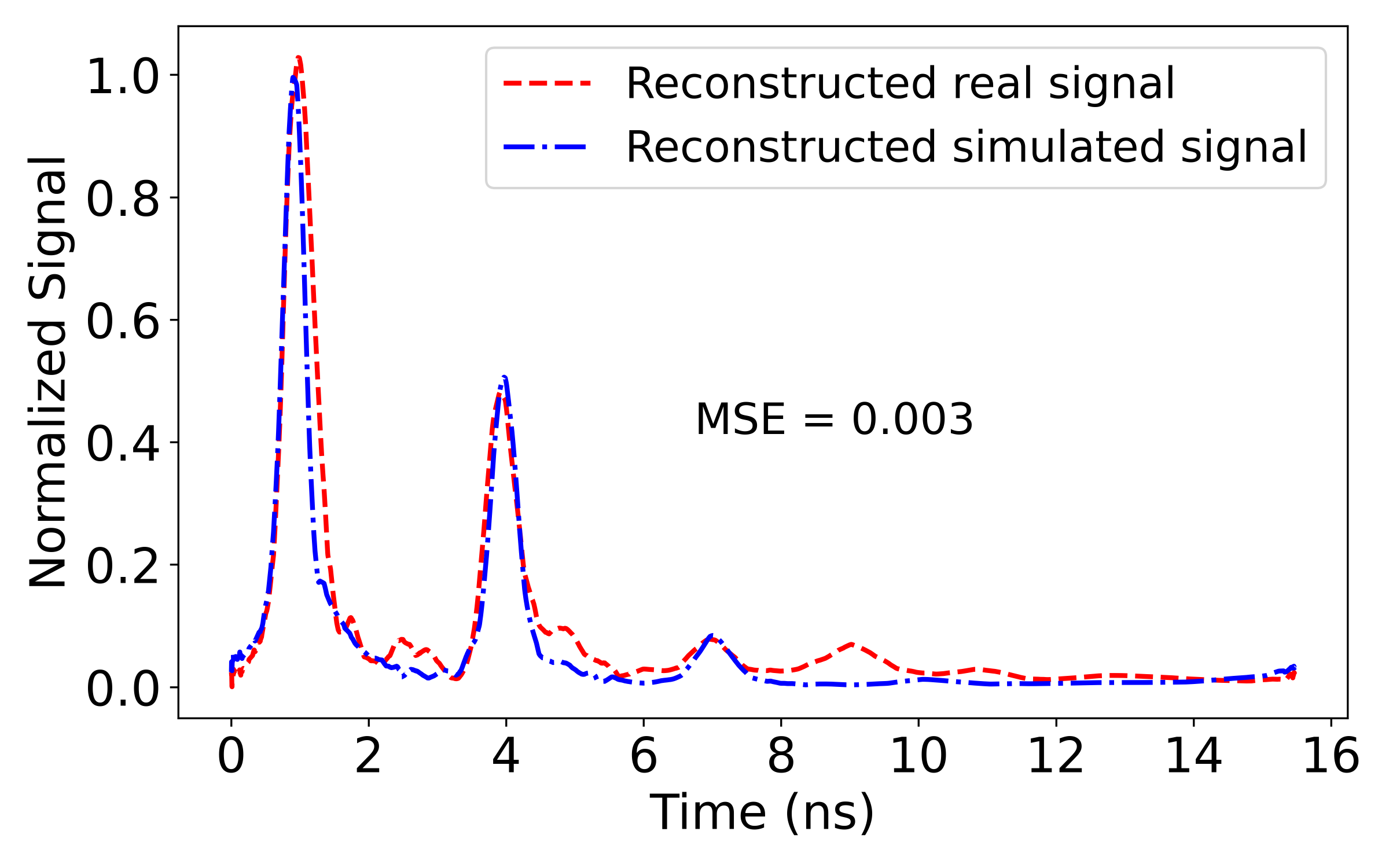}
         \captionsetup{justification=centering}
         \caption{HierPhyDANN-1.}
         \label{Fig_18c}
     \end{subfigure}
     \begin{subfigure}[t]{0.4\textwidth}
         \centering
         \includegraphics[width=1\linewidth]{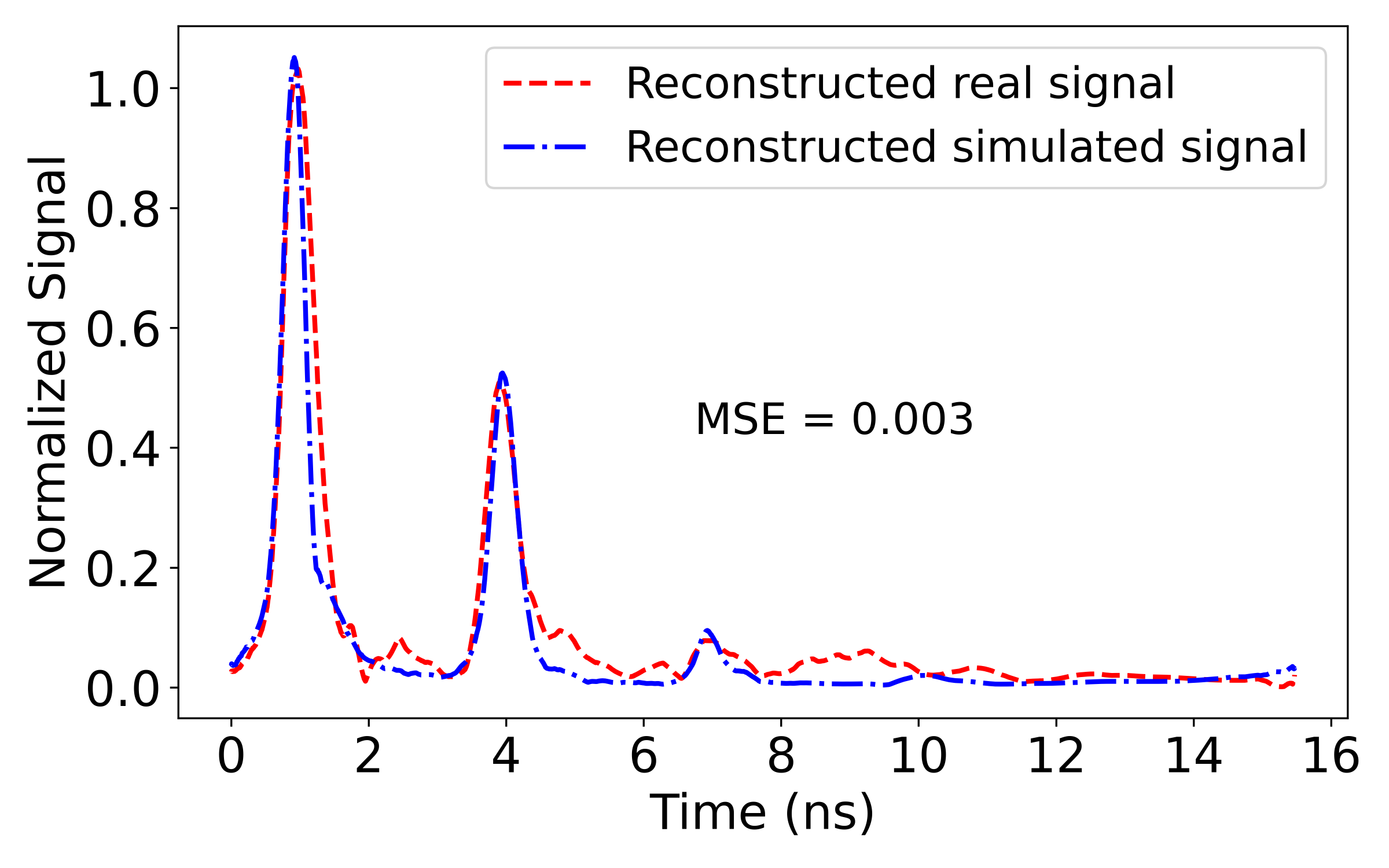}
         \captionsetup{justification=centering}
         \caption{PhyDANN-1.}
         \label{Fig_18d}
     \end{subfigure}
     \begin{subfigure}[t]{0.4\textwidth}
         \centering
         \includegraphics[width=1\linewidth]{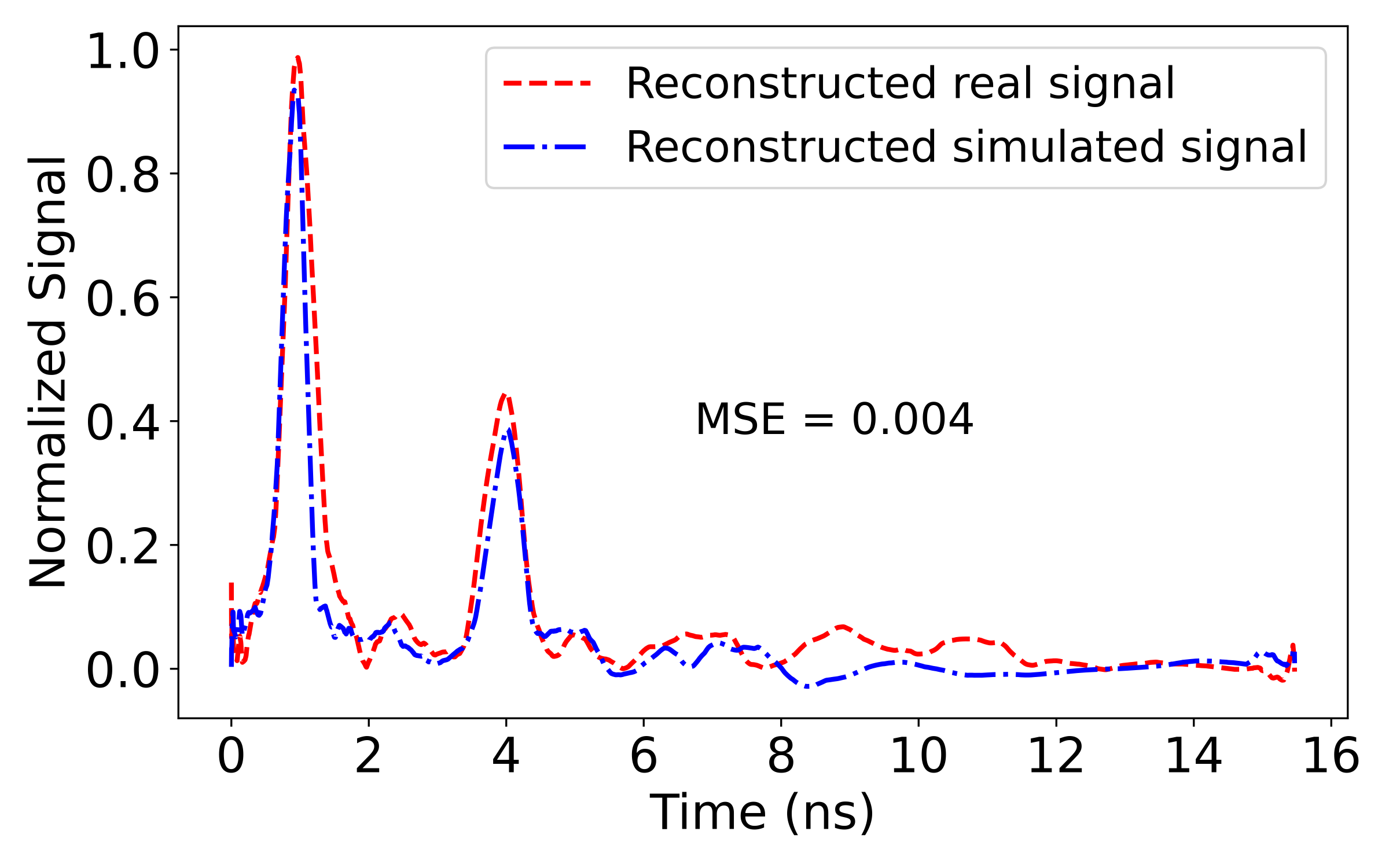}
         \captionsetup{justification=centering}
         \caption{PhyDANN-2.}
         \label{Fig_18e}
     \end{subfigure}
     \caption{\centering Radar signal reconstruction with and without physics-guided hierarchical domain adaptation.}
     \label{Fig_18}
\end{figure*}

As demonstrated in Section \ref{DANN}, DANN learns domain-invariant features through adversarial training between the feature extractor and the domain discriminator. These features are expected to enable the reconstruction of radar signals with similar radar-based physical features in both the source and target domains. For comparison, an autoencoder (AE) with an encoder architecture identical to DANN’s feature extractor is trained without domain adaptation. The reconstructed signals produced by the signal reconstructors of PhyDANN-1, PhyDANN-2, HierPhyDANN-1, and HierPhyDANN-2 can then be compared with those produced by the AE, illustrating the impact of incorporating physics-guided machine learning, hierarchical domain adaptation, and deep adversarial learning. As shown in Figure \ref{Fig_18}, physics-guided domain adaptation yields domain-invariant and material-discriminative features that result in smaller MSE values between the reconstructed simulated and real signals corresponding to the same material parameter values. In contrast, the AE without domain adaptation, shown in Figure \ref{Fig_18a}, exhibits a larger discrepancy between the reconstructed simulated and real signals because it lacks domain-invariant features in its bottleneck layer. Furthermore, by incorporating hierarchical domain adaptation, HierPhyDANN-1 and HierPhyDANN-2 yield smaller MSEs than their counterparts, PhyDANN-1 and PhyDANN-2. Specifically, HierPhyDANN-2 achieves the lowest MSE (Figure \ref{Fig_18b}). Since the radar-based physical features essentially remain consistent between the source and target domains, HierPhyDANN-2 attains the highest $R$ values for permittivity and conductivity estimation compared to other approaches (Table \ref{T8}).

\subsection{Laboratory Test for Two-Layer Material}
The two-layer material represents soil with a vegetation or biomass layer on top of it. Since a metal plate is embedded at the bottom of the soil layer to enhance radar signal reflection, the soil depth $d_1=0.1$ m is assumed to be known. The objective is to estimate the soil permittivity ($p_1$), soil conductivity ($c_1$), wood shavings permittivity ($p_2$), wood shavings conductivity ($c_2$), and wood shavings layer depth ($d_2$) from the radar scans. As discussed in Section \ref{HierDANN}, Sobol’s method suggests that the optimal order for parameter estimation in a single-layer material is permittivity, conductivity, and depth. For the two-layer material (soil–wood shavings), since the soil layer exhibits greater variability in permittivity and conductivity than the wood shavings layer, it is recommended to first estimate the parameters of the more complex layer (soil) to facilitate parameter estimation in the wood shavings layer. Accordingly, the estimation order is $p_1$, $c_1$, $p_2$, $c_2$, and $d_2$. The laboratory test results for the two-layer, five-parameter estimation obtained from different approaches are summarized in Tables \ref{T6} (soil layer) and \ref{T13} (wood shavings layer). Overall, the proposed physics-guided hierarchical domain adaptation approaches, HierDANN, PhyDANN-1, HierPhyDANN-1, PhyDANN-2, and HierPhyDANN-2, outperform the baseline methods, 1D CNN and DANN. Table \ref{T10} shows the computational costs of different data-driven models. The 1D CNN and DANN models each train a single network to predict all five material parameters simultaneously, requiring 7.4 and 14.9 minutes, respectively. In contrast, the HierDANN trains five individual networks to predict each parameter sequentially, with training times of 14.0, 1.5, 0.4, 0.2, and 0.1 minutes. As the simulated training dataset is progressively pruned during the hierarchical domain adaptation process, the corresponding training time decreases accordingly. Due to the hierarchical domain adaptation, the computational time required for HierDANN, HierPhyDANN-1, and HierPhyDANN-2 to make a prediction includes both training and inference, totaling 16.2, 23.0, and 26.5 minutes, respectively. In contrast, PhyDANN-1 and PhyDANN-2 can make a prediction in $2.7 \times 10^{-4}$ seconds and $2.3 \times 10^{-4}$ seconds, respectively, which is significantly faster than iterative model updating, which requires 14.7 minutes per prediction.

\begin{table*}[ht]
    \small\sf\centering
    \caption{\centering Laboratory test results of the two-layer, five-parameter estimation (soil layer).}
    \label{T6}
    \begin{threeparttable}
    \begin{tabular}{@{\extracolsep{\fill}}lcccccccc}
        \toprule
        \multirow{2}{*}{Approach}
            & \multicolumn{4}{c}{$p_1$}
            & \multicolumn{4}{c}{$c_1$} \\
        \cmidrule(lr){2-5} \cmidrule(lr){6-9}
            & $R$ & Bias & RMSE & ubRMSE
            & $R$ & Bias & RMSE & ubRMSE \\
        \midrule
        1D CNN (baseline)        & 0.939 & -0.201 & 1.506 & 1.492 & 0.794 & -0.038 & 0.048 & 0.030 \\
        DANN (baseline)          & 0.973 & -2.313 & 2.454 & 0.821 & 0.758 & -0.061 & 0.067 & 0.026 \\
        HierDANN (ours)          & 0.966 &  0.164 & 1.240 & 1.229 & 0.904 & -0.002 & 0.018 & 0.018 \\
        PhyDANN-1 (ours)         & \textit{\textbf{\underline{0.987}}} &  1.688 & 2.105 & 1.258 & 0.869 & -0.022 & 0.032 & 0.023 \\
        HierPhyDANN-1 (ours)     & 0.954 & -0.801 & 1.499 & 1.267 & 0.958 &  0.000 & 0.015 & 0.015 \\
        PhyDANN-2 (ours)         & 0.974 &  2.519 & 2.818 & 1.263 & 0.936 & -0.005 & 0.021 & 0.020 \\
        HierPhyDANN-2 (ours)     & 0.949 & -0.452 & 1.383 & 1.308 & \textit{\textbf{\underline{0.968}}} &  0.004 & 0.013 & 0.013 \\
        \bottomrule
    \end{tabular}
    \begin{tablenotes}
        \item[1] The bold values indicate the best correlation coefficient $R$ among the compared approaches.
    \end{tablenotes}
    \end{threeparttable}
\end{table*}

\begin{table*}[ht]
    \small\sf\centering
    \caption{\centering Laboratory test results of the two-layer, five-parameter estimation (wood shavings layer).}
    \label{T13}
    \begin{threeparttable}
    \begin{tabular}{@{\extracolsep{\fill}}lccccccccc}
        \toprule
        \multirow{2}{*}{Approach}
            & \multicolumn{3}{c}{$p_2$}
            & \multicolumn{3}{c}{$c_2$}
            & \multicolumn{3}{c}{$d_2$} \\
        \cmidrule(lr){2-4} \cmidrule(lr){5-7} \cmidrule(lr){8-10}
            & Bias & RMSE & ubRMSE
            & Bias & RMSE & ubRMSE
            & Bias & RMSE & ubRMSE \\
        \midrule
        1D CNN (baseline)    & -0.641 & 0.853 & 0.563 & -0.001 & 0.001 & 0.000 & 0.023 & 0.033 & 0.024 \\
        DANN (baseline)      & -1.132 & 1.290 & 0.619 &  0.004 & 0.004 & 0.001 & 0.020 & 0.034 & 0.028 \\
        HierDANN (ours)      & -0.595 & 0.730 & 0.423 &  0.000 & 0.001 & 0.001 & 0.004 & 0.022 & 0.022 \\
        PhyDANN-1 (ours)     & -1.320 & 1.430 & 0.550 & -0.001 & 0.005 & 0.005 & 0.062 & 0.070 & 0.031 \\
        HierPhyDANN-1 (ours) & -0.891 & 1.198 & 0.800 &  0.003 & 0.003 & 0.000 & -0.009 & 0.024 & 0.023 \\
        PhyDANN-2 (ours)     & -0.988 & 1.128 & 0.545 &  0.003 & 0.003 & 0.003 & 0.060 & 0.070 & 0.037 \\
        HierPhyDANN-2 (ours) & -0.161 & 0.539 & 0.515 &  0.001 & 0.001 & 0.001 & 0.000 & 0.024 & 0.024 \\
        \bottomrule
    \end{tabular}
    \end{threeparttable}
\end{table*}

\begin{table*}[ht]
    \small\sf\centering
    \caption{\centering Computational costs of data-driven models.}
    \label{T10}
    \begin{threeparttable}
    \begin{tabular}{lcc}
        \toprule
        Approach & Training time (min) & \begin{tabular}[c]{@{}c@{}}Inference time \\ per prediction ($10^{-4}$ sec)\end{tabular}  \\ 
        \midrule
        Iterative model updating (baseline) & --- & $8.8 \times 10^{6}$\\
        1D CNN (baseline) & 7.4 & $1.2 \times 10^{-1}$ \\
        DANN (baseline) & 14.9 & 2.3 \\
        HierDANN (ours) & 16.2 & 12.8\\
        PhyDANN-1 (ours) & 25.4 & 2.7 \\
        HierPhyDANN-1 (ours) & 23.0 & 12.3 \\
        PhyDANN-2 (ours) & 33.0 & 2.3 \\
        HierPhyDANN-2 (ours) & 26.5 & 13.3 \\
        \bottomrule
    \end{tabular}
    \begin{tablenotes}
        \item GPU: GeForce RTX 5080
    \end{tablenotes}
    \end{threeparttable}
\end{table*}

\begin{figure*}[ht]
     \centering
     \begin{subfigure}[t]{0.32\textwidth}
         \centering
         \includegraphics[width=1\linewidth]{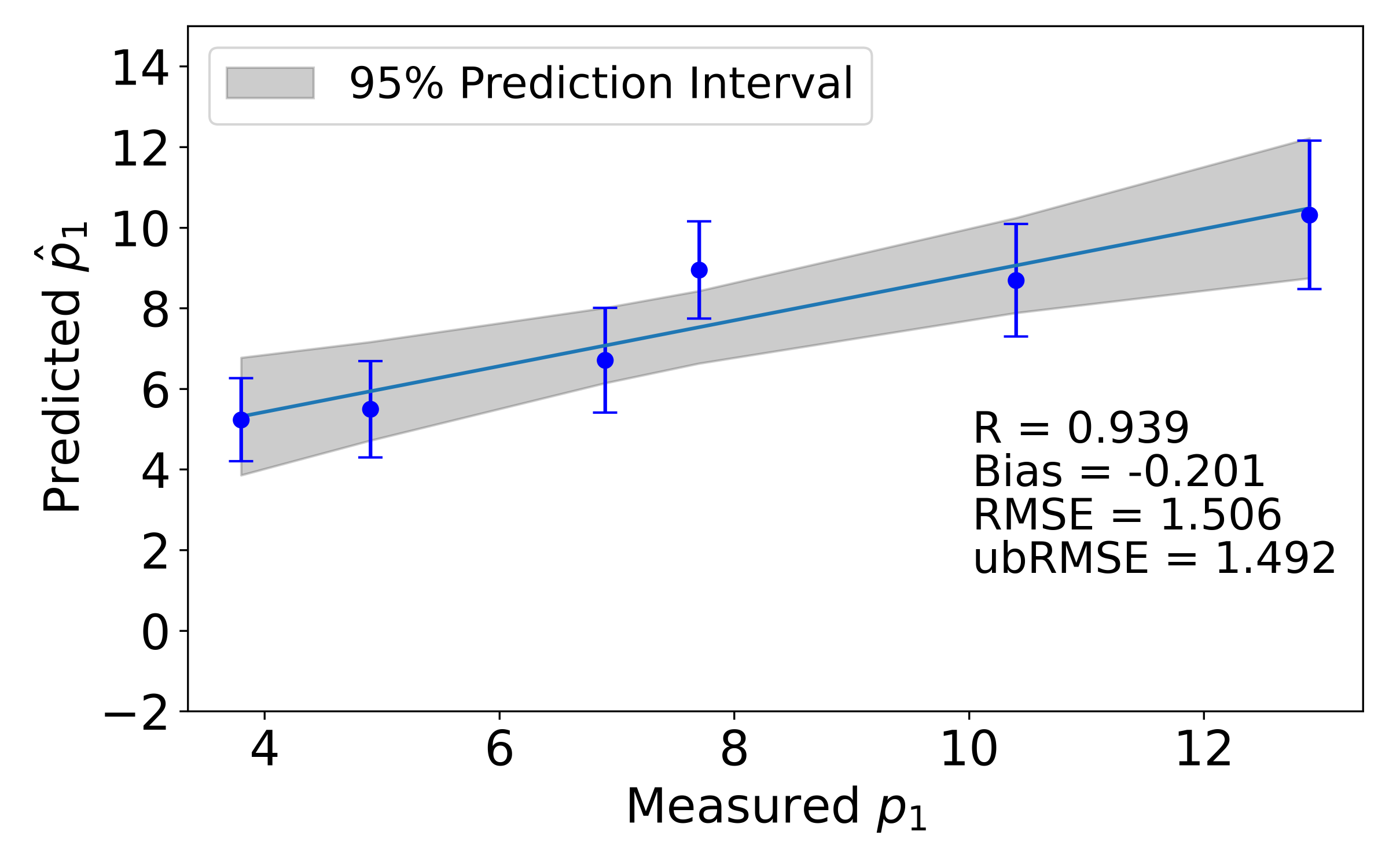}
         \captionsetup{justification=centering}
         \caption{1D CNN (all specimens).}
         \label{Fig_11a}
     \end{subfigure}
     \begin{subfigure}[t]{0.32\textwidth}
         \centering
         \includegraphics[width=1\linewidth]{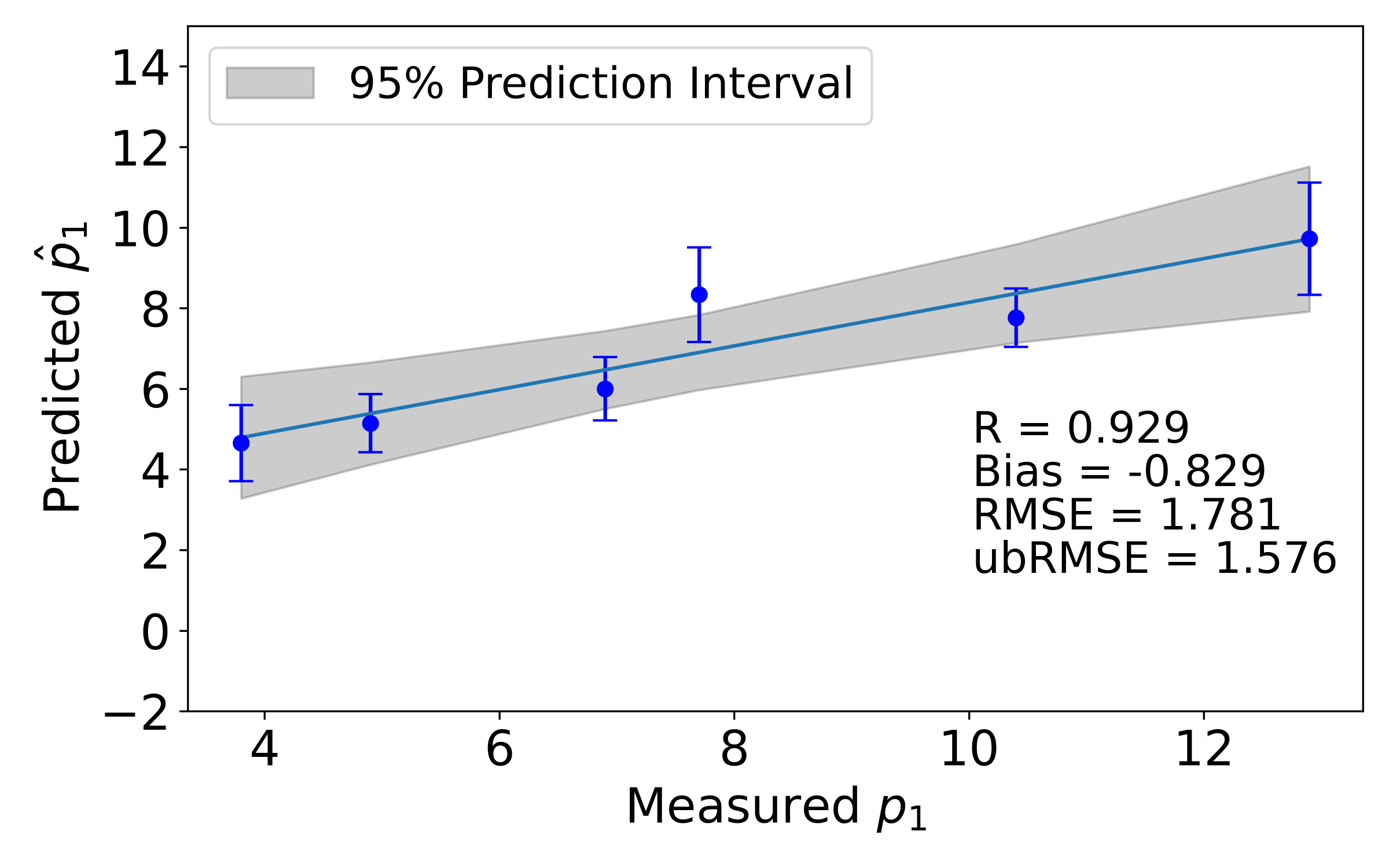}
         \captionsetup{justification=centering}
         \caption{1D CNN (0.15 m soil + 0.1 m dry wood shavings).}
         \label{Fig_11b}
     \end{subfigure}
     \begin{subfigure}[t]{0.32\textwidth}
         \centering
         \includegraphics[width=1\linewidth]{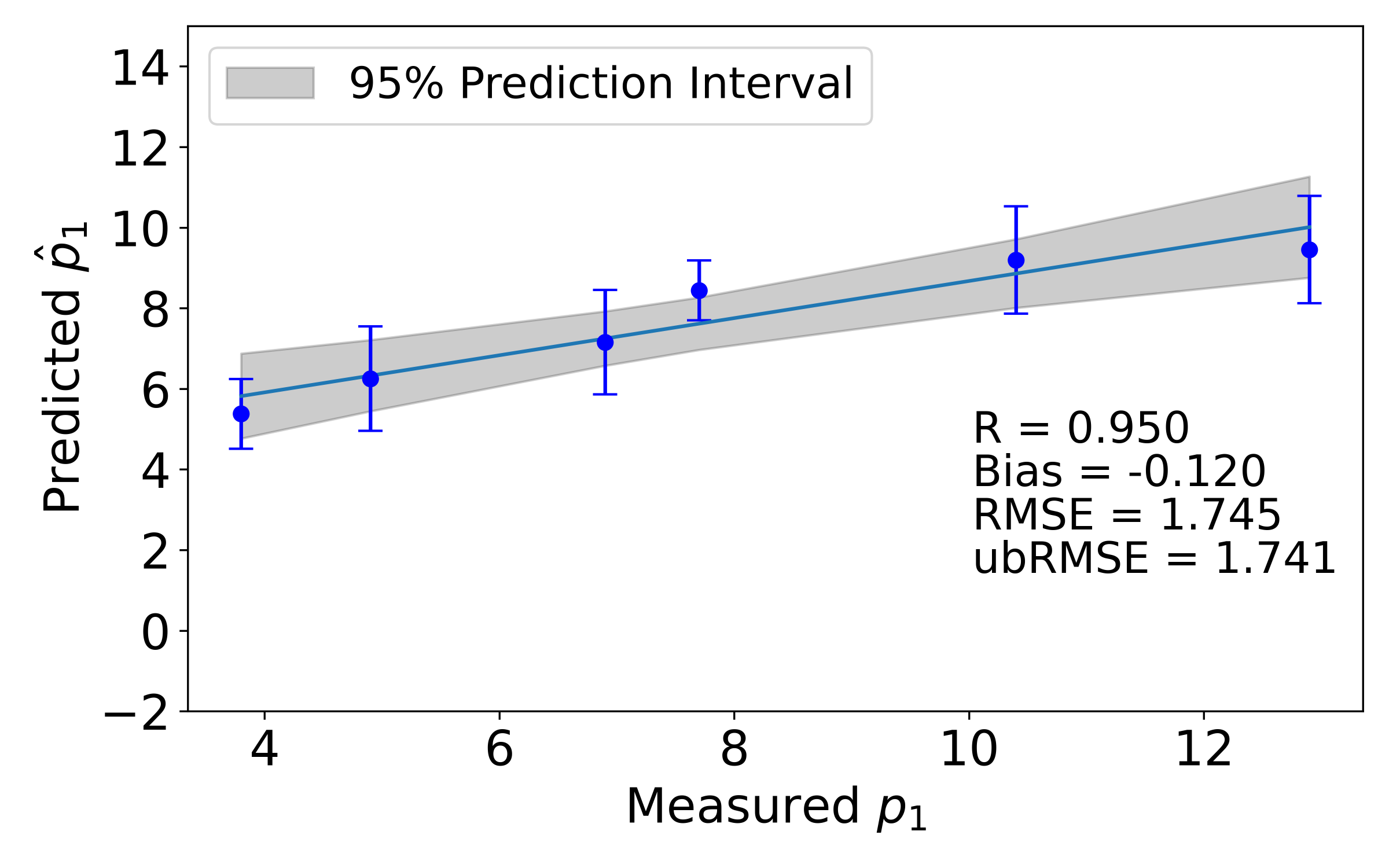}
         \captionsetup{justification=centering}
         \caption{1D CNN (0.15 m soil + 0.15 m dry wood shavings).}
         \label{Fig_11c}
     \end{subfigure}
     \begin{subfigure}[t]{0.32\textwidth}
         \centering
         \includegraphics[width=1\linewidth]{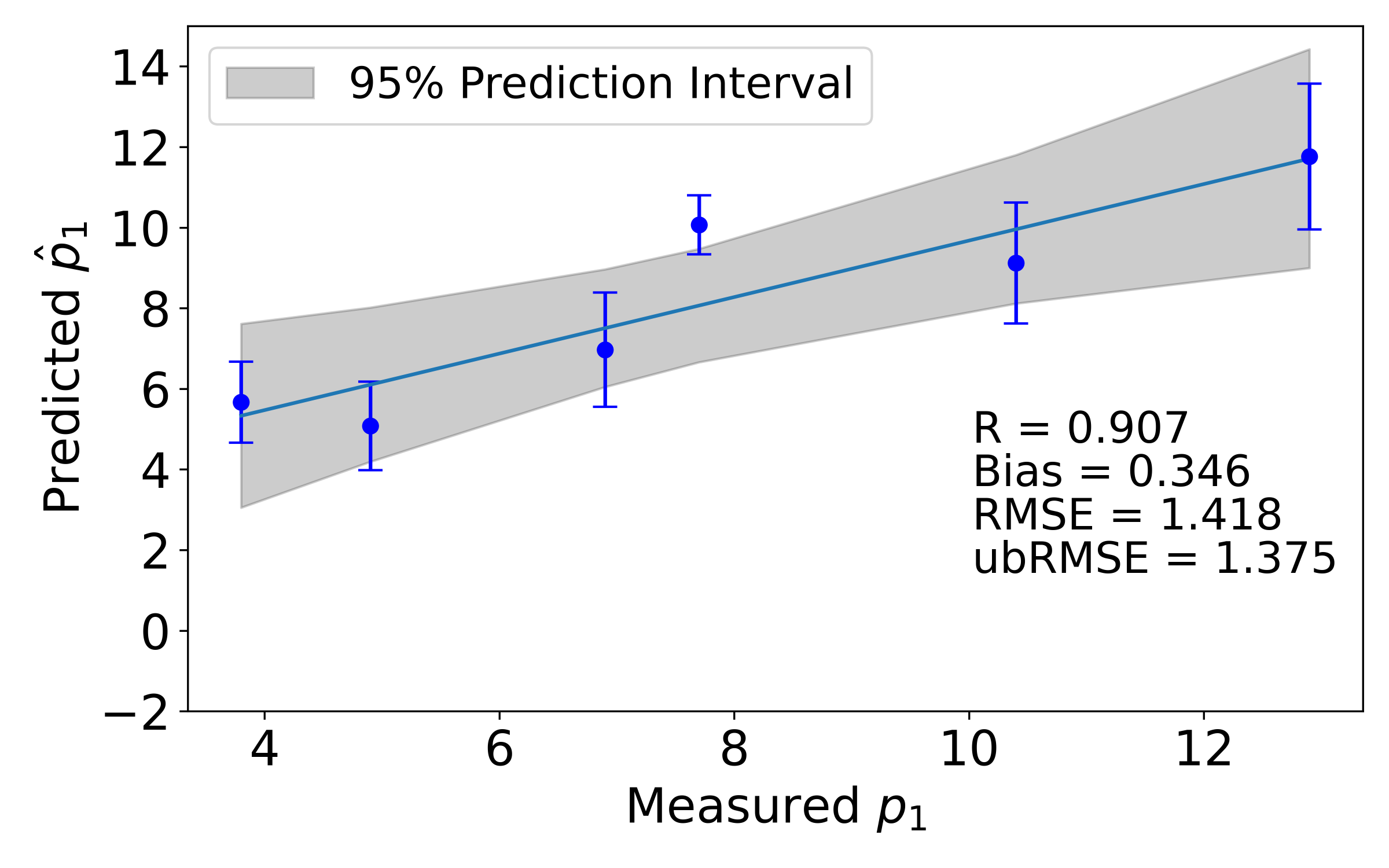}
         \captionsetup{justification=centering}
         \caption{1D CNN (0.15 m soil + 0.1 m wet wood shavings).}
         \label{Fig_11d}
     \end{subfigure}
     \begin{subfigure}[t]{0.32\textwidth}
         \centering
         \includegraphics[width=1\linewidth]{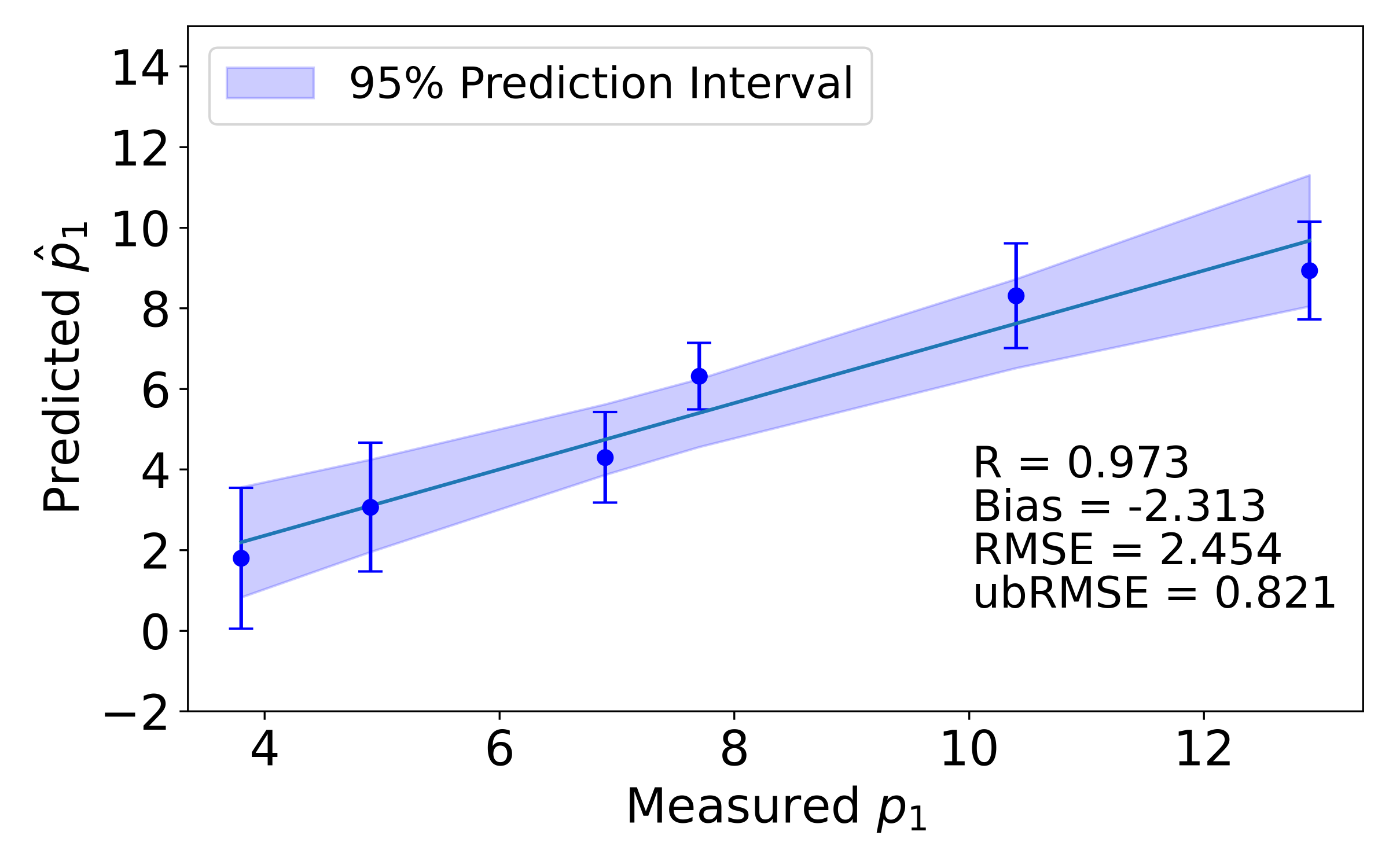}
         \captionsetup{justification=centering}
         \caption{DANN (all specimens).}
         \label{Fig_11e}
     \end{subfigure}
     \begin{subfigure}[t]{0.32\textwidth}
         \centering
         \includegraphics[width=1\linewidth]{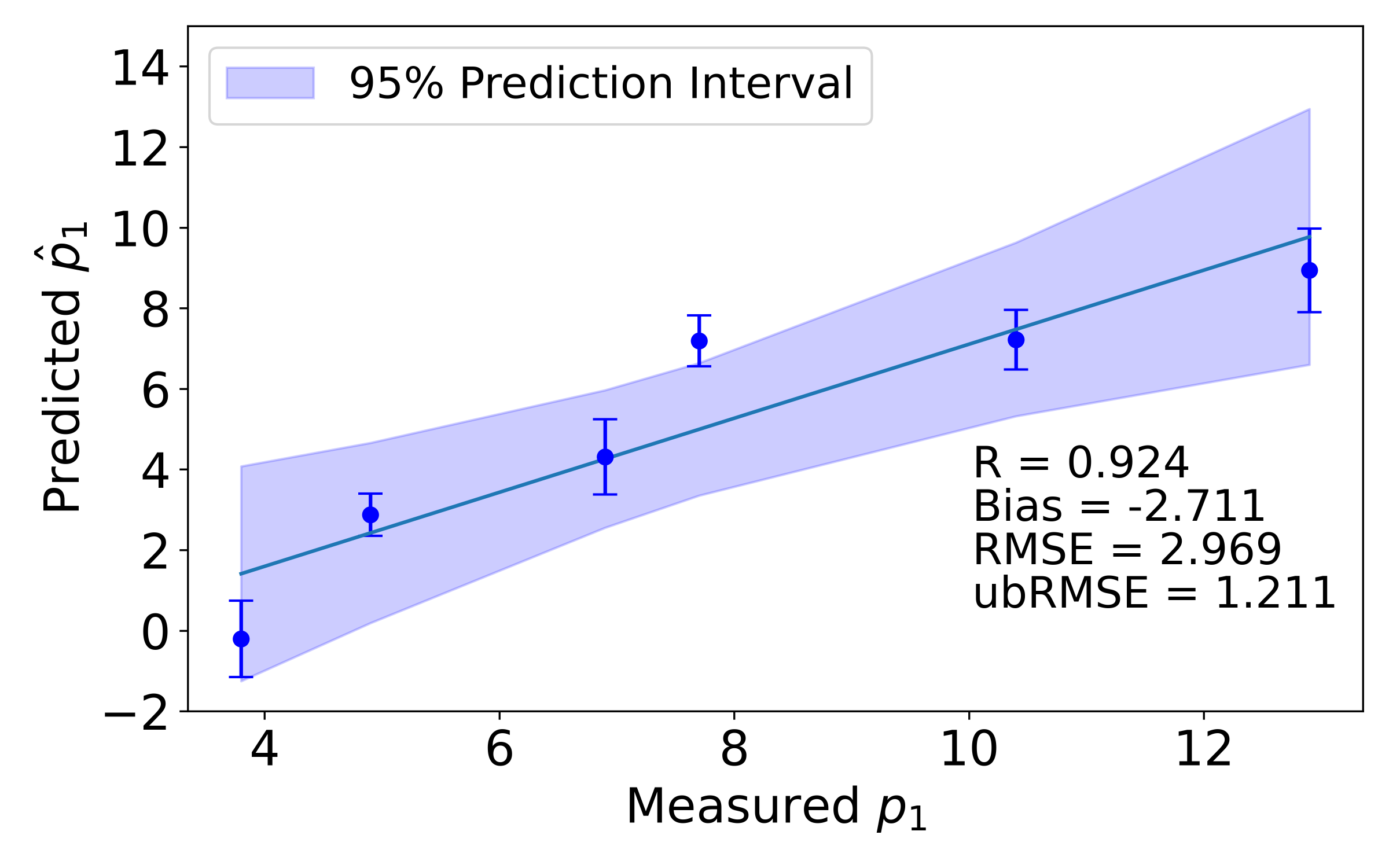}
         \captionsetup{justification=centering}
         \caption{DANN (0.15 m soil + 0.1 m dry wood shavings).}
         \label{Fig_11f}
     \end{subfigure}
     \begin{subfigure}[t]{0.32\textwidth}
         \centering
         \includegraphics[width=1\linewidth]{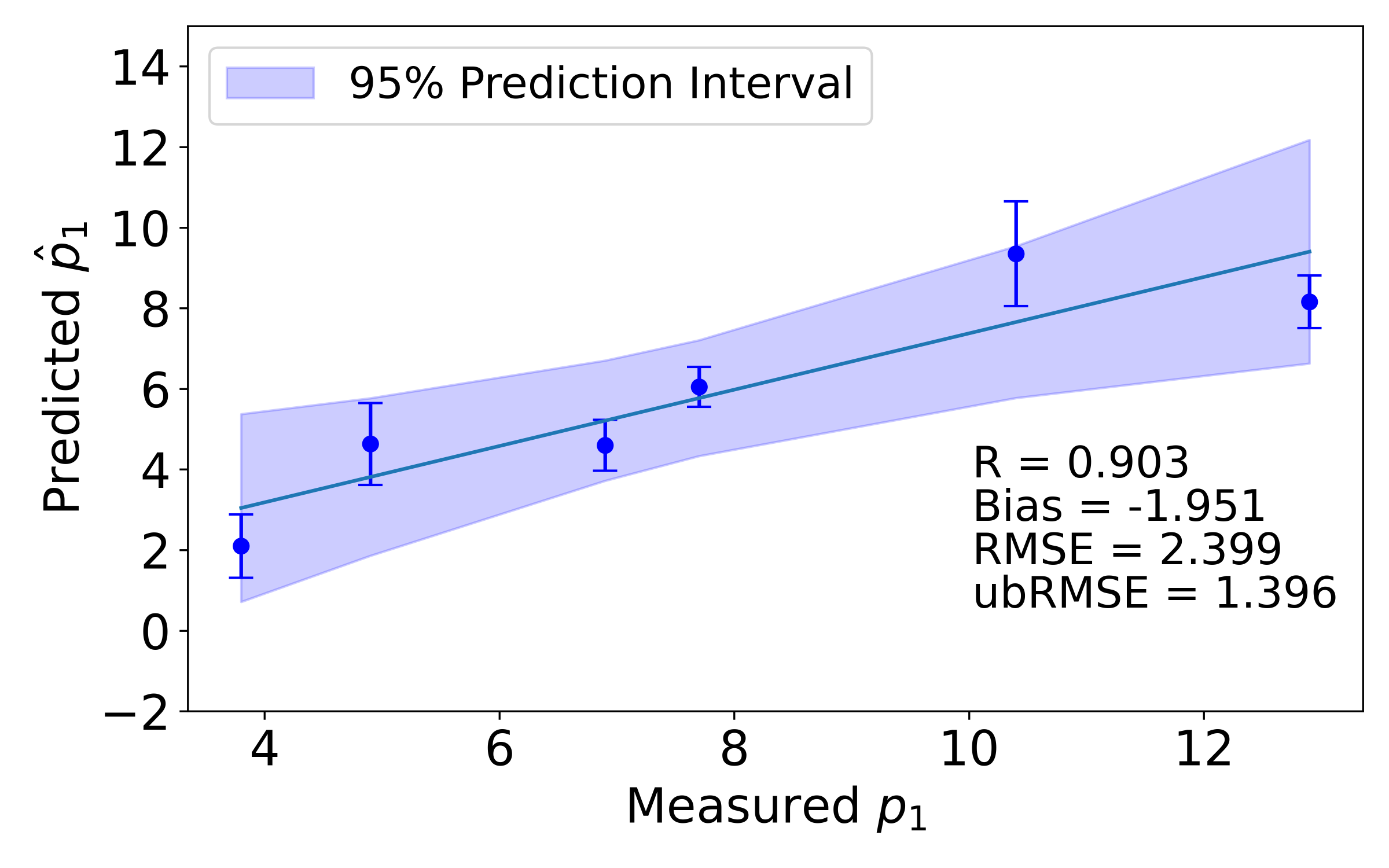}
         \captionsetup{justification=centering}
         \caption{DANN (0.15 m soil + 0.15 m dry wood shavings).}
         \label{Fig_11g}
     \end{subfigure}
     \begin{subfigure}[t]{0.32\textwidth}
         \centering
         \includegraphics[width=1\linewidth]{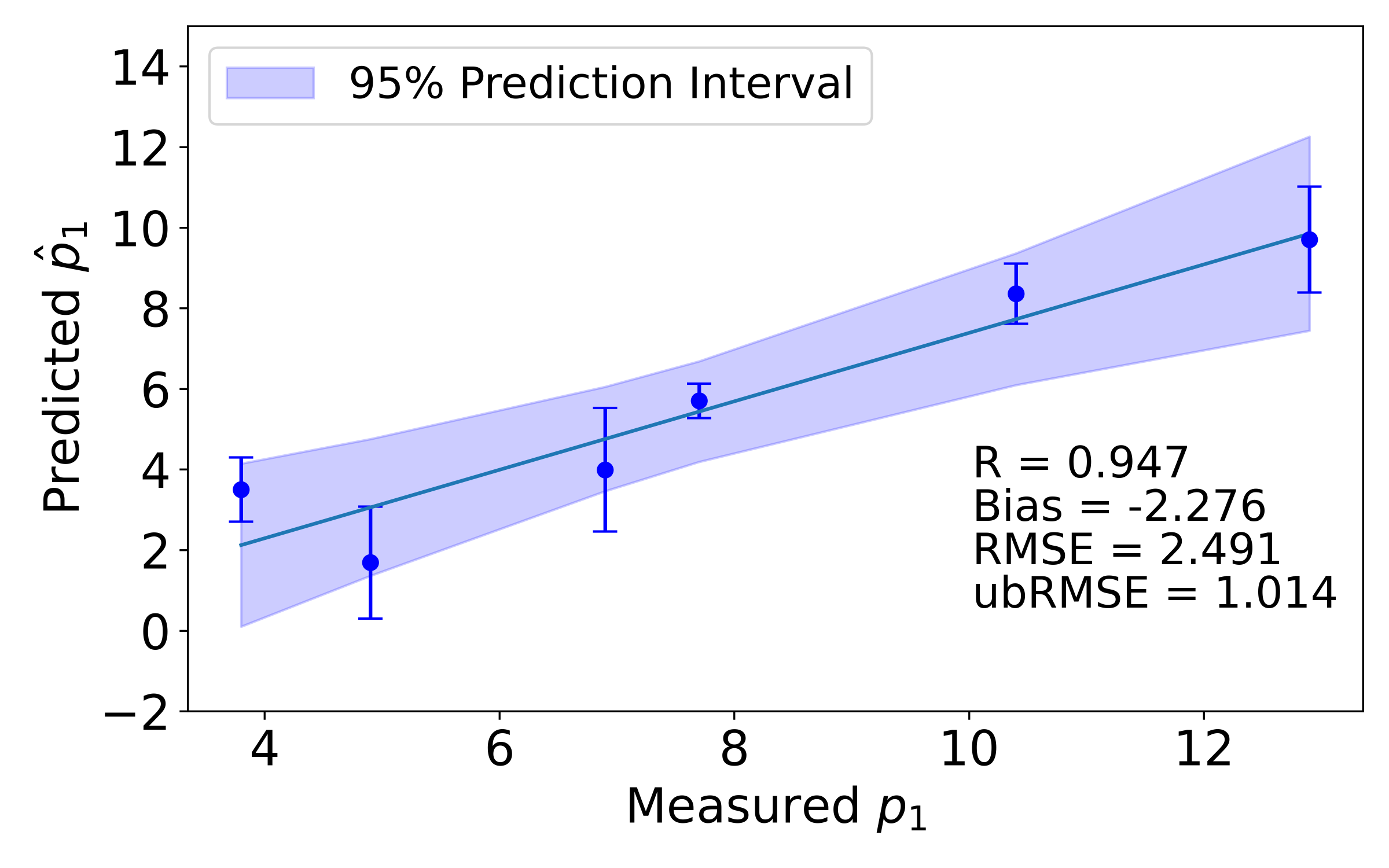}
         \captionsetup{justification=centering}
         \caption{DANN (0.15 m soil + 0.1 m wet wood shavings).}
         \label{Fig_11h}
     \end{subfigure}
     \begin{subfigure}[t]{0.32\textwidth}
         \centering
         \includegraphics[width=1\linewidth]{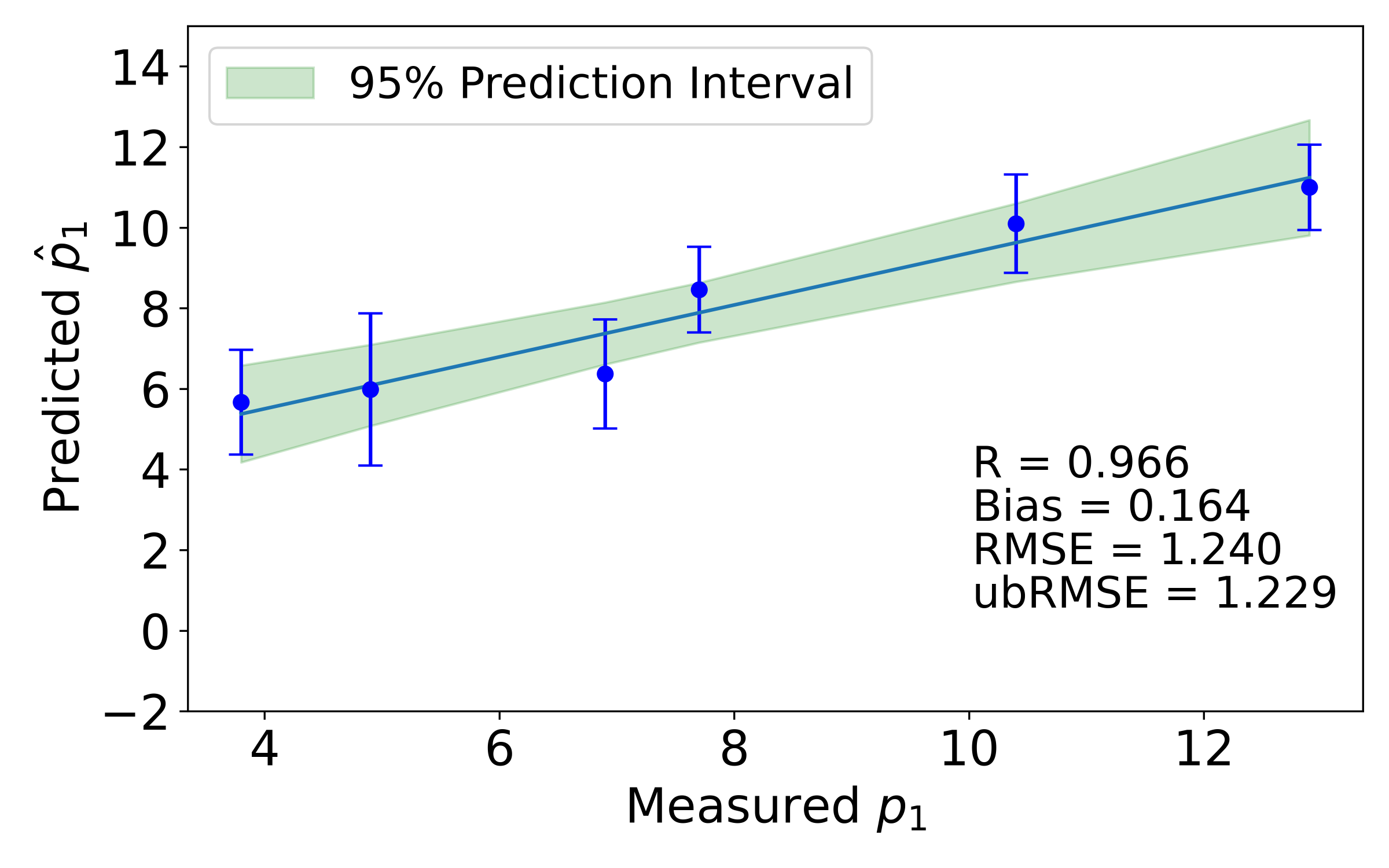}
         \captionsetup{justification=centering}
         \caption{HierDANN (all specimens).}
         \label{Fig_11i}
     \end{subfigure}
     \begin{subfigure}[t]{0.32\textwidth}
         \centering
         \includegraphics[width=1\linewidth]{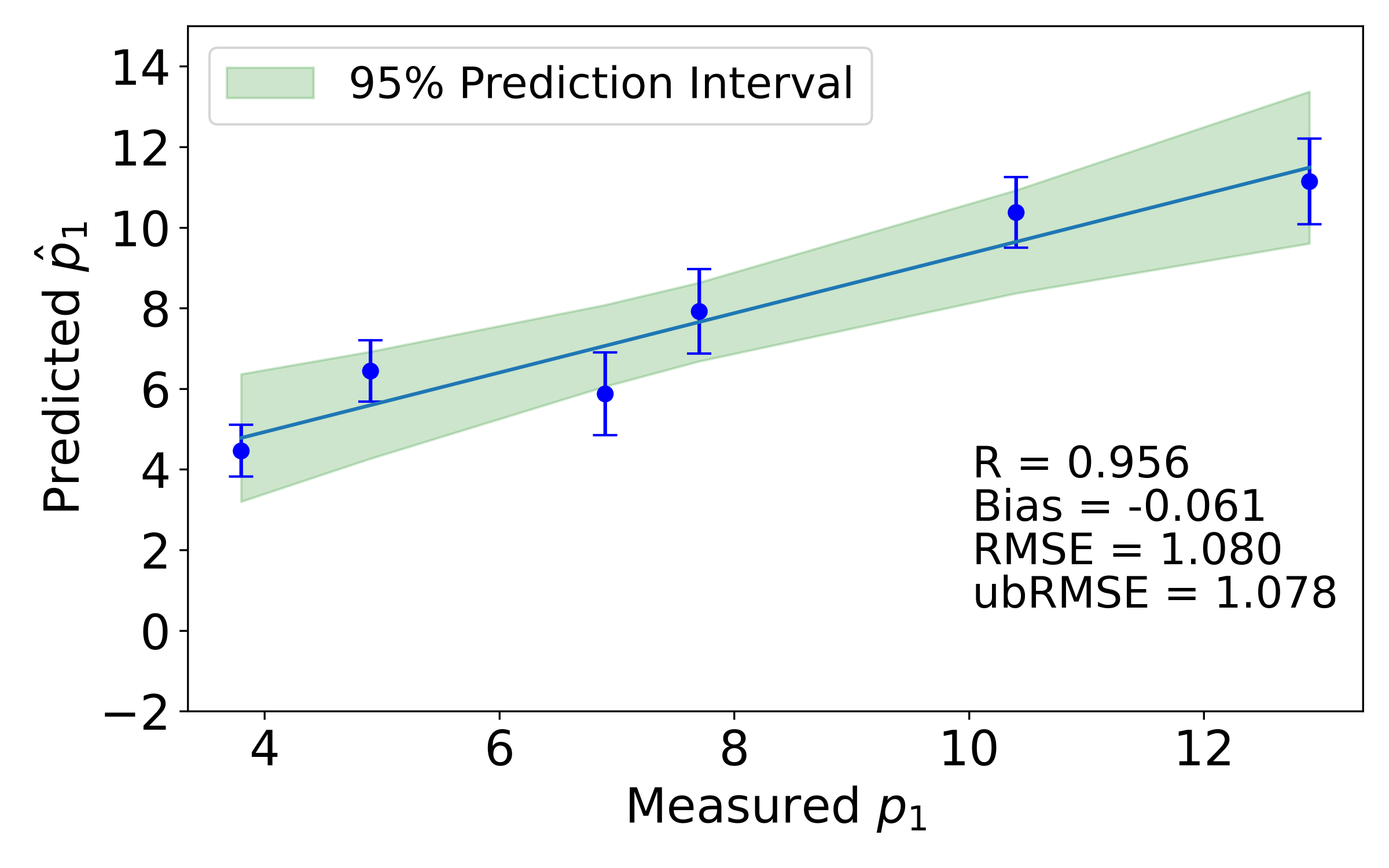}
         \captionsetup{justification=centering}
         \caption{HierDANN (0.15 m soil + 0.1 m dry wood shavings).}
         \label{Fig_11j}
     \end{subfigure}
     \begin{subfigure}[t]{0.32\textwidth}
         \centering
         \includegraphics[width=1\linewidth]{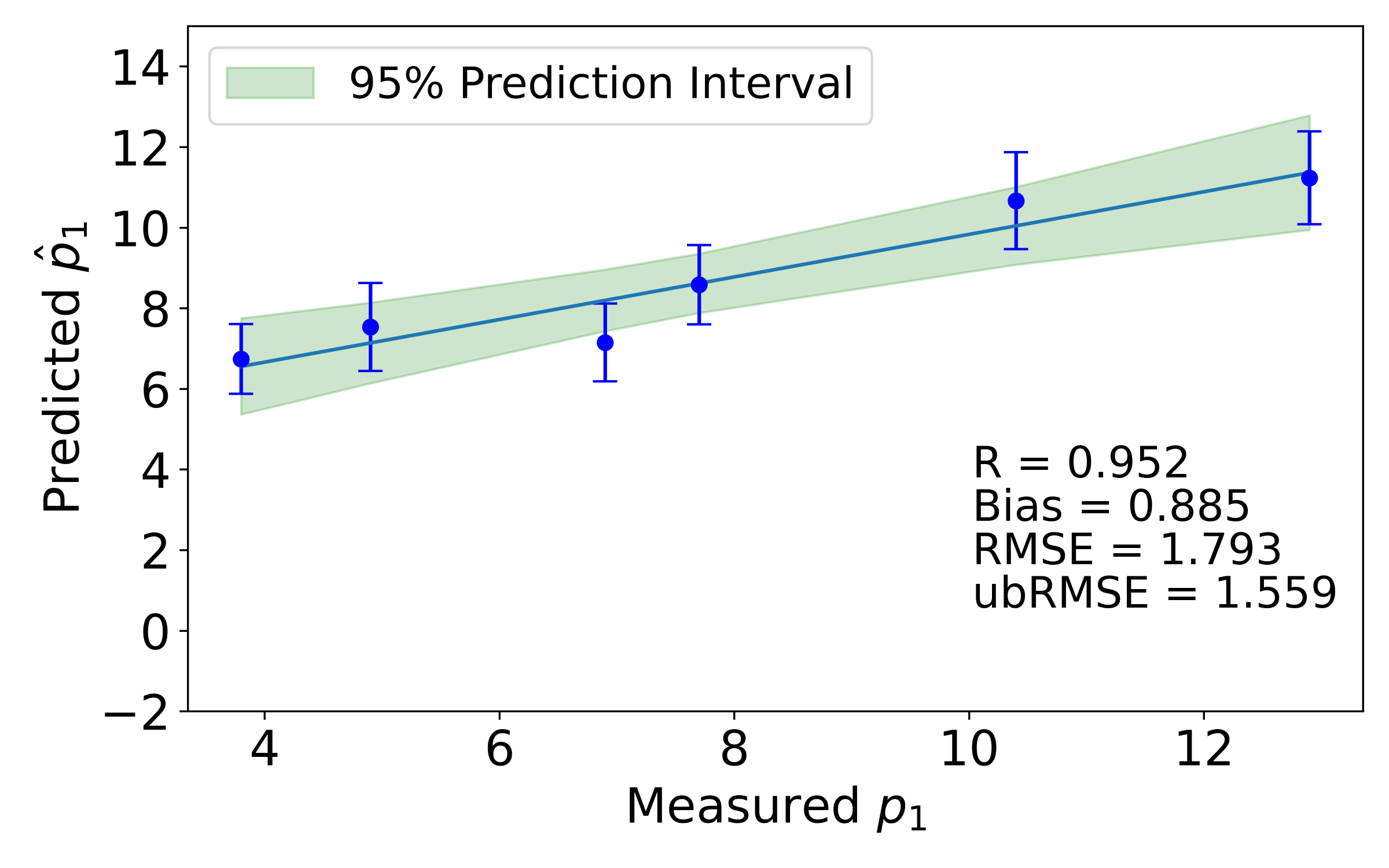}
         \captionsetup{justification=centering}
         \caption{HierDANN (0.15 m soil + 0.15 m dry wood shavings).}
         \label{Fig_11k}
     \end{subfigure}
     \begin{subfigure}[t]{0.32\textwidth}
         \centering
         \includegraphics[width=1\linewidth]{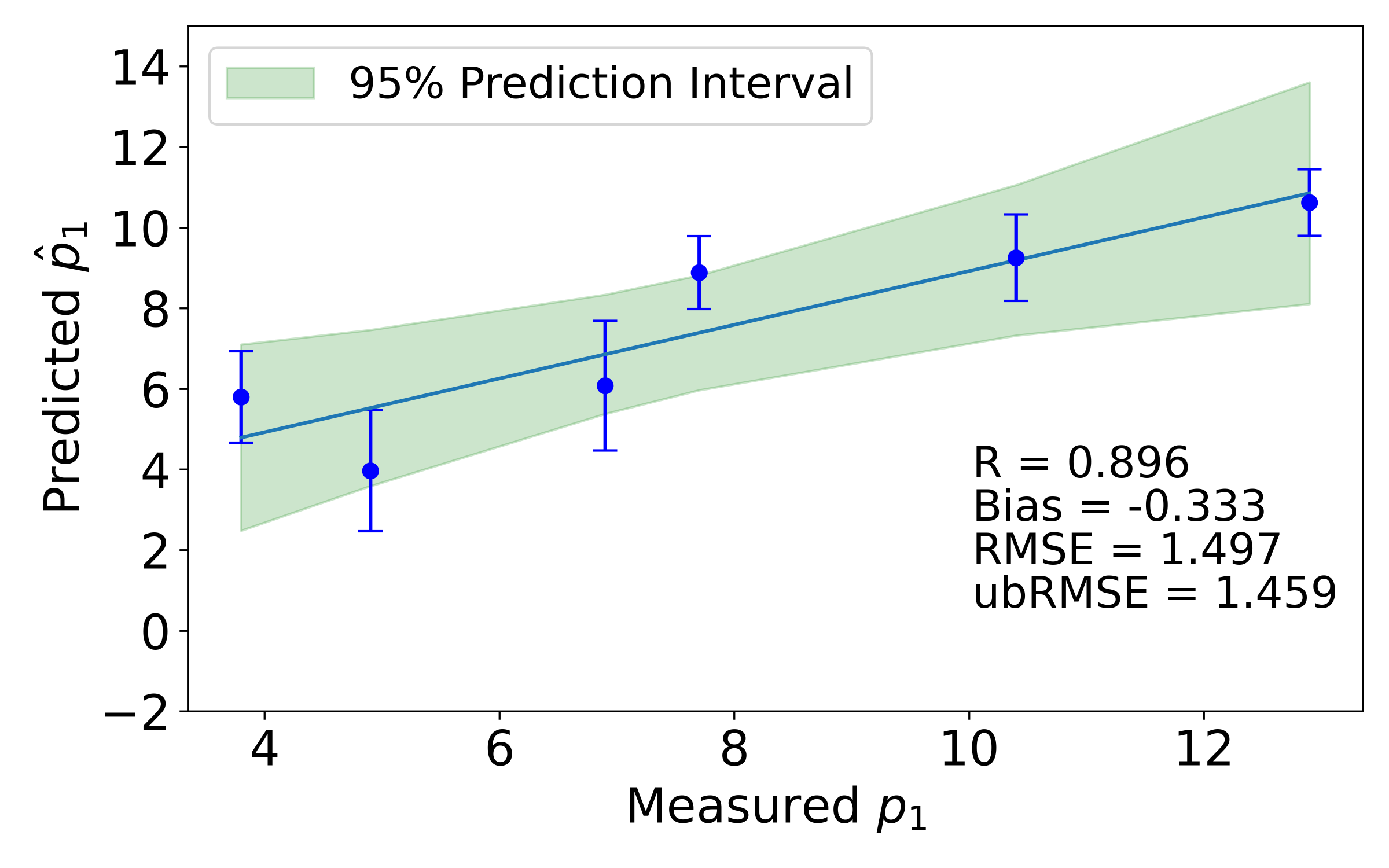}
         \captionsetup{justification=centering}
         \caption{HierDANN (0.15 m soil + 0.1 m wet wood shavings).}
         \label{Fig_11l}
     \end{subfigure}
     \caption{\centering Laboratory test results for soil permittivity prediction in the two-layer, five-parameter estimation.}
     \label{Fig_11}
\end{figure*}

For soil permittivity estimation, as shown in Table \ref{T6}, the 1D CNN without domain adaptation achieves an $R$ value of 0.939, whereas with domain adaptation, DANN improves the $R$ value to 0.973 at the cost of a higher Bias. PhyDANN-1 produces the highest $R$ value, albeit with a slightly higher Bias. HierDANN achieves the lowest Bias while maintaining an $R$ value comparable to those of DANN and PhyDANN-1. The prediction results for soil permittivity are presented in Figure \ref{Fig_11}. An ablation study was conducted to assess the impact of the depth and moisture level of the top wood shavings layer on soil moisture estimation. As shown in Figure \ref{Fig_11}, the standard deviation of the soil moisture predictions for the 0.15 m soil + 0.1 m dry wood shavings case is smaller than that for the 0.15 m soil + 0.1 m wet wood shavings case. This can be explained by the higher moisture content in the wet wood shavings, which attenuates the radar signal and increases estimation uncertainty. For the same reason, HierDANN yields a lower $R$ value in the 0.15 m soil + 0.1 m wet wood shavings case. Additionally, for both HierDANN and DANN, the $R$ values for soil moisture estimation in the 0.15 m soil + 0.15 m dry wood shavings case are lower than those in the 0.15 m soil + 0.1 m dry wood shavings case. This is due to the greater depth of the wood shavings layer, which attenuates the radar signal and reduces the information available for parameter estimation.

\begin{figure*}[ht]
     \centering
     \begin{subfigure}[t]{0.32\textwidth}
         \centering
         \includegraphics[width=1\linewidth]{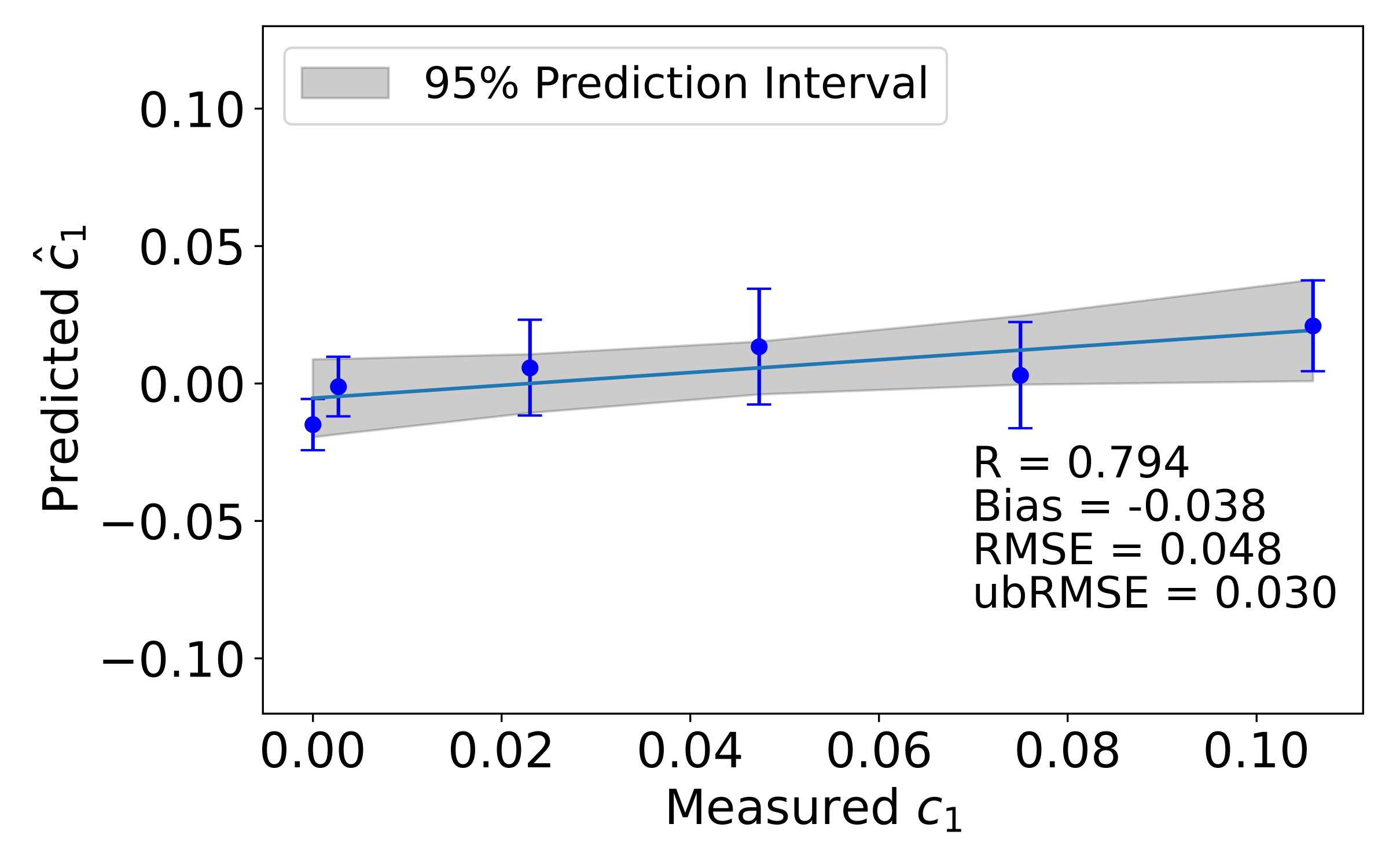}
         \captionsetup{justification=centering}
         \caption{Soil conductivity estimated by 1D CNN.}
         \label{Fig_12a}
     \end{subfigure}
     \begin{subfigure}[t]{0.32\textwidth}
         \centering
         \includegraphics[width=1\linewidth]{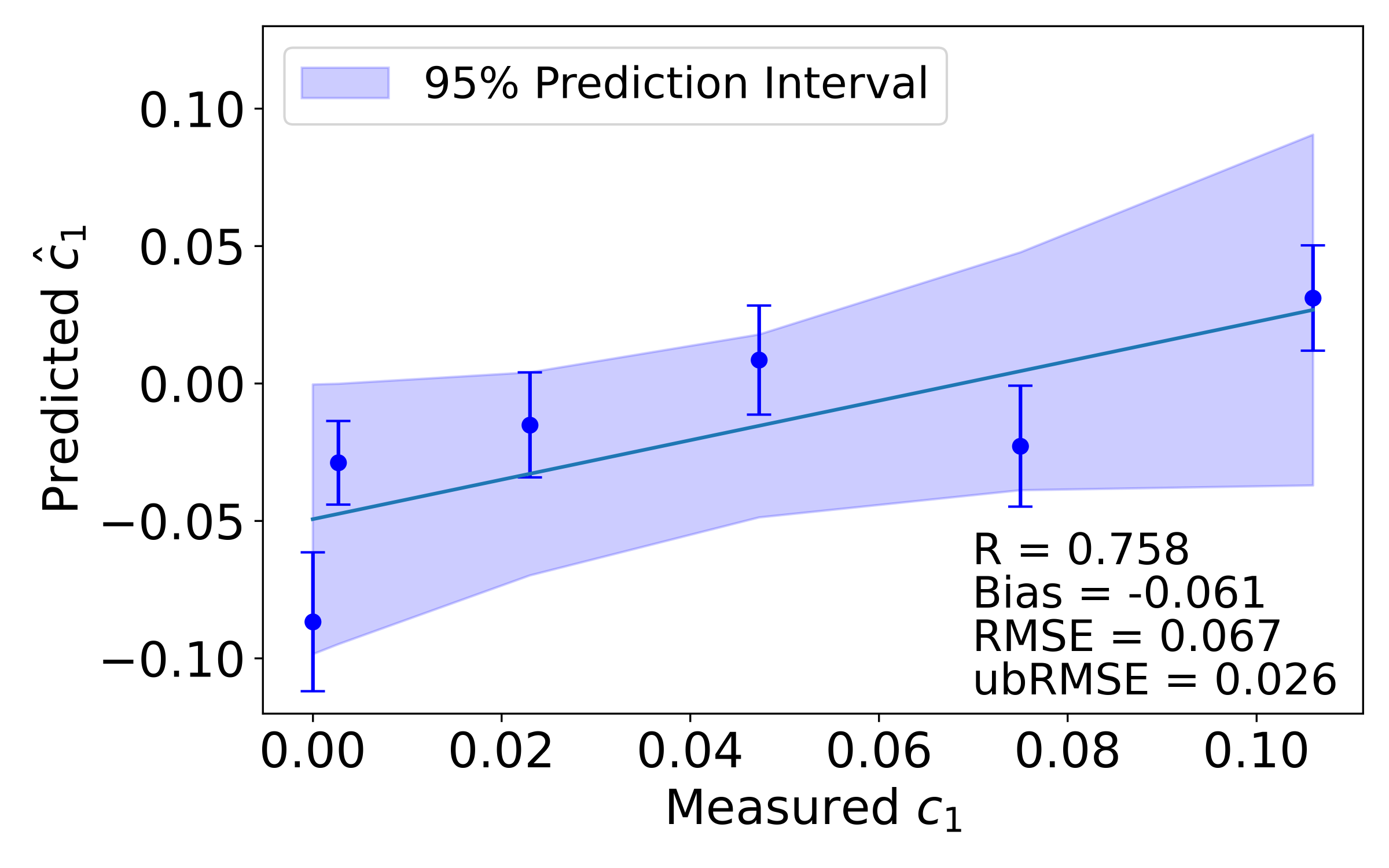}
         \captionsetup{justification=centering}
         \caption{Soil conductivity estimated by DANN.}
         \label{Fig_12b}
     \end{subfigure}
     \begin{subfigure}[t]{0.32\textwidth}
         \centering
         \includegraphics[width=1\linewidth]{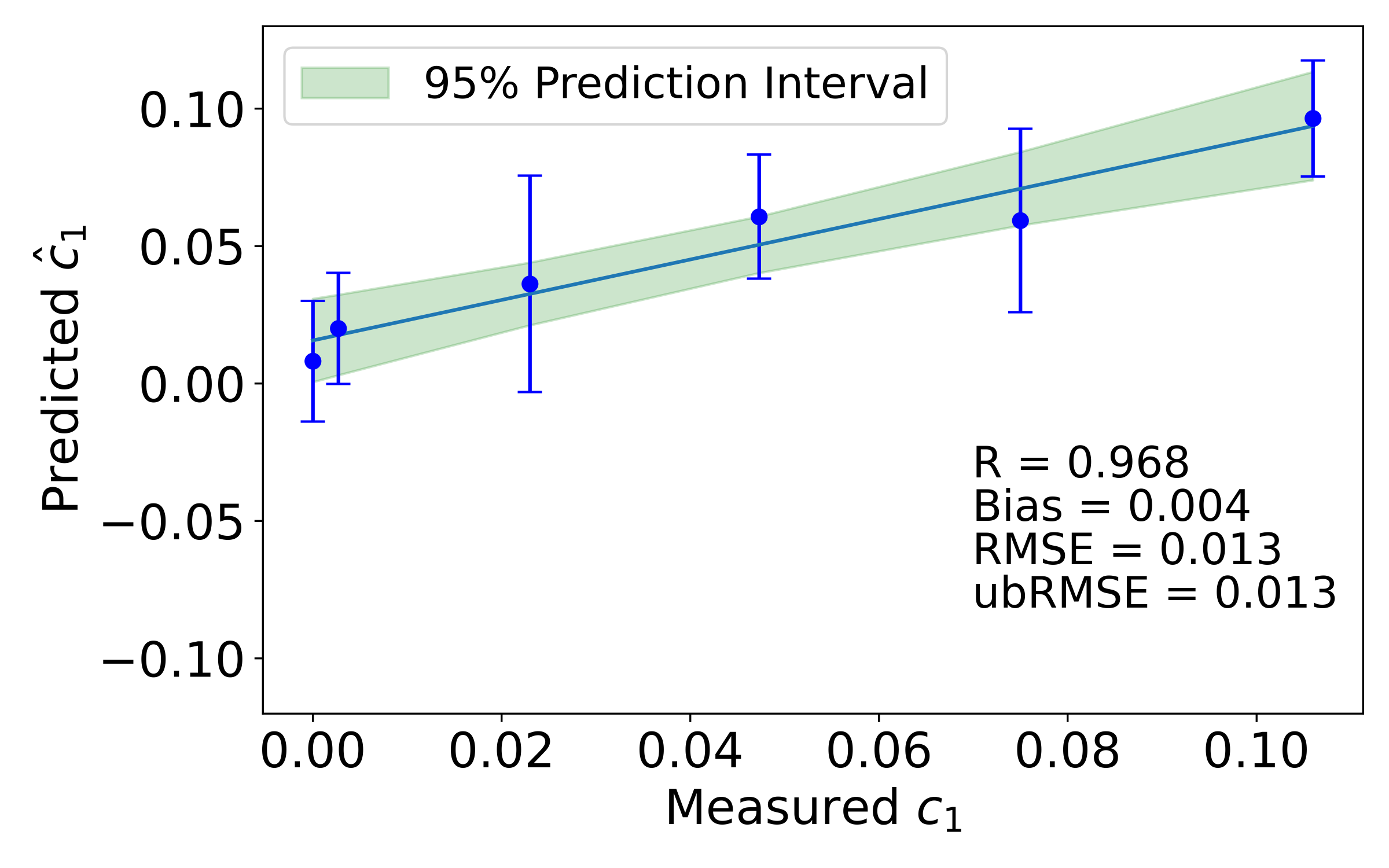}
         \captionsetup{justification=centering}
         \caption{Soil conductivity estimated by HierPhyDANN-2.}
         \label{Fig_12c}
     \end{subfigure}
     \caption{\centering Laboratory test results for soil conductivity in the two-layer, five-parameter estimation.}
     \label{Fig_12}
\end{figure*}

\begin{figure*}[ht]
     \centering
     \begin{subfigure}[t]{0.32\textwidth}
         \centering
         \includegraphics[width=1\linewidth]{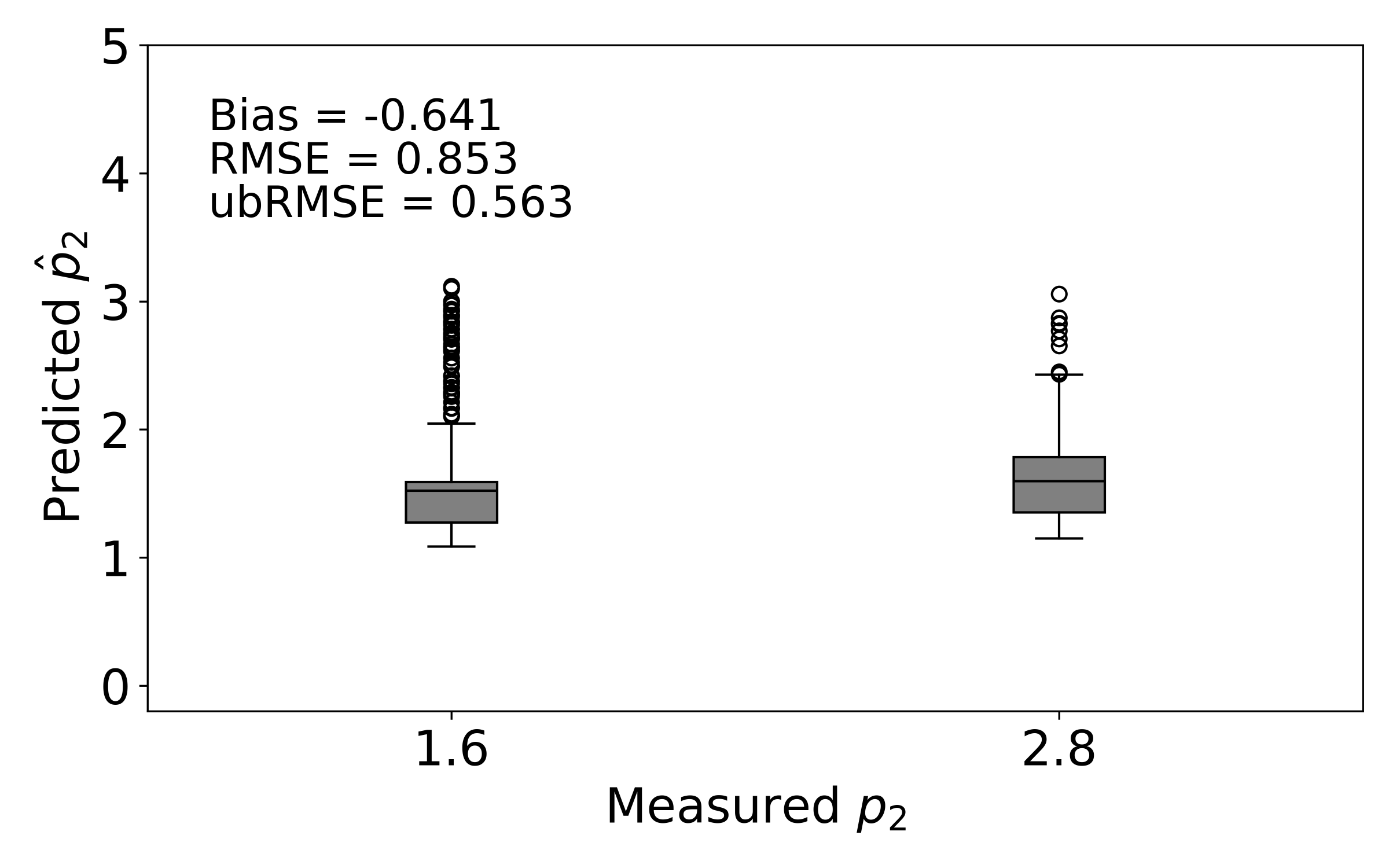}
         \captionsetup{justification=centering}
         \caption{Wood shavings permittivity estimated by 1D CNN.}
         \label{Fig_13a}
     \end{subfigure}
     \begin{subfigure}[t]{0.32\textwidth}
         \centering
         \includegraphics[width=1\linewidth]{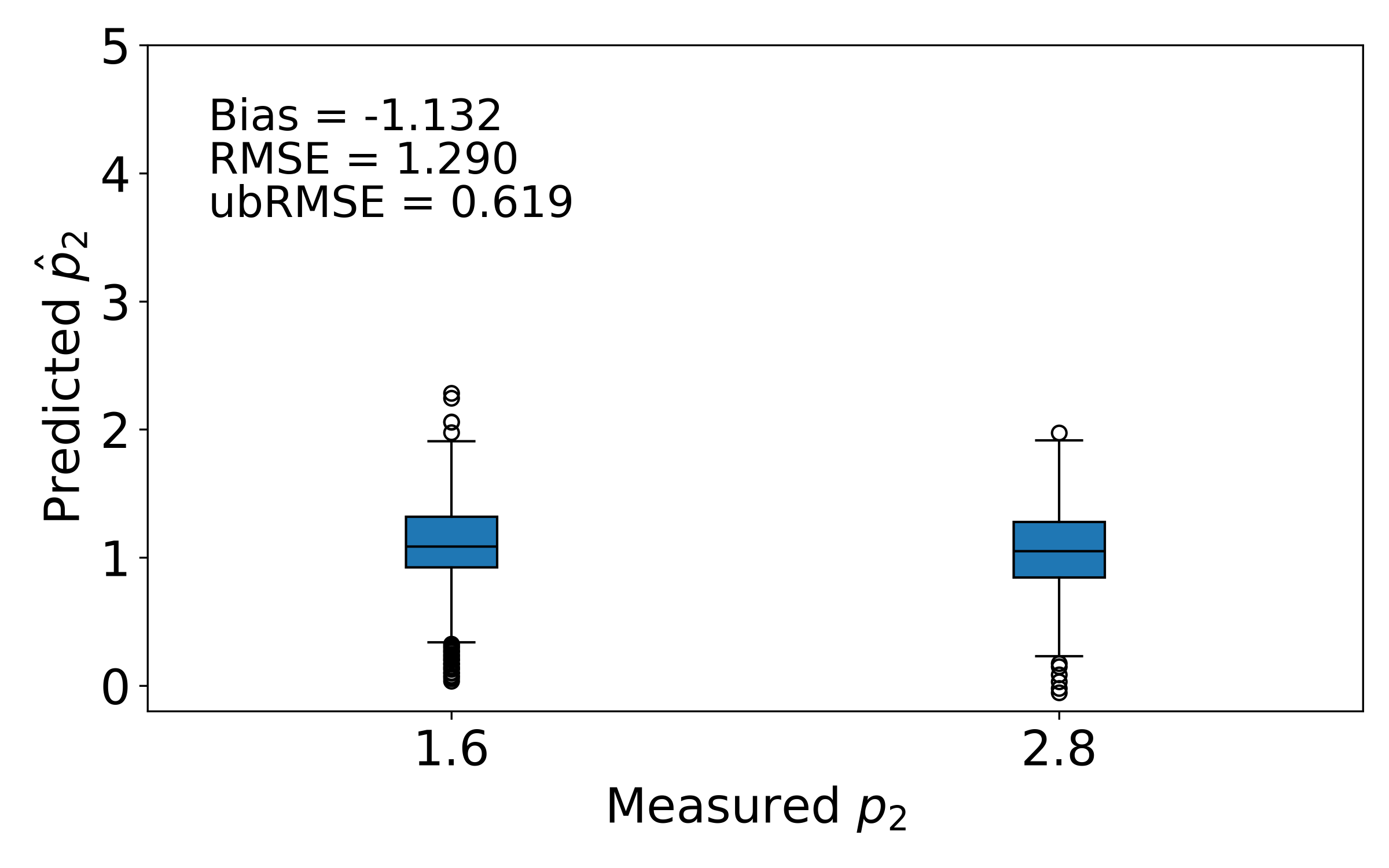}
         \captionsetup{justification=centering}
         \caption{Wood shavings permittivity estimated by DANN.}
         \label{Fig_13b}
     \end{subfigure}
     \begin{subfigure}[t]{0.32\textwidth}
         \centering
         \includegraphics[width=1\linewidth]{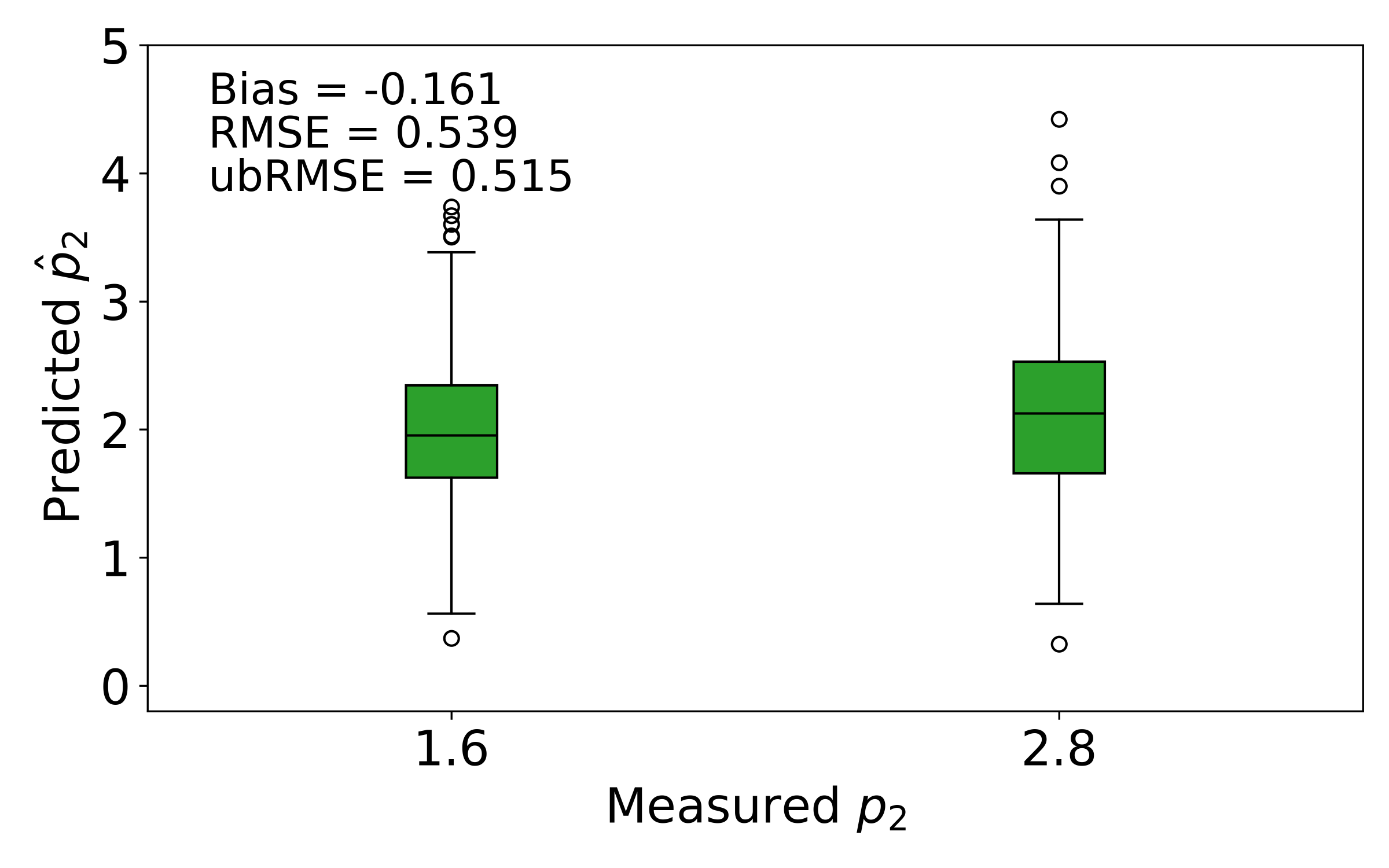}
         \captionsetup{justification=centering}
         \caption{Wood shavings permittivity estimated by HierPhyDANN-2.}
         \label{Fig_13c}
     \end{subfigure}
     \begin{subfigure}[t]{0.32\textwidth}
         \centering
         \includegraphics[width=1\linewidth]{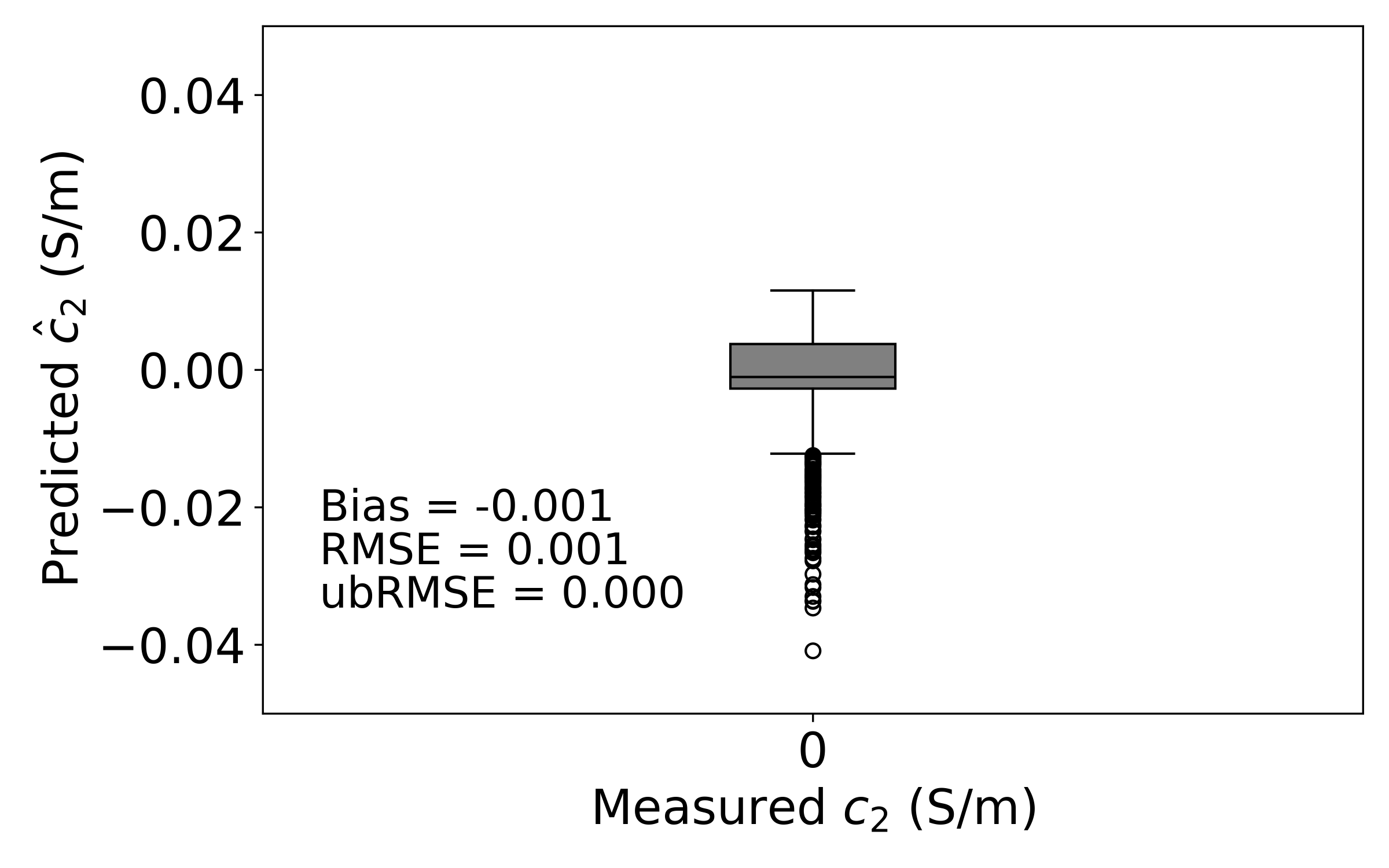}
         \captionsetup{justification=centering}
         \caption{Wood shavings conductivity estimated by 1D CNN.}
         \label{Fig_13d}
     \end{subfigure}
     \begin{subfigure}[t]{0.32\textwidth}
         \centering
         \includegraphics[width=1\linewidth]{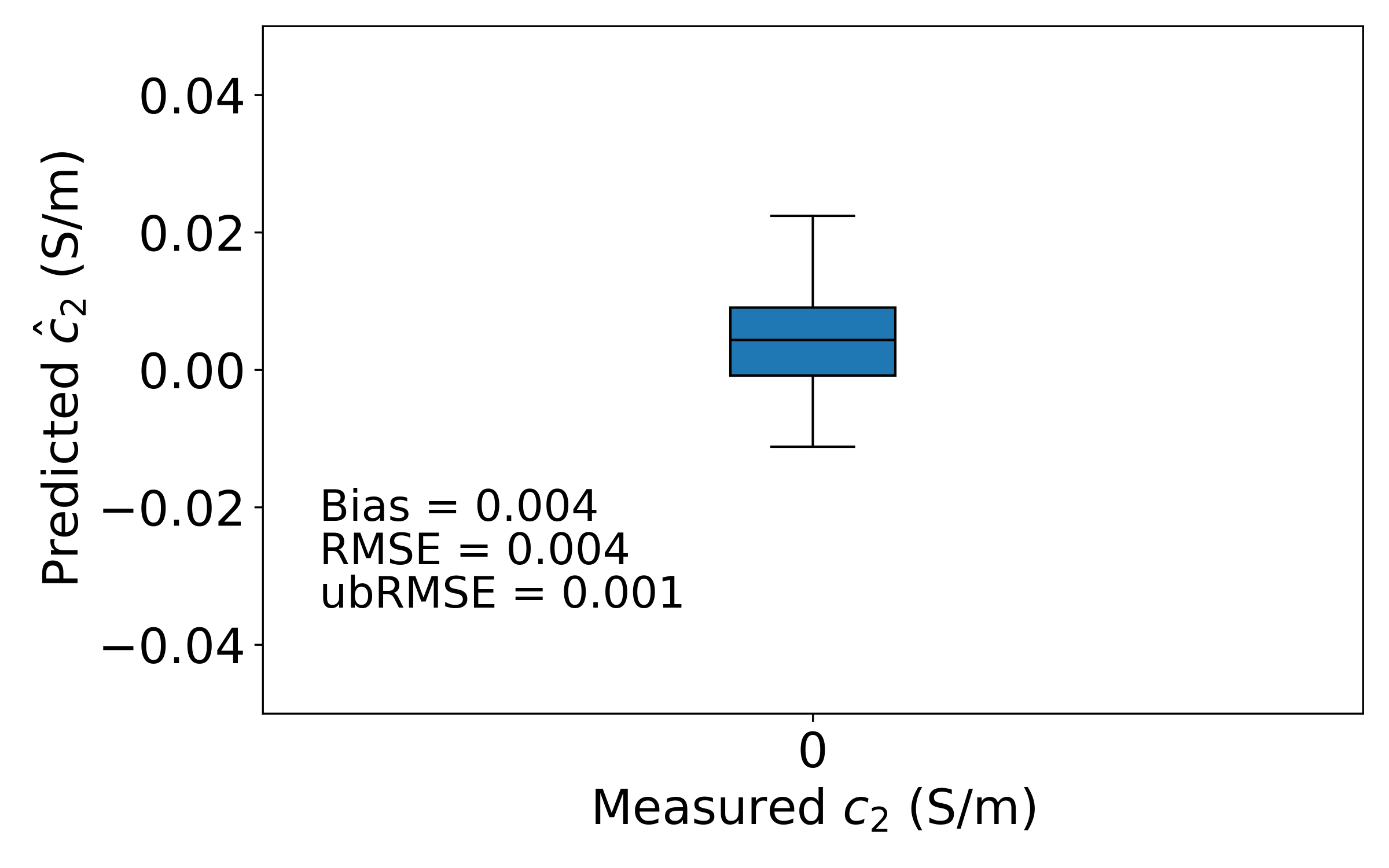}
         \captionsetup{justification=centering}
         \caption{Wood shavings conductivity estimated by DANN.}
         \label{Fig_13e}
     \end{subfigure}
     \begin{subfigure}[t]{0.32\textwidth}
         \centering
         \includegraphics[width=1\linewidth]{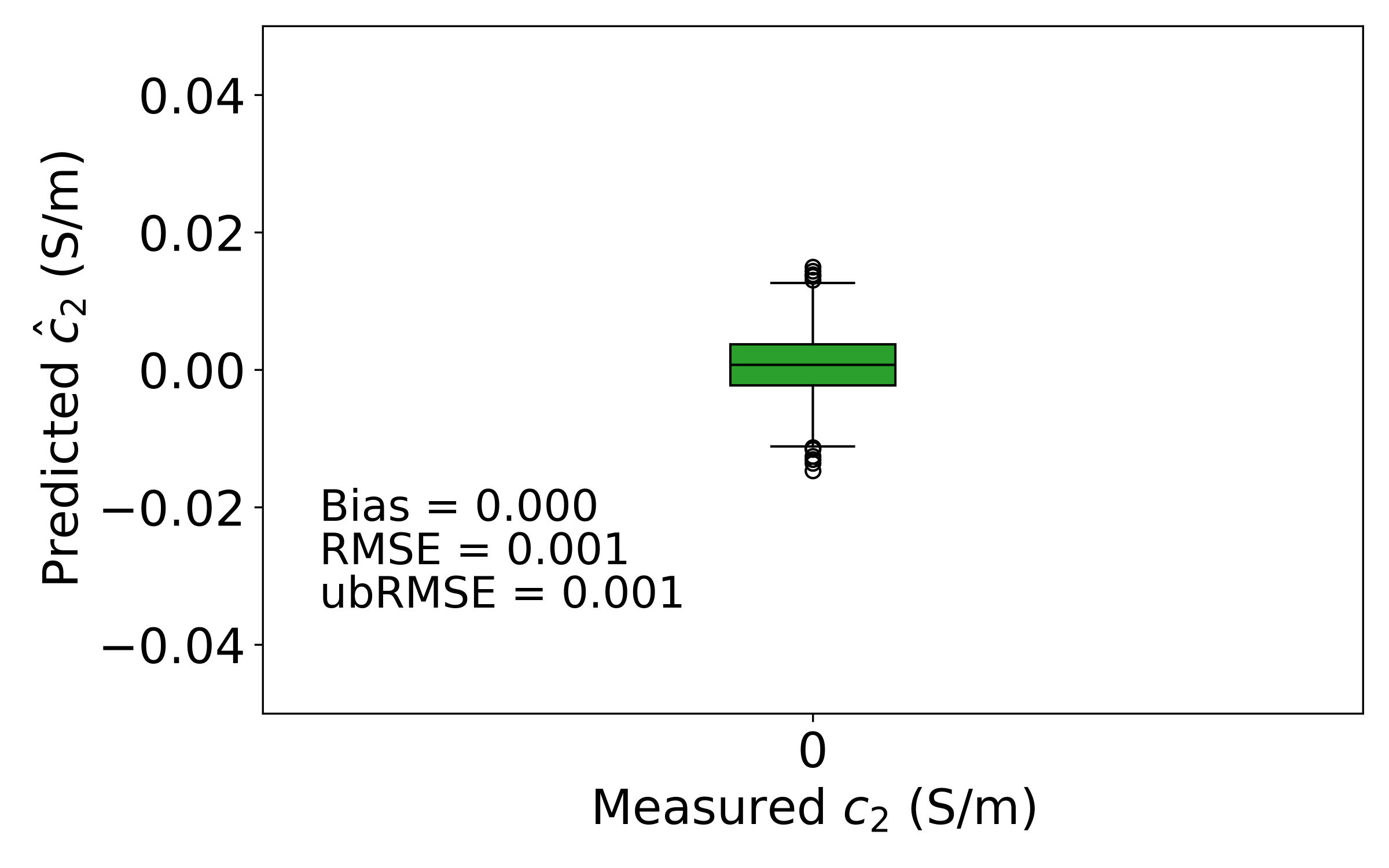}
         \captionsetup{justification=centering}
         \caption{Wood shavings conductivity estimated by HierDANN.}
         \label{Fig_13f}
     \end{subfigure}
     \begin{subfigure}[t]{0.32\textwidth}
         \centering
         \includegraphics[width=1\linewidth]{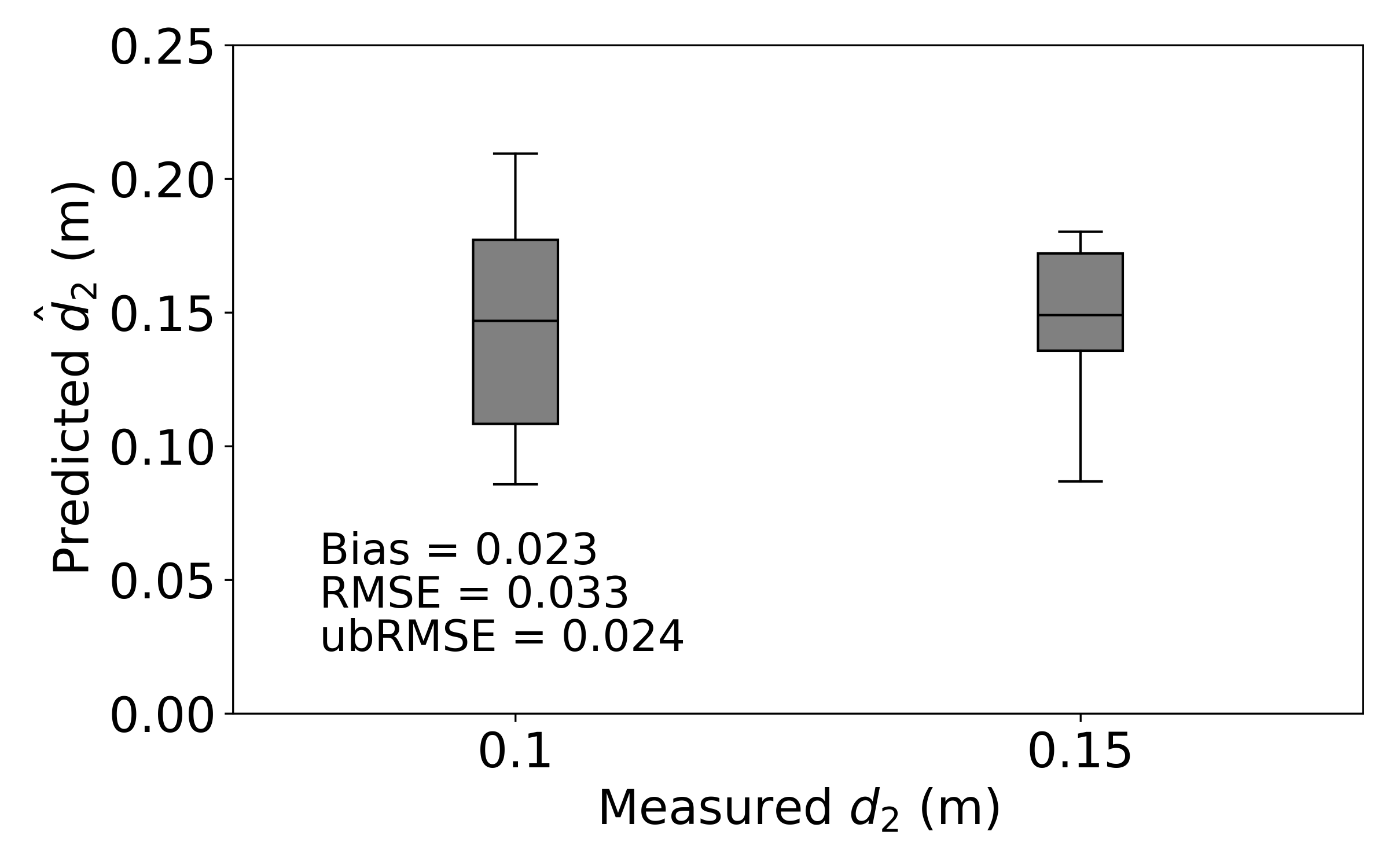}
         \captionsetup{justification=centering}
         \caption{Wood shavings depth estimated by 1D CNN.}
         \label{Fig_13g}
     \end{subfigure}
     \begin{subfigure}[t]{0.32\textwidth}
         \centering
         \includegraphics[width=1\linewidth]{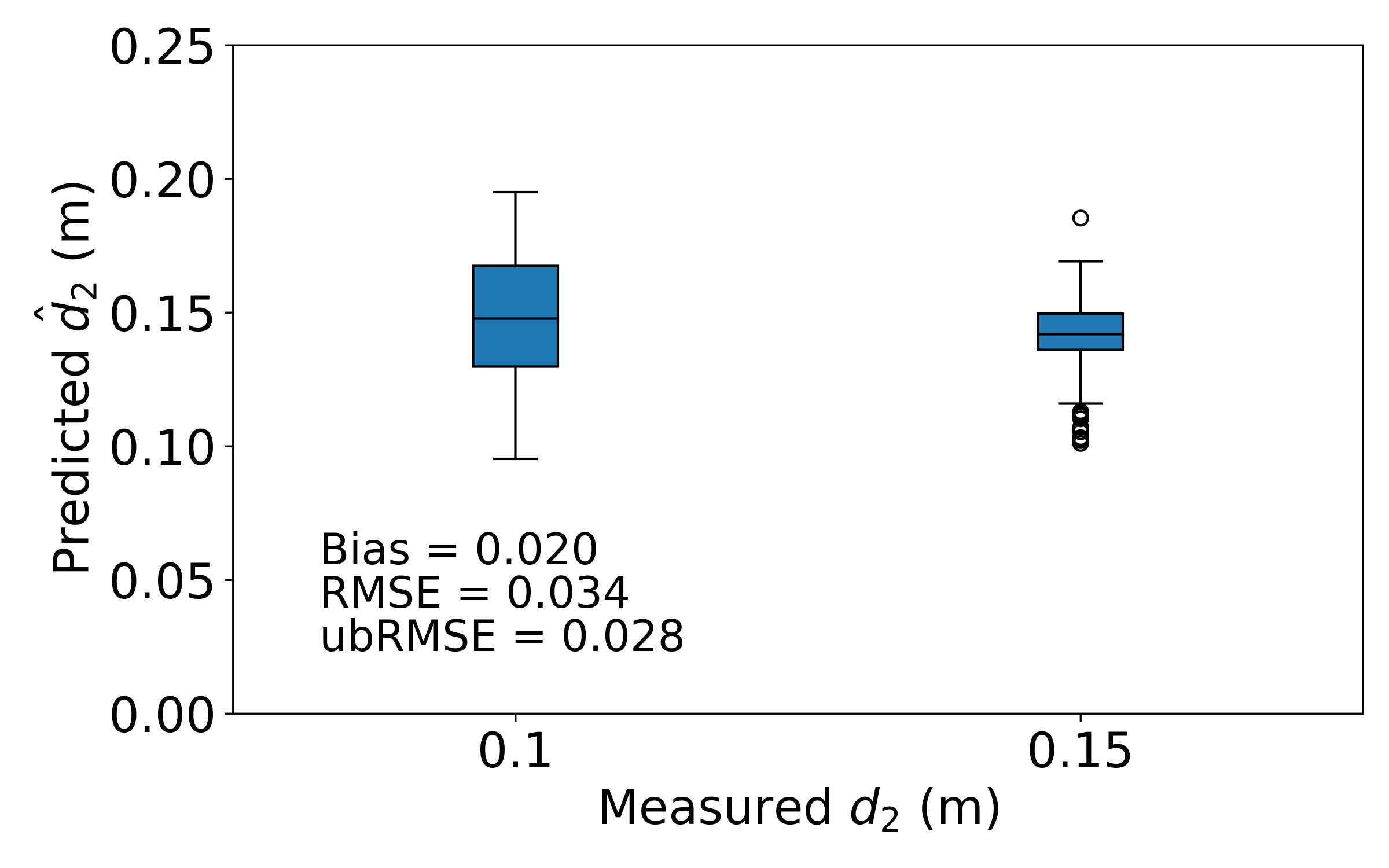}
         \captionsetup{justification=centering}
         \caption{Wood shavings depth estimated by DANN.}
         \label{Fig_13h}
     \end{subfigure}
     \begin{subfigure}[t]{0.32\textwidth}
         \centering
         \includegraphics[width=1\linewidth]{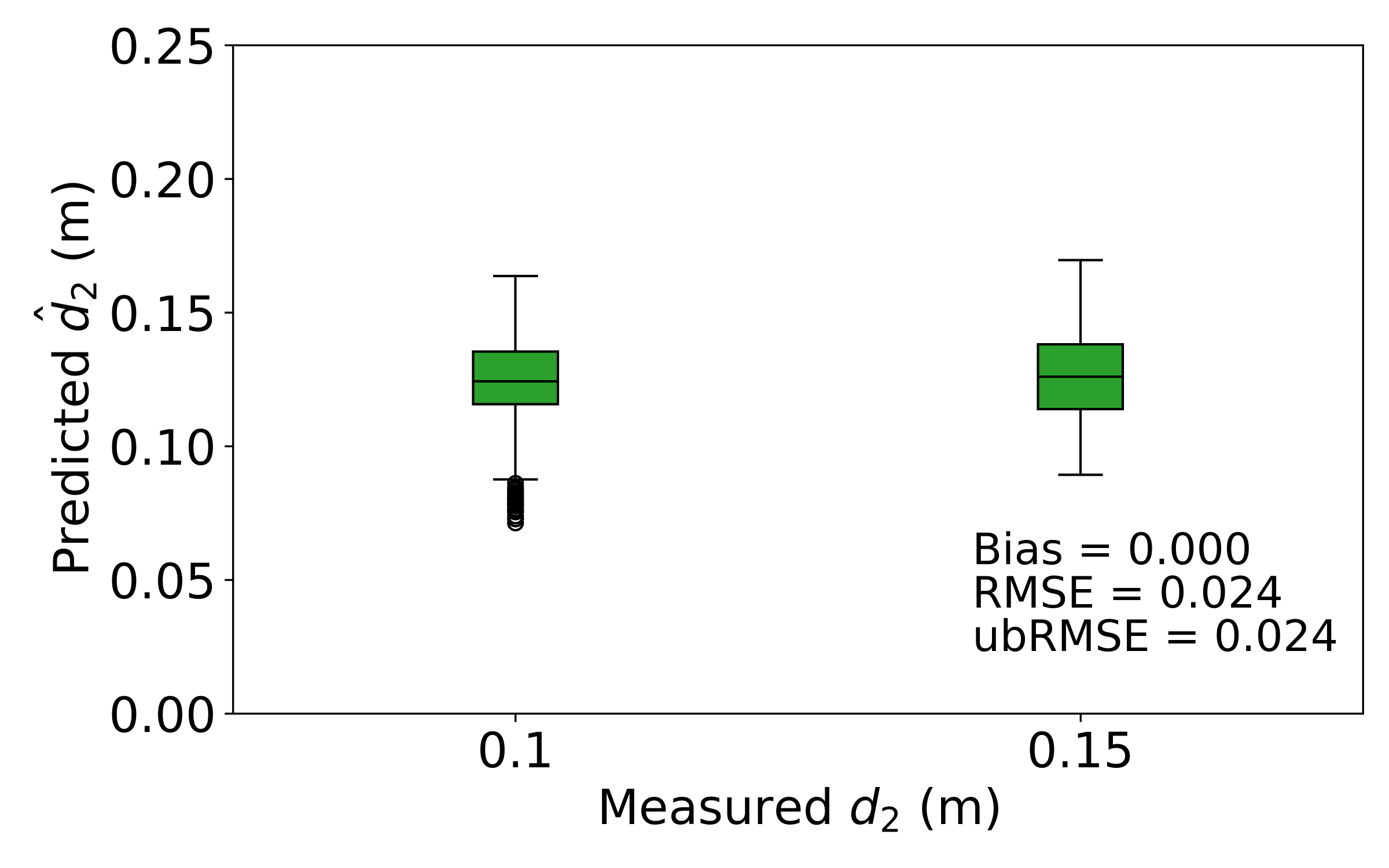}
         \captionsetup{justification=centering}
         \caption{Wood shavings depth estimated by HierPhyDANN-2.}
         \label{Fig_13i}
     \end{subfigure}
     \caption{\centering Laboratory test results for wood shavings permittivity, conductivity, and depth in the two-layer, five-parameter estimation.}
     \label{Fig_13}
\end{figure*}

For soil conductivity estimation, as shown in Table \ref{T6}, the 1D CNN and DANN yield comparable $R$ values and Bias, whereas HierPhyDANN-2 delivers the best performance, achieving the highest $R$ value along with a lower Bias. The corresponding prediction results are presented in Figure \ref{Fig_12}. HierPhyDANN-2 utilizes the estimated soil permittivity values ($\hat{p}_1 = 5.5, 4.5, 6.5, 7.5, 9.5, 10.0$) to refine the simulated dataset by removing radar scans with permittivity values deviating from the estimated ones. The updated dataset is then used for soil conductivity estimation. The laboratory test results for wood shavings permittivity, conductivity, and depth estimation are summarized in Table \ref{T13}. HierDANN attains the lowest Bias for wood shavings conductivity, while HierPhyDANN-2 achieves the lowest Bias for wood shavings permittivity and depth. The corresponding prediction results are shown in Figure \ref{Fig_13}. Because there is only one variation for wood shavings conductivity and two variations for wood shavings permittivity and depth, correlation plots are not informative. Instead, box plots are generated based on the estimation results for each parameter. In HierPhyDANN-2, the estimated soil permittivity ($\hat{p}_1 = 5.5, 4.5, 6.5, 7.5, 9.5, 10.0$), soil conductivity ($\hat{c}_1 = 0.010, 0.020, 0.035, 0.060, 0.060, 0.095$ S/m), wood shavings permittivity ($\hat{p}_2 = 2.0, 2.0$), and wood shavings conductivity ($\hat{c}_2 = 0$ S/m) values are employed for their corresponding subsequent parameter estimations by pruning the simulated dataset according to the known parameter values. In addition, HierDANN achieves zero Bias and the lowest standard deviation for wood shavings conductivity estimation compared with the baseline 1D CNN and DANN.

\subsection{Field Test for Single- and Two-Layer Materials}
The field test focuses on tracking variations in soil permittivity and volumetric water content (VWC), during natural rainfall and subsequent drying periods. A comparison of the soil permittivity estimation results obtained using different approaches for the single-layer material (soil) and the two-layer materials (soil–wood chips and soil–leaves) is presented in Tables \ref{T7} and \ref{T14}, respectively.

\begin{table*}[ht]
    \small\sf\centering
    \caption{\centering Field test results for soil permittivity estimation (Single-layer material).}
    \label{T7}
    \begin{threeparttable}
    \begin{tabular}{@{\extracolsep{\fill}}lcccccccc}
        \toprule
        \multirow{2}{*}{Approach}
            & \multicolumn{4}{c}{Soil (Calibrated FDTD)}
            & \multicolumn{4}{c}{Soil (Uncalibrated FDTD)} \\
        \cmidrule(lr){2-5} \cmidrule(lr){6-9}
            & $R$ & Bias & RMSE & ubRMSE
            & $R$ & Bias & RMSE & ubRMSE \\
        \midrule
        1D CNN (baseline)    & 0.892 & -0.524 & 1.918 & 1.845 & 0.471 & 2.925 & 4.113 & 2.892 \\
        DANN (baseline)      & 0.922 &  1.736 & 2.208 & 1.363 & 0.822 & 3.614 & 4.107 & 1.950 \\
        HierDANN (ours)      & 0.978 &  0.358 & 0.796 & 0.711 & \textit{\textbf{\underline{0.945}}} & 9.865 & 10.635 & 3.972 \\
        PhyDANN-1 (ours)     & 0.978 &  0.704 & 0.979 & 0.680 & 0.616 & 4.146 & 4.873 & 2.560 \\
        HierPhyDANN-1 (ours) & 0.972 &  1.716 & 1.940 & 0.905 & 0.614 & 3.228 & 4.183 & 2.660 \\
        PhyDANN-2 (ours)     & 0.962 &  0.266 & 1.016 & 0.980 & 0.758 & 3.447 & 4.127 & 2.268 \\
        HierPhyDANN-2 (ours) & \textit{\textbf{\underline{0.987}}} & 1.053 & 1.180 & 0.533 & 0.663 & 4.080 & 4.763 & 2.457 \\
        \bottomrule
    \end{tabular}
    \begin{tablenotes}
        \item[1] The bold values indicate the best correlation coefficient $R$ among the compared approaches.
    \end{tablenotes}
    \end{threeparttable}
\end{table*}

\begin{table*}[ht]
    \small\sf\centering
    \caption{\centering Field test results for soil permittivity estimation (Two-layer material).}
    \label{T14}
    \begin{threeparttable}
    \begin{tabular}{@{\extracolsep{\fill}}lcccccccc}
        \toprule
        \multirow{2}{*}{Approach}
            & \multicolumn{4}{c}{Soil-wood chips}
            & \multicolumn{4}{c}{Soil-leaves} \\
        \cmidrule(lr){2-5} \cmidrule(lr){6-9}
            & $R$ & Bias & RMSE & ubRMSE
            & $R$ & Bias & RMSE & ubRMSE \\
        \midrule
        1D CNN (baseline)    
            & 0.827 & -2.333 & 3.298 & 2.332
            & 0.613 & -1.673 & 3.104 & 2.615 \\
        DANN (baseline)      
            & 0.678 & 2.980 & 4.759 & 3.710
            & 0.597 & -1.239 & 2.927 & 2.652 \\
        HierDANN (ours)      
            & 0.888 & 0.568 & 1.602 & 1.498
            & \textit{\textbf{\underline{0.835}}} & -0.648 & 1.954 & 1.843 \\
        PhyDANN-1 (ours)     
            & 0.863 & 1.865 & 2.509 & 1.678
            & 0.589 & 1.113 & 2.852 & 2.626 \\
        HierPhyDANN-1 (ours) 
            & 0.908 & 0.227 & 1.793 & 1.778
            & 0.734 & 0.093 & 2.213 & 2.211 \\
        PhyDANN-2 (ours)     
            & \textit{\textbf{\underline{0.975}}} & -1.361 & 1.707 & 1.030
            & 0.648 & -1.213 & 2.825 & 2.551 \\
        HierPhyDANN-2 (ours) 
            & 0.895 & 0.691 & 1.860 & 1.727
            & 0.719 & 2.430 & 3.399 & 2.377 \\
        \bottomrule
    \end{tabular}
    \begin{tablenotes}
        \item[1] The bold values indicate the best correlation coefficient $R$ among the compared approaches.
        \item[2] The FDTD model used for the two-layer material is calibrated.
    \end{tablenotes}
    \end{threeparttable}
\end{table*}

\begin{figure*}[ht]
     \centering
     \begin{subfigure}[t]{0.32\textwidth}
         \centering
         \includegraphics[width=1\linewidth]{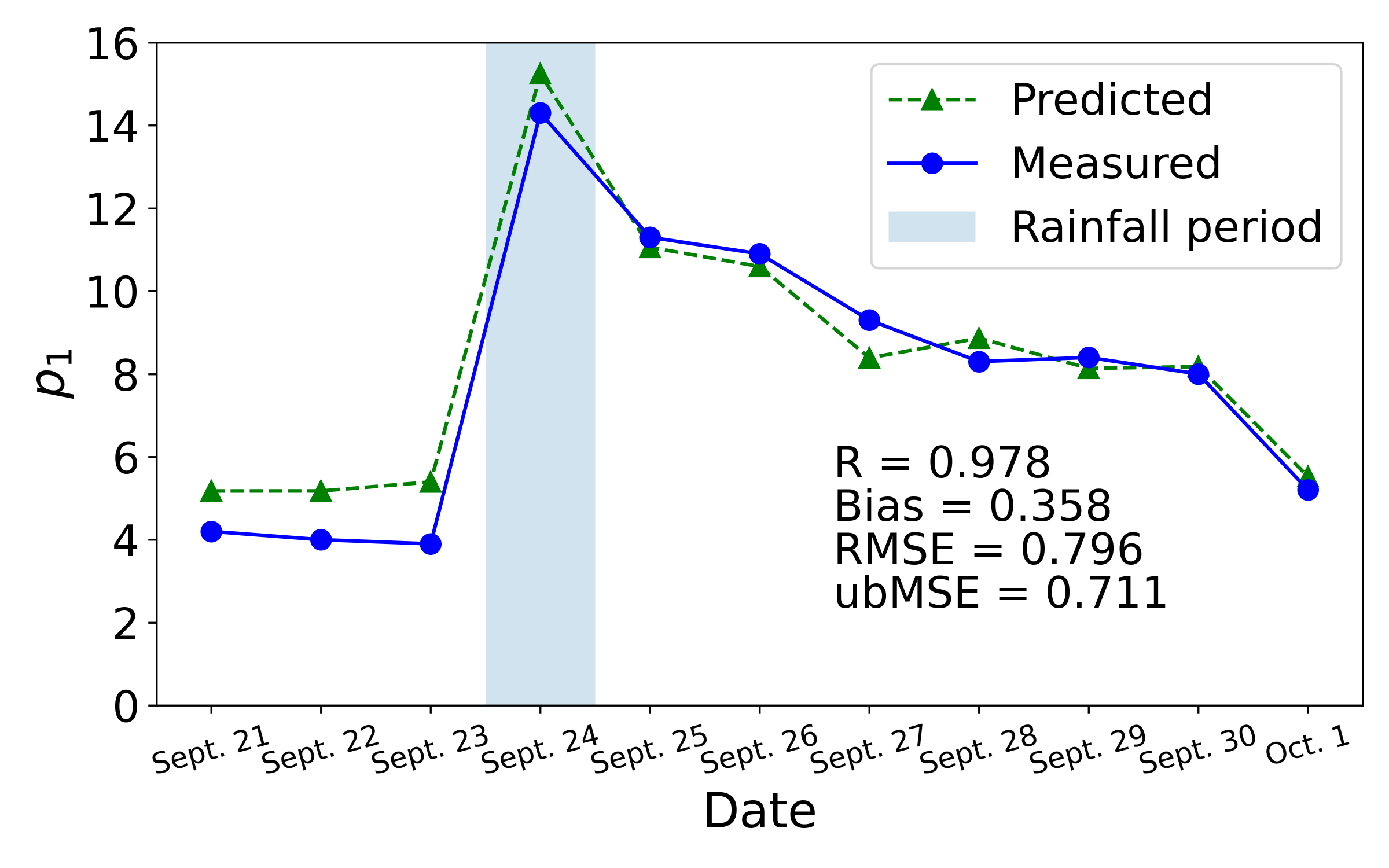}
         \captionsetup{justification=centering}
         \caption{Soil permittivity in single-layer material (soil) estimated by HierDANN.}
         \label{Fig_14a}
     \end{subfigure}
    \begin{subfigure}[t]{0.32\textwidth}
         \centering
         \includegraphics[width=1\linewidth]{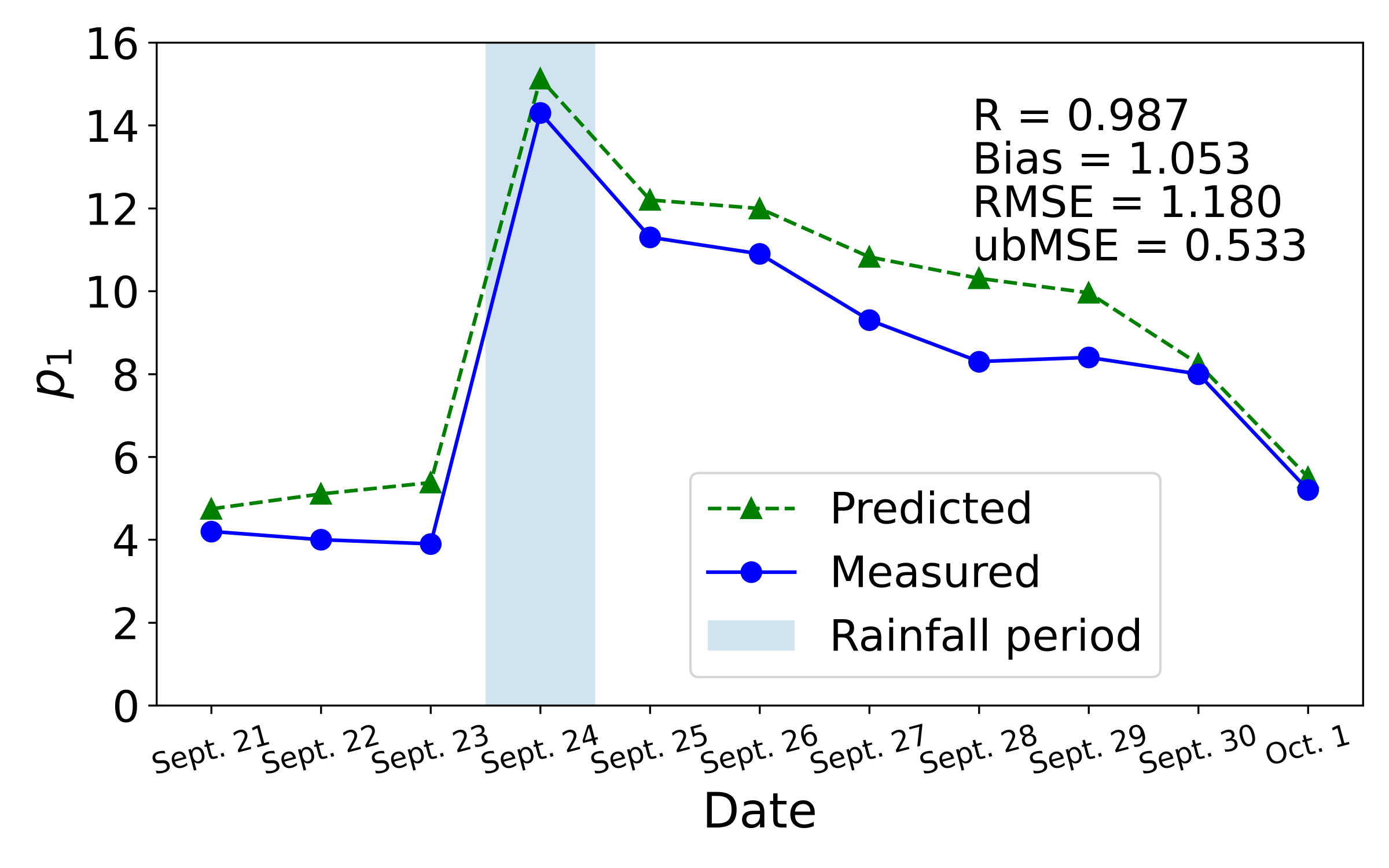}
         \captionsetup{justification=centering}
         \caption{Soil permittivity in single-layer material (soil) estimated by HierPhyDANN-2.}
         \label{Fig_14b}
     \end{subfigure}
     \begin{subfigure}[t]{0.32\textwidth}
         \centering
         \includegraphics[width=1\linewidth]{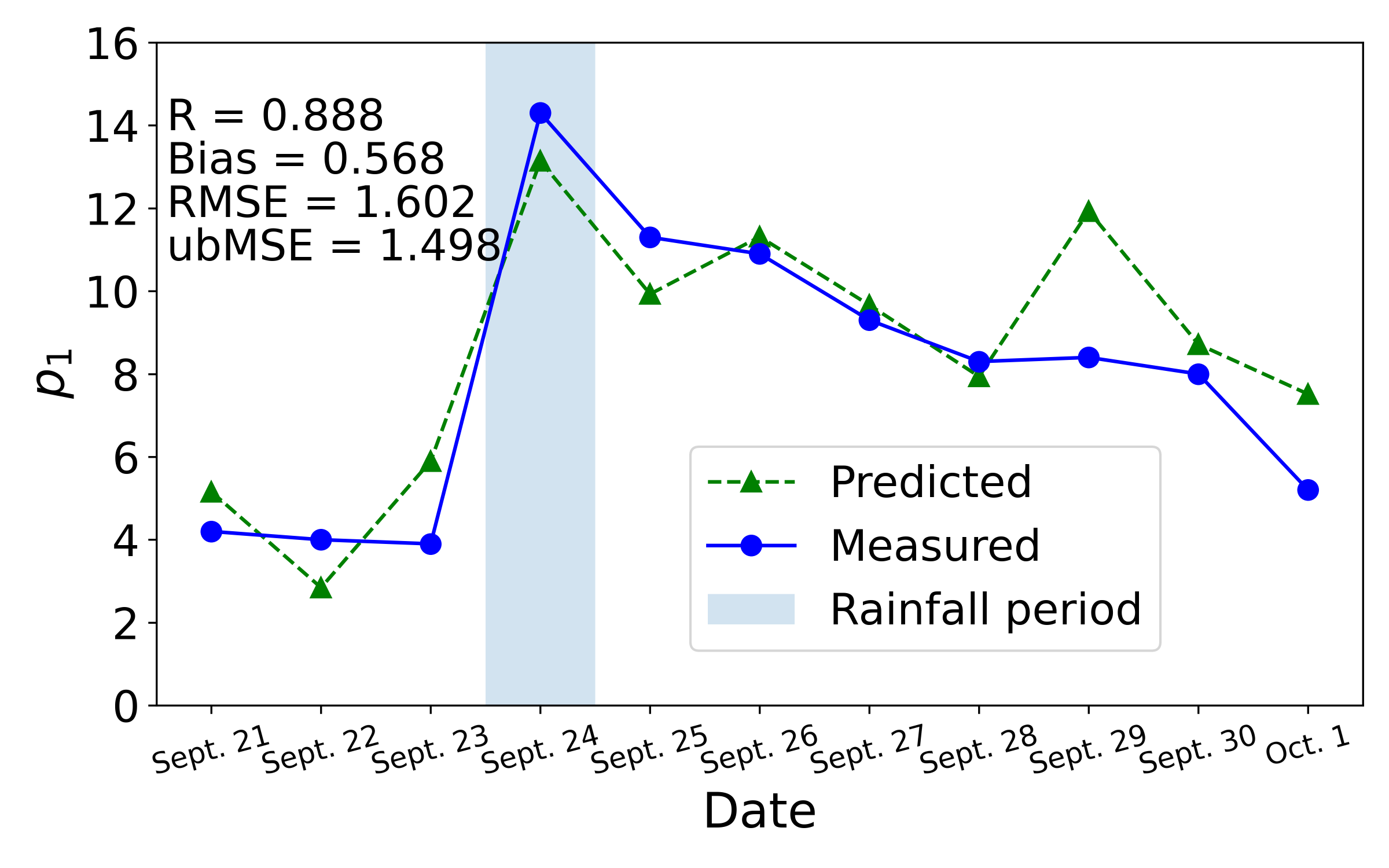}
         \captionsetup{justification=centering}
         \caption{Soil permittivity in two-layer material (soil-wood chips) estimated by HierDANN.}
         \label{Fig_14c}
    \end{subfigure}
    \begin{subfigure}[t]{0.32\textwidth}
         \centering
         \includegraphics[width=1\linewidth]{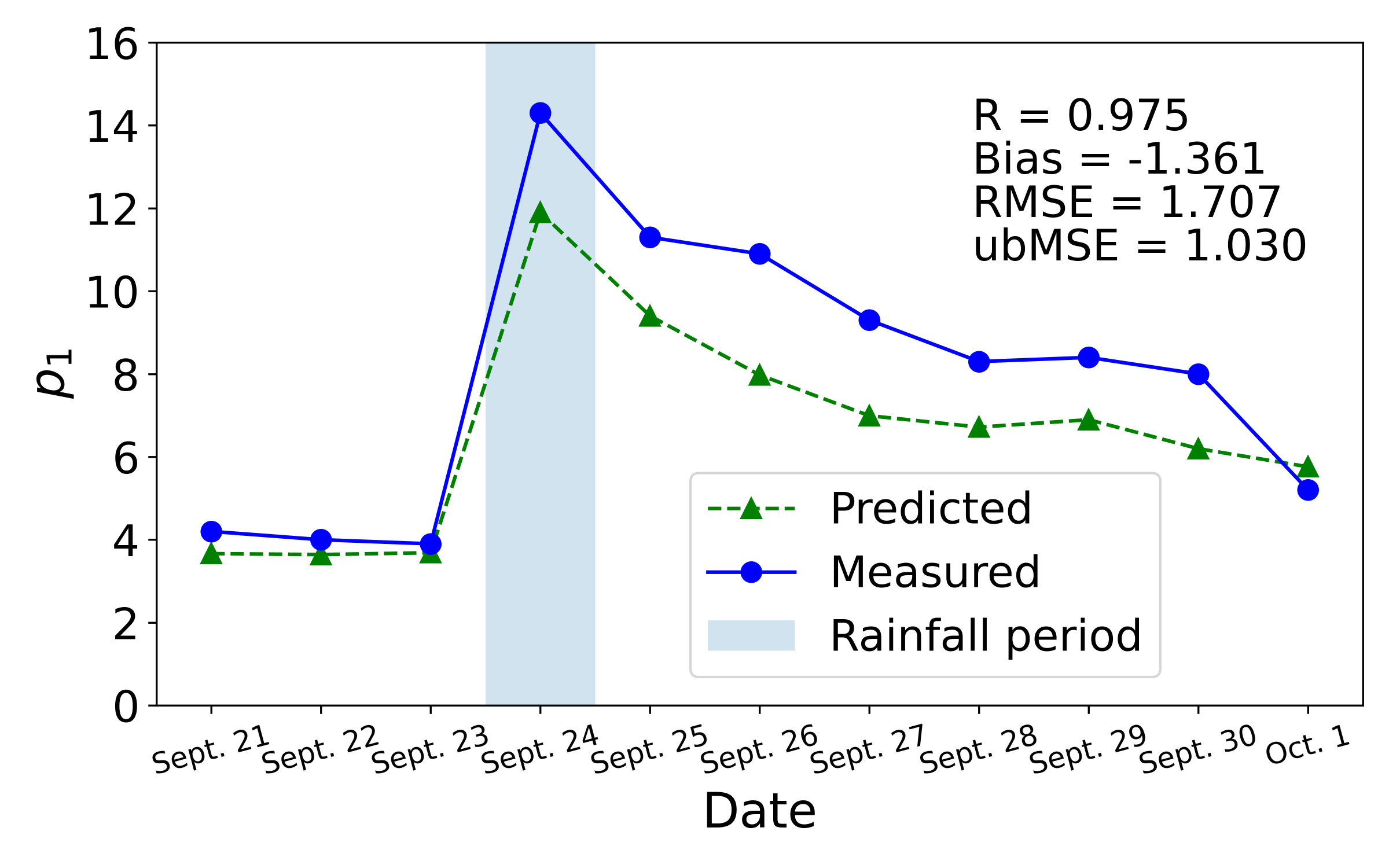}
         \captionsetup{justification=centering}
         \caption{Soil permittivity in two-layer material (soil-wood chips) estimated by PhyDANN-2.}
         \label{Fig_14d}
     \end{subfigure}
     \begin{subfigure}[t]{0.32\textwidth}
         \centering
         \includegraphics[width=1\linewidth]{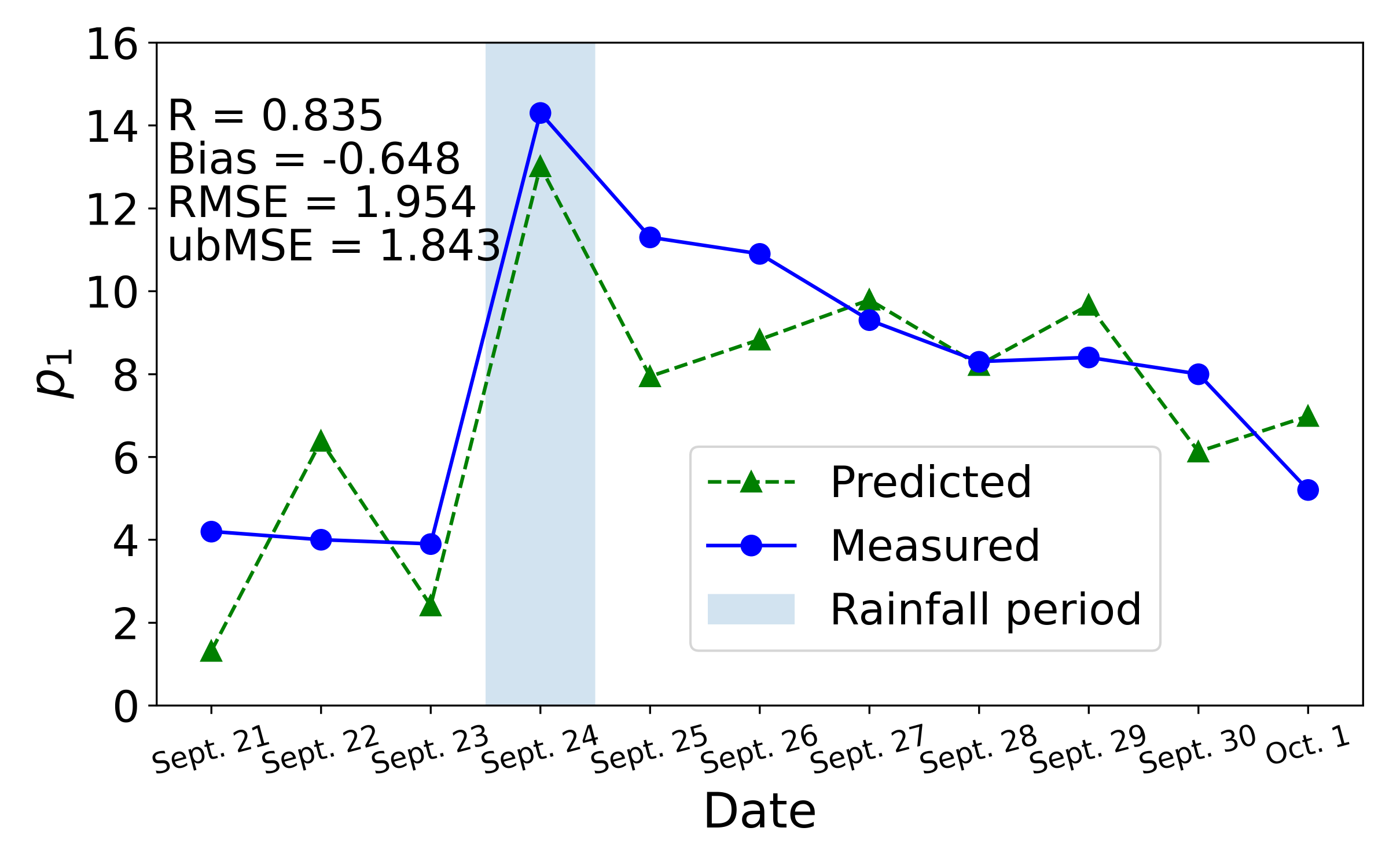}
         \captionsetup{justification=centering}
         \caption{Soil permittivity in two-layer material (soil-leaves) estimated by HierDANN.}
         \label{Fig_14e}
     \end{subfigure}
     \caption{\centering Field test results for soil permittivity estimation.}
     \label{Fig_14}
\end{figure*}

\begin{figure*}[ht]
     \centering
     \begin{subfigure}[t]{0.32\textwidth}
         \centering
         \includegraphics[width=1\linewidth]{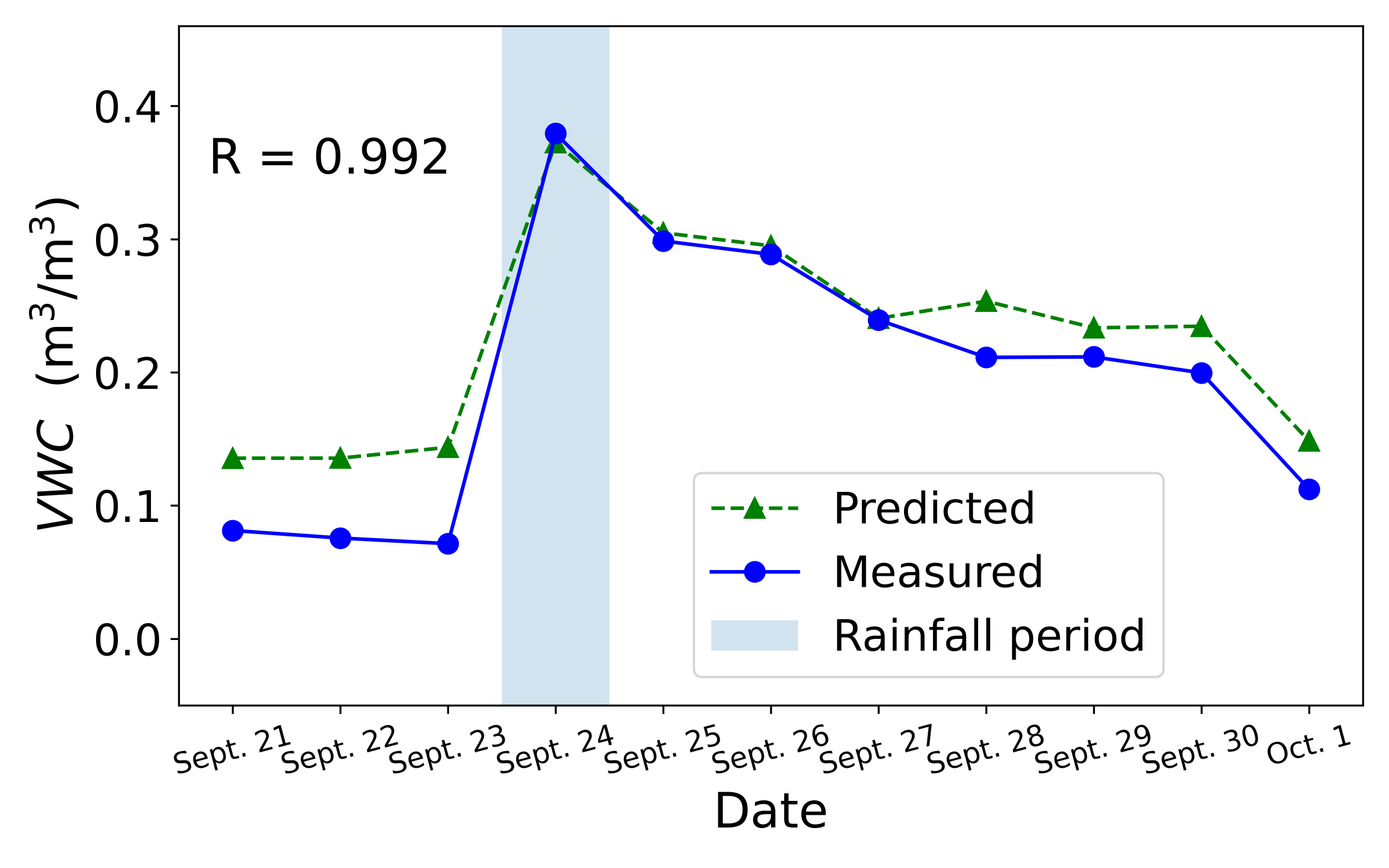}
         \captionsetup{justification=centering}
         \caption{Soil VWC in single-layer material (soil) estimated by HierDANN.}
         \label{Fig_Field_VWC_a}
     \end{subfigure}
     \begin{subfigure}[t]{0.32\textwidth}
         \centering
         \includegraphics[width=1\linewidth]{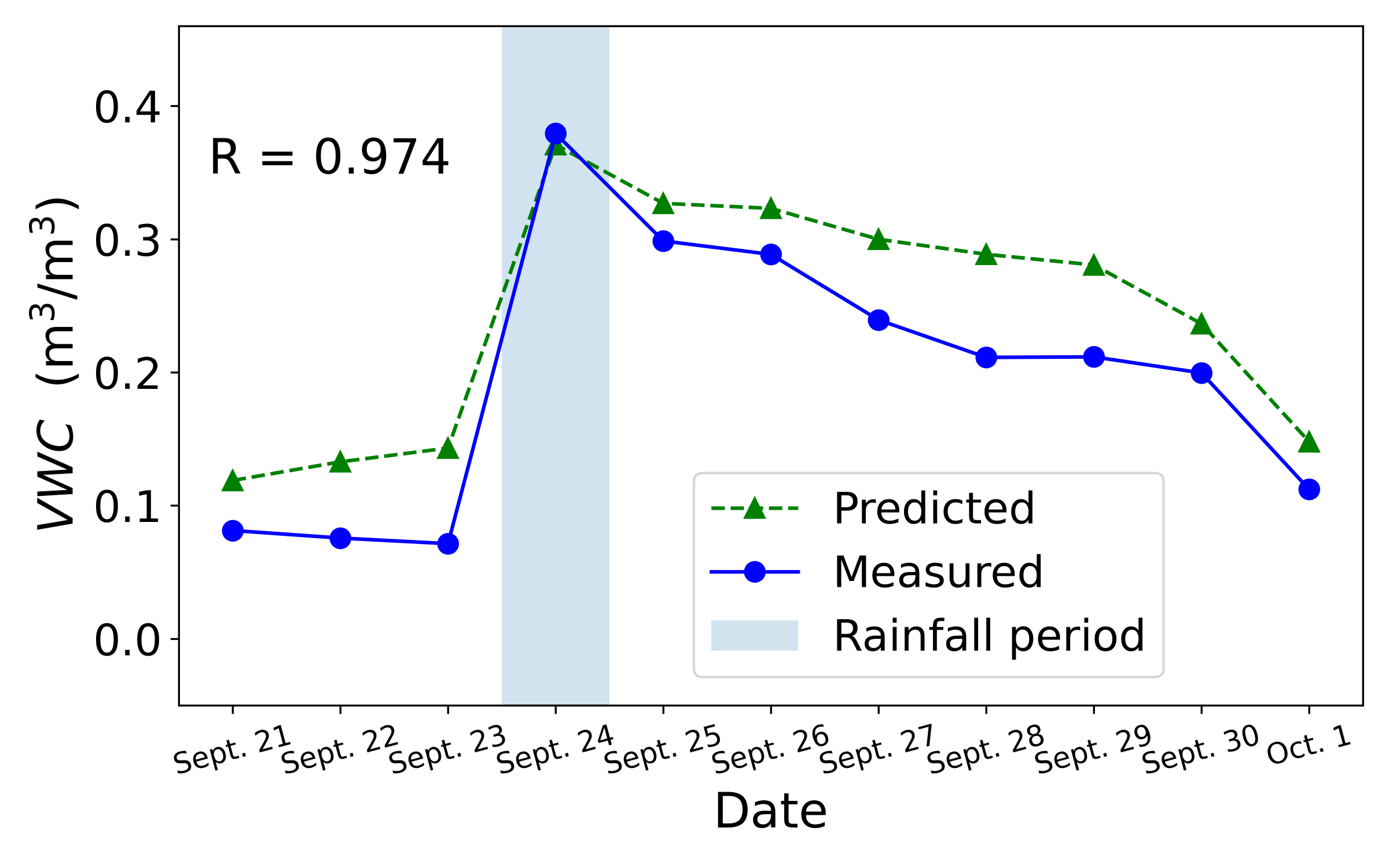}
         \captionsetup{justification=centering}
         \caption{Soil VWC in single-layer material (soil) estimated by HierPhyDANN-2.}
         \label{Fig_Field_VWC_b}
     \end{subfigure}
     \begin{subfigure}[t]{0.32\textwidth}
         \centering
         \includegraphics[width=1\linewidth]{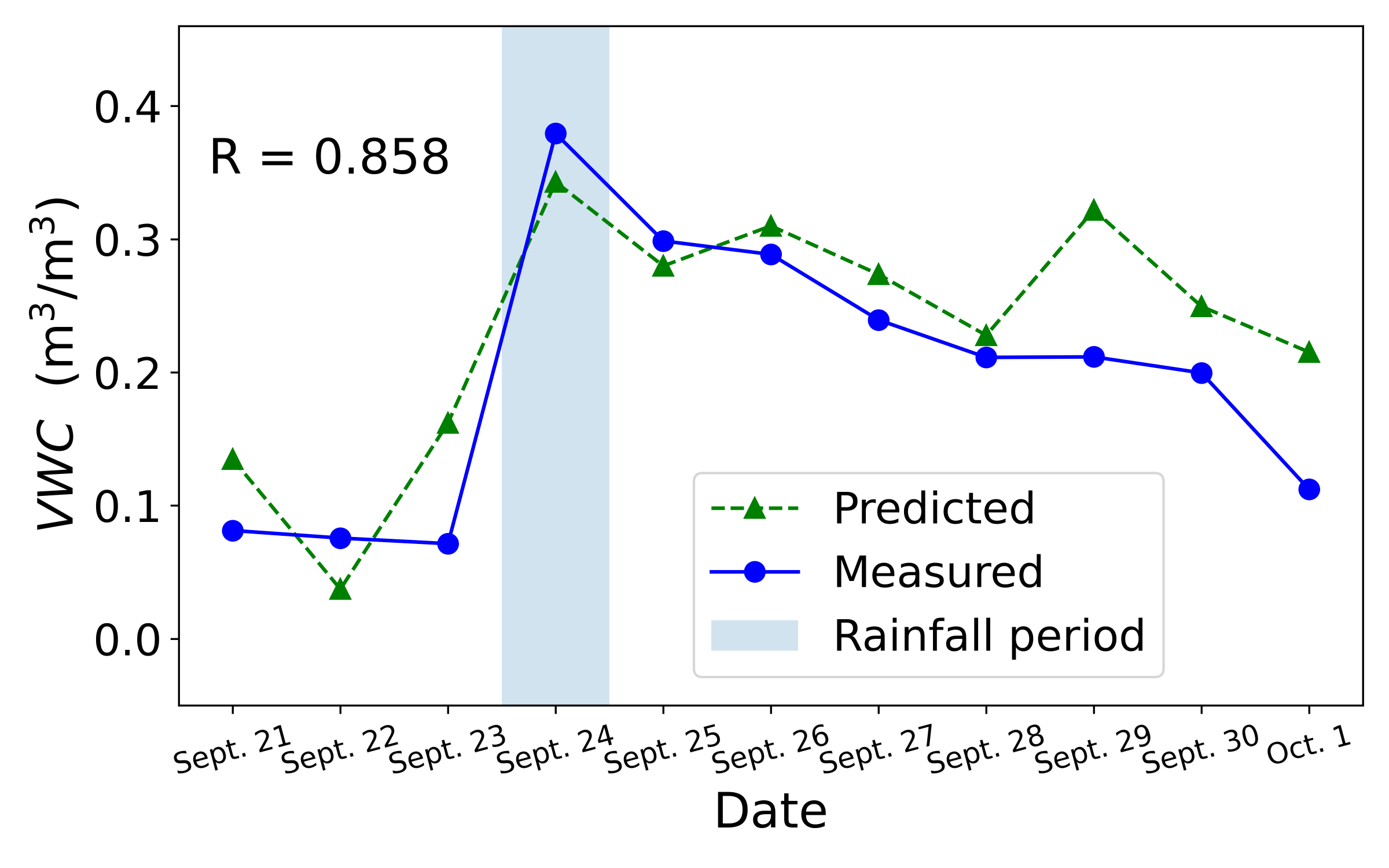}
         \captionsetup{justification=centering}
         \caption{Soil VWC in two-layer material (soil-wood chips) estimated by HierDANN.}
         \label{Fig_Field_VWC_c}
     \end{subfigure}
     \begin{subfigure}[t]{0.32\textwidth}
         \centering
         \includegraphics[width=1\linewidth]{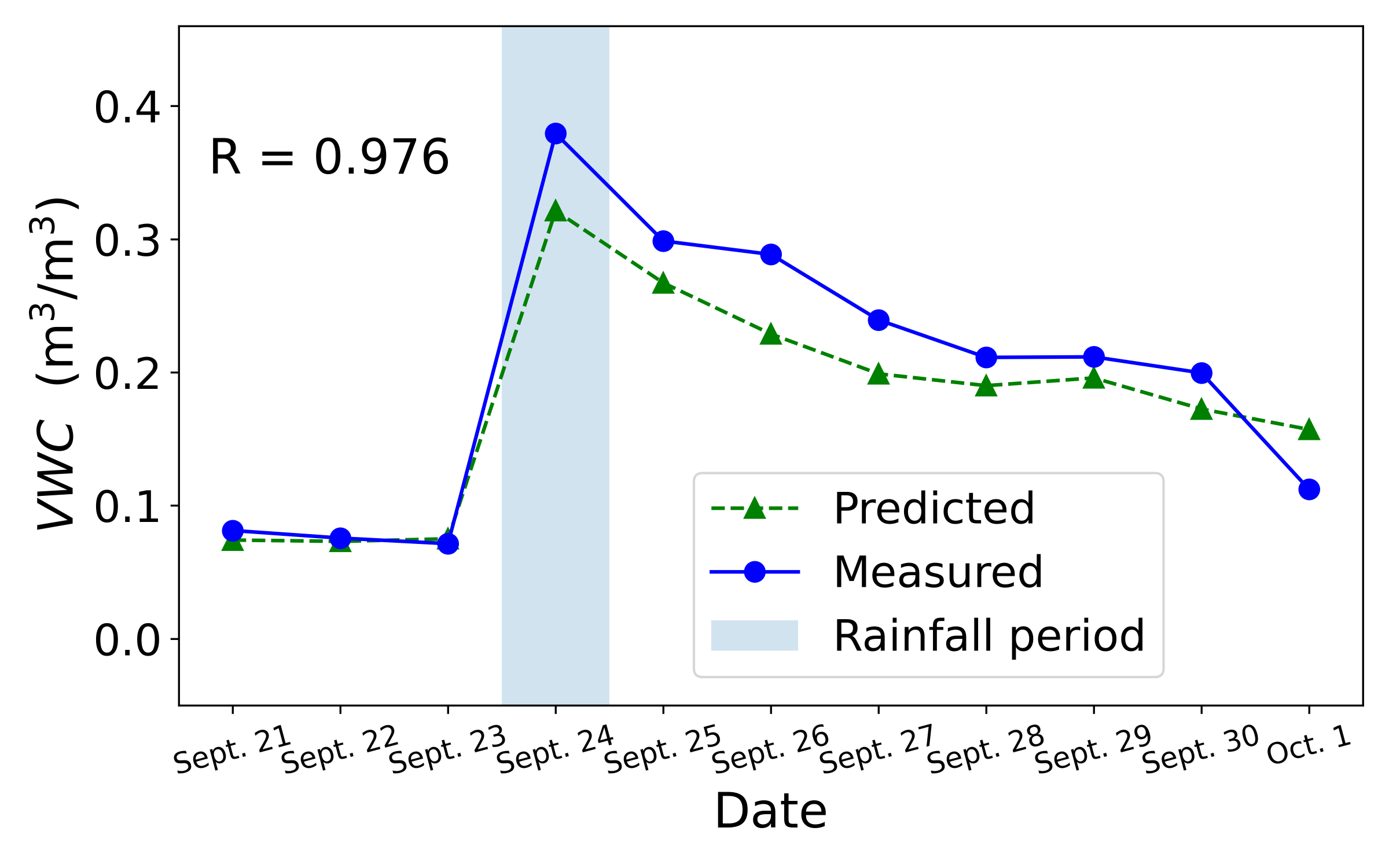}
         \captionsetup{justification=centering}
         \caption{Soil VWC in two-layer material (soil-wood chips) estimated by PhyDANN-2.}
         \label{Fig_Field_VWC_d}
     \end{subfigure}
     \begin{subfigure}[t]{0.32\textwidth}
         \centering
         \includegraphics[width=1\linewidth]{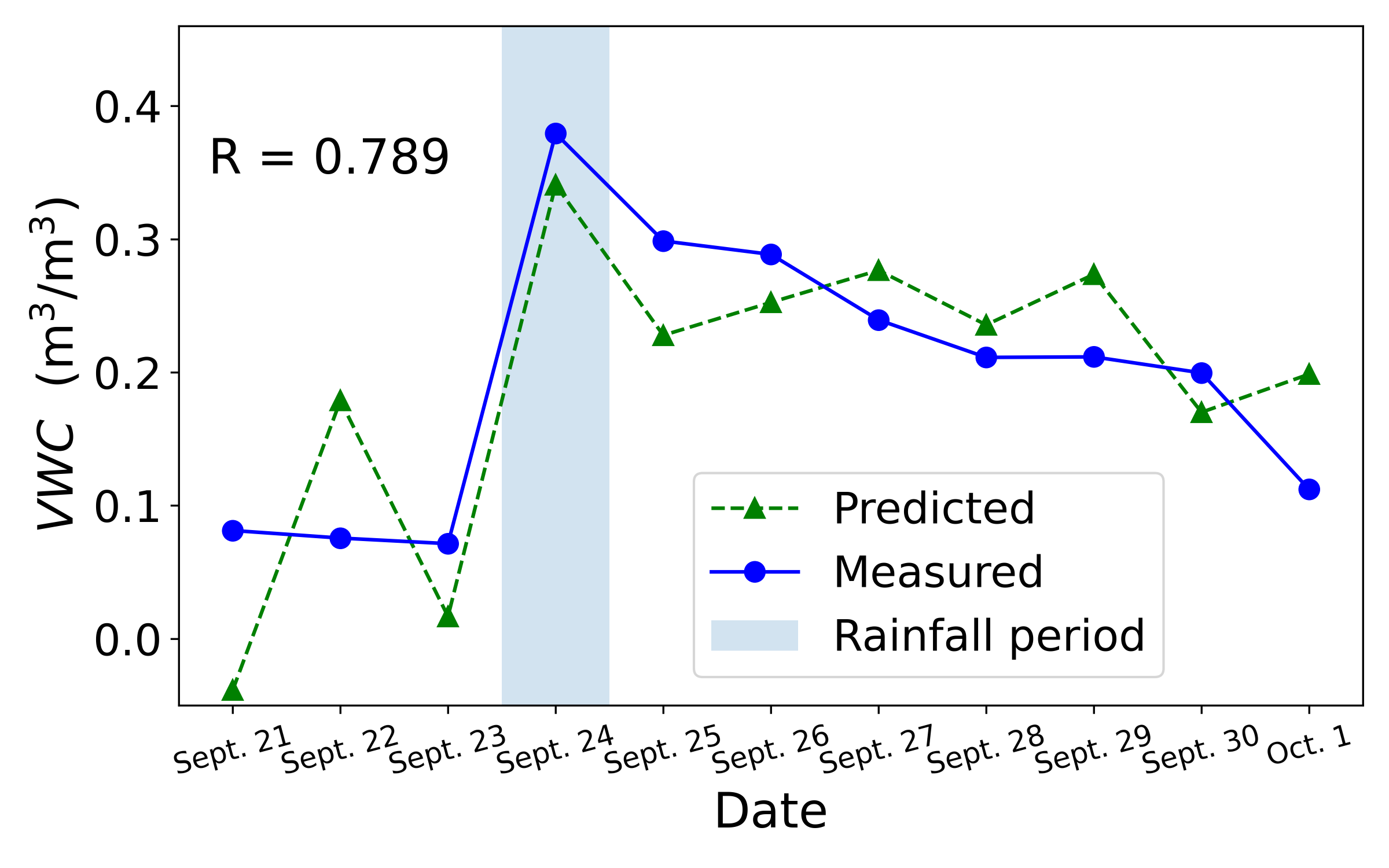}
         \captionsetup{justification=centering}
         \caption{Soil VWC in two-layer material (soil-leaves) estimated by HierDANN.}
         \label{Fig_Field_VWC_e}
     \end{subfigure}
     \caption{\centering Field test results for volumetric water content (VWC) of soil.}
     \label{Fig_VWC_Field}
\end{figure*}

With the calibrated FDTD model, the proposed HierPhyDANN-2, PhyDANN-2, and HierDANN achieve the highest $R$ values in soil permittivity estimation for the single-layer material (soil) and the two-layer materials (soil–wood chips and soil–leaves), respectively. For the soil and soil–wood chips materials, HierDANN attains $R$ values comparable to those of HierPhyDANN-2 and PhyDANN-2 while achieving lower Bias, respectively. The comparison between the measured and predicted soil permittivity values is shown in Figure \ref{Fig_14}, and the predictions of VWC are shown in Figure \ref{Fig_VWC_Field}. The peak soil permittivity and the corresponding VWC observed on September 24 resulted from rainfall on that day, followed by a gradual decrease due to subsequent drying under sunlight. The proposed approaches exhibit superior estimation accuracy in the single-layer material compared with the two-layer cases, as the presence of the top vegetation layer attenuates the radar signal penetrating into the soil. Nonetheless, the proposed approaches with physics-guided hierarchical domain adaptation accurately capture the variation trend of soil permittivity before and after the rainfall for both the single- and two-layer materials. In contrast, the baseline 1D CNN and DANN models yield lower $R$ values for both the single-layer and two-layer materials compared with the proposed approaches. The trends in soil moisture content (VWCs) estimated by HierPhyDANN-2, PhyDANN-2, and HierDANN also agree with the true VWC-trends obtained from gravimetric oven-drying tests (Figure \ref{Fig_VWC_Field}). Overall, the proposed physics-guided hierarchical domain adaptation approaches demonstrate superior performance in estimating material properties in both single- and two-layer configurations.

If the FDTD model is not properly calibrated, all approaches exhibit a substantial decline in $R$ values and an increase in Bias, as shown in Table \ref{T7}, compared with the cases where the FDTD model is calibrated. Although the proposed HierDANN achieves the highest $R$ value of 0.945, the corresponding Bias remains considerably high. It is well recognized that unsupervised domain adaptation assumes the source and target domains share similar distributions \cite{vanTulder_2023,Jiménez-Guarneros_2021,Mehra_2021,Wang_2018}. When a significant discrepancy exists between the two domains, negative transfer may occur, leading to degraded performance of unsupervised domain adaptation models. Therefore, the results indicate that the FDTD model should first be calibrated to ensure consistency between the simulated and measured radar responses, after which the proposed physics-guided hierarchical domain adaptation approaches can be employed to mitigate the residual gap that cannot be fully eliminated through the model calibration.

\section{Conclusions} \label{Conclusions}
This study presents a novel sim-to-real domain adaptation framework that leverages physics-guided machine learning, hierarchical domain adaptation, and deep adversarial learning to robustly retrieve subsurface material properties, including permittivity, conductivity, and depth, from GPR signals. The proposed physics-guided hierarchical domain adaptation approaches, HierDANN, PhyDANN-1, HierPhyDANN-1, PhyDANN-2, and HierPhyDANN-2, are comprehensively benchmarked against state-of-the-art methods, namely the 1D CNN and DANN. Based on systematic laboratory and field tests on one- and two-layer materials, the following conclusions can be drawn:

(1) The proposed physics-guided hierarchical domain adaptation approaches exhibit superior material estimation performance compared with the baseline 1D CNN and DANN methods, achieving higher correlation coefficients $R$ and lower Bias between the predicted and measured parameter values in almost all laboratory and field tests. This improvement results from incorporating physical constraints through radar signal reconstruction, which preserves essential physics-based features needed for accurate material property estimation. Moreover, the hierarchical domain adaptation decomposes the complex parameter estimation task into sequential steps, leveraging previously estimated parameters to guide the prediction of subsequent ones, thereby reducing task complexity and enhancing estimation accuracy. The order of parameters to be estimated in the hierarchical domain adaptation is determined using variance-based sensitivity analysis (i.e., Sobol’s method).

(2) Better parameter estimation performance is achieved by the proposed approaches for the single-layer material (soil) than for the two-layer material (biomass over soil), which can be attributed to the attenuation of radar signals by the top biomass layer as they penetrate the underlying soil. Moreover, increased moisture in the biomass layer leads to a slight degradation in estimation performance and greater uncertainty in the predicted parameters, as the higher moisture content further attenuates the GPR signals and limits their penetration into the soil beneath. Overall, the proposed physics-guided hierarchical domain adaptation framework effectively extracts domain-invariant and material-discriminative features, bridges the gap between simulated and real GPR signals, and successfully captures rainfall-induced variations in soil moisture for both single-layer and two-layer materials.

(3) Compared with the iterative model updating approach previously used for subsurface material property retrieval, the proposed physics-guided domain adaptation methods (i.e., PhyDANN-1, PhyDANN-2) achieve significantly faster inference. Although the hierarchical domain adaptation approaches, HierDANN, HierPhyDANN-1, and HierPhyDANN-2, incur higher computational costs than their non-hierarchical counterparts due to the multi-stage adversarial learning process, they generally achieve superior material property estimation performance. Furthermore, proper calibration of the FDTD model is shown to be essential for effective domain adaptation, as an uncalibrated model can introduce a substantial domain gap between simulated and real GPR signals, potentially resulting in negative transfer from the source to the target domain. Since radar model calibration is a one-time process, it is recommended to update the FDTD model before applying the physics-guided hierarchical domain adaptation approaches for subsurface material property estimation.

The proposed physics-guided hierarchical domain adaptation framework is a physics-guided machine learning paradigm that leverages a physics-based FDTD model to extract representative features from the source domain and reconstruct the radar signal to preserve essential physics-based features during adversarial learning. Developing a physics-informed domain adaptation pipeline that further constrains network training through the incorporation of explicit physical constraints (e.g., satisfying Maxwell’s equations) is of particular interest for future work. Moreover, in this study, a metal plate was embedded beneath the soil to generate a strong reflection of the radar signal. While this is consistent with previous studies in the literature \cite{fang2025metalplate,anbazhagan2020metalplate,ding2023metalplate,Aziz_2024_RSE}, in future research, it is crucial to develop novel radar sensing techniques and advanced data processing strategies that eliminate the need for the metal plate, thereby enhancing the applicability of the proposed approach to a wider range of engineering scenarios.

\vspace{4mm}

\noindent \textbf {\fontsize{11}{12}\fontfamily{qhv}\selectfont Data Availability Statement}\\
Some or all data, models, or code that support the findings of this study are available from the corresponding author upon reasonable request.

\vspace{4mm}

\noindent \textbf {\fontsize{11}{12}\fontfamily{qhv}\selectfont Acknowledgements}\\
This research was funded by the NASA FireSense Technology Program grant 121509. The authors sincerely thank graduate student Tianyi Zhong for assisting with the data collection conducted in this study.

\bibliographystyle{ieeetr}
\small
\bibliography{references}

@article{abufares_2025asphalt,
  title={{Asphalt concrete density monitoring during compaction using roller-mounted GPR}},
  author={Abufares, Lama and Chen, Yihan and Al-Qadi, Imad L},
  journal={Automation in Construction},
  volume={174},
  pages={106158},
  year={2025},
  publisher={Elsevier}
}

@article{fang2025metalplate,
  title={Study on automated detection methods of shallow surface soil water content based on GPR signal level},
  author={Fang, Yunfeng and Hei, Tianqing and Tong, Zheng and Ma, Tao},
  journal={Remote Sensing of Environment},
  volume={331},
  pages={115003},
  year={2025},
  publisher={Elsevier}
}

@article{ding2023metalplate,
  title={Soil moisture sensing with uav-mounted ir-uwb radar and deep learning},
  author={Ding, Rong and Jin, Haiming and Xiang, Dong and Wang, Xiaocheng and Zhang, Yongkui and Shen, Dingman and Su, Lu and Hao, Wentian and Tao, Mingyuan and Wang, Xinbing and others},
  journal={Proceedings of the ACM on Interactive, Mobile, Wearable and Ubiquitous Technologies},
  volume={7},
  number={1},
  pages={1--25},
  year={2023},
  publisher={ACM New York, NY, USA}
}

@article{anbazhagan2020metalplate,
  title={Comparison of soil water content estimation equations using ground penetrating radar},
  author={Anbazhagan, Pa and Bittelli, Marco and Pallepati, Rao Raghuveer and Mahajan, Puskar},
  journal={Journal of Hydrology},
  volume={588},
  pages={125039},
  year={2020},
  publisher={Elsevier}
}

@article{Kim_2025,
title = {{Investigating the impact of data normalization methods on predicting electricity consumption in a building using different artificial neural network models}},
journal = {Sustainable Cities and Society},
volume = {118},
pages = {105570},
year = {2025},
issn = {2210-6707},
doi = {https://doi.org/10.1016/j.scs.2024.105570},
url = {https://www.sciencedirect.com/science/article/pii/S2210670724003962},
author = {Yang-Seon Kim and Moon Keun Kim and Nuodi Fu and Jiying Liu and Junqi Wang and Jelena Srebric},
}

@article{Singh_2020,
title = {{Investigating the impact of data normalization on classification performance}},
journal = {Applied Soft Computing},
volume = {97},
pages = {105524},
year = {2020},
issn = {1568-4946},
doi = {https://doi.org/10.1016/j.asoc.2019.105524},
url = {https://www.sciencedirect.com/science/article/pii/S1568494619302947},
author = {Dalwinder Singh and Birmohan Singh},
}

@article{Alzubaidi_2021,
  title={{Review of deep learning: concepts, CNN architectures, challenges, applications, future directions}},
  author={Alzubaidi, Laith and Zhang, Jinglan and Humaidi, Amjad J and Al-Dujaili, Ayad and Duan, Ye and Al-Shamma, Omran and Santamar{\'\i}a, Jos{\'e} and Fadhel, Mohammed A and Al-Amidie, Muthana and Farhan, Laith},
  journal={Journal of big Data},
  volume={8},
  number={1},
  pages={53},
  year={2021},
  publisher={Springer},
  doi={https://doi.org/10.1186/s40537-021-00444-8}
}

@article{Kiranyaz_2021,
title = {{1D convolutional neural networks and applications: A survey}},
journal = {Mechanical Systems and Signal Processing},
volume = {151},
pages = {107398},
year = {2021},
issn = {0888-3270},
doi = {https://doi.org/10.1016/j.ymssp.2020.107398},
url = {https://www.sciencedirect.com/science/article/pii/S0888327020307846},
author = {Serkan Kiranyaz and Onur Avci and Osama Abdeljaber and Turker Ince and Moncef Gabbouj and Daniel J. Inman},
}

@article{Vahidi_2025,
title = {{Multi-depth soil moisture estimation via 1D convolutional neural networks from drone-mounted ground penetrating Radar data}},
journal = {Computers and Electronics in Agriculture},
volume = {232},
pages = {110104},
year = {2025},
issn = {0168-1699},
doi = {https://doi.org/10.1016/j.compag.2025.110104},
url = {https://www.sciencedirect.com/science/article/pii/S0168169925002108},
author = {Milad Vahidi and Sanaz Shafian and William Hunter Frame},
}

@article{Gardner_2020,
title = {{On the application of domain adaptation in structural health monitoring}},
journal = {Mechanical Systems and Signal Processing},
volume = {138},
pages = {106550},
year = {2020},
issn = {0888-3270},
doi = {https://doi.org/10.1016/j.ymssp.2019.106550},
url = {https://www.sciencedirect.com/science/article/pii/S088832701930771X},
author = {P. Gardner and X. Liu and K. Worden},
}

@ARTICLE{Kouw_2018,
       author = {{Kouw}, Wouter M. and {Loog}, Marco},
        title = "{{An introduction to domain adaptation and transfer learning}}",
      journal = {arXiv e-prints},
         year = 2018,
        month = dec,
          eid = {arXiv:1812.11806},
        pages = {arXiv:1812.11806},
          doi = {10.48550/arXiv.1812.11806},
archivePrefix = {arXiv},
       eprint = {1812.11806},
 primaryClass = {cs.LG},
       adsurl = {https://ui.adsabs.harvard.edu/abs/2018arXiv181211806K}
}

@article{Ganin_2016,
author = {Ganin, Yaroslav and Ustinova, Evgeniya and Ajakan, Hana and Germain, Pascal and Larochelle, Hugo and Laviolette, Fran\c{c}ois and Marchand, Mario and Lempitsky, Victor},
title = {{Domain-adversarial training of neural networks}},
year = {2016},
issue_date = {January 2016},
publisher = {JMLR.org},
volume = {17},
number = {1},
issn = {1532-4435},
journal = {J. Mach. Learn. Res.},
month = jan,
pages = {2096–2030},
numpages = {35},
url     = {http://jmlr.org/papers/v17/15-239.html}
}

@article{Borchani_2015,
author = {Borchani, Hanen and Varando, Gherardo and Bielza, Concha and Larrañaga, Pedro},
title = {{A survey on multi-output regression}},
journal = {WIREs Data Mining and Knowledge Discovery},
volume = {5},
number = {5},
pages = {216-233},
doi = {https://doi.org/10.1002/widm.1157},
url = {https://wires.onlinelibrary.wiley.com/doi/abs/10.1002/widm.1157},
eprint = {https://wires.onlinelibrary.wiley.com/doi/pdf/10.1002/widm.1157},
year = {2015}
}

@article{Sobol_2001,
title = {{Global sensitivity indices for nonlinear mathematical models and their Monte Carlo estimates}},
journal = {Mathematics and Computers in Simulation},
volume = {55},
number = {1},
pages = {271-280},
year = {2001},
note = {The Second IMACS Seminar on Monte Carlo Methods},
issn = {0378-4754},
doi = {https://doi.org/10.1016/S0378-4754(00)00270-6},
url = {https://www.sciencedirect.com/science/article/pii/S0378475400002706},
author = {I.M Sobol}
}

@article{Homma_1996,
title = {{Importance measures in global sensitivity analysis of nonlinear models}},
journal = {Reliability Engineering \& System Safety},
volume = {52},
number = {1},
pages = {1-17},
year = {1996},
issn = {0951-8320},
doi = {https://doi.org/10.1016/0951-8320(96)00002-6},
url = {https://www.sciencedirect.com/science/article/pii/0951832096000026},
author = {Toshimitsu Homma and Andrea Saltelli},
}

@article{Saltelli_2010,
title = {{Variance based sensitivity analysis of model output. Design and estimator for the total sensitivity index}},
journal = {Computer Physics Communications},
volume = {181},
number = {2},
pages = {259-270},
year = {2010},
issn = {0010-4655},
doi = {https://doi.org/10.1016/j.cpc.2009.09.018},
url = {https://www.sciencedirect.com/science/article/pii/S0010465509003087},
author = {Andrea Saltelli and Paola Annoni and Ivano Azzini and Francesca Campolongo and Marco Ratto and Stefano Tarantola},
}

@article{Ojha_2022,
title = {{Assessing ranking and effectiveness of evolutionary algorithm hyperparameters using global sensitivity analysis methodologies}},
journal = {Swarm and Evolutionary Computation},
volume = {74},
pages = {101130},
year = {2022},
issn = {2210-6502},
doi = {https://doi.org/10.1016/j.swevo.2022.101130},
url = {https://www.sciencedirect.com/science/article/pii/S2210650222001006},
author = {Varun Ojha and Jon Timmis and Giuseppe Nicosia},
}

@article{Warren_2016,
title = {{gprMax: Open source software to simulate electromagnetic wave propagation for Ground Penetrating Radar}},
journal = {Computer Physics Communications},
volume = {209},
pages = {163-170},
year = {2016},
issn = {0010-4655},
doi = {https://doi.org/10.1016/j.cpc.2016.08.020},
url = {https://www.sciencedirect.com/science/article/pii/S0010465516302533},
author = {Craig Warren and Antonios Giannopoulos and Iraklis Giannakis},
}

@article{vanTulder_2023,
title = {{Unpaired, unsupervised domain adaptation assumes your domains are already similar}},
journal = {Medical Image Analysis},
volume = {87},
pages = {102825},
year = {2023},
issn = {1361-8415},
doi = {https://doi.org/10.1016/j.media.2023.102825},
url = {https://www.sciencedirect.com/science/article/pii/S1361841523000865},
author = {Gijs {van Tulder} and Marleen {de Bruijne}},
}

@article{Jiménez-Guarneros_2021,
title = {{A study of the effects of negative transfer on deep unsupervised domain adaptation methods}},
journal = {Expert Systems with Applications},
volume = {167},
pages = {114088},
year = {2021},
issn = {0957-4174},
doi = {https://doi.org/10.1016/j.eswa.2020.114088},
url = {https://www.sciencedirect.com/science/article/pii/S0957417420308459},
author = {Magdiel Jiménez-Guarneros and Pilar Gómez-Gil},
}

@inproceedings{Mehra_2021,
  title={{Understanding the Limits of Unsupervised Domain Adaptation via Data Poisoning}},
  author={Akshay Mehra and Bhavya Kailkhura and Pin-Yu Chen and Jihun Hamm},
  booktitle={Neural Information Processing Systems},
  year={2021},
  url={https://api.semanticscholar.org/CorpusID:235765580}
}

@article{Wang_2018,
  title={{Characterizing and Avoiding Negative Transfer}},
  author={Zirui Wang and Zihang Dai and Barnab{\'a}s P{\'o}czos and Jaime G. Carbonell},
  journal={2019 IEEE/CVF Conference on Computer Vision and Pattern Recognition (CVPR)},
  year={2018},
  pages={11285-11294},
  url={https://api.semanticscholar.org/CorpusID:53748459}
}

@article{Wang_2025_JEM,
author = {Zixin Wang  and Mohammad R. Jahanshahi  and Alana Lund  and Adnan Shahriar  and Arturo Montoya },
title = {{Physics-Informed Machine Learning for Hybrid Digital Twin–Enhanced Damage Detection and Localization}},
journal = {Journal of Engineering Mechanics},
volume = {151},
number = {12},
pages = {04025080},
year = {2025},
doi = {10.1061/JENMDT.EMENG-8325},
URL = {https://ascelibrary.org/doi/abs/10.1061/JENMDT.EMENG-8325},
eprint = {https://ascelibrary.org/doi/pdf/10.1061/JENMDT.EMENG-8325},
}

@inproceedings{Benigmim_2023,
  title={{One-shot unsupervised domain adaptation with personalized diffusion models}},
  author={Benigmim, Yasser and Roy, Subhankar and Essid, Slim and Kalogeiton, Vicky and Lathuili{\`e}re, St{\'e}phane},
  booktitle={Proceedings of the IEEE/CVF conference on computer vision and pattern recognition},
  pages={698--708},
  year={2023},
  doi={10.1109/CVPRW59228.2023.00077}
}

@inproceedings{Bousmalis_2018,
author = {Bousmalis, Konstantinos and Irpan, Alex and Wohlhart, Paul and Bai, Yunfei and Kelcey, Matthew and Kalakrishnan, Mrinal and Downs, Laura and Ibarz, Julian and Pastor, Peter and Konolige, Kurt and Levine, Sergey and Vanhoucke, Vincent},
title = {{Using Simulation and Domain Adaptation to Improve Efficiency of Deep Robotic Grasping}},
year = {2018},
publisher = {IEEE Press},
url = {https://doi.org/10.1109/ICRA.2018.8460875},
doi = {10.1109/ICRA.2018.8460875},
booktitle = {2018 IEEE International Conference on Robotics and Automation (ICRA)},
pages = {4243–4250},
numpages = {8},
location = {Brisbane, Australia}
}

@ARTICLE{DeBortoli_2021,
  author={DeBortoli, Robert and Fuxin, Li and Kapoor, Ashish and Hollinger, Geoffrey A.},
  journal={IEEE Robotics and Automation Letters}, 
  title={{Adversarial Training on Point Clouds for Sim-to-Real 3D Object Detection}}, 
  year={2021},
  volume={6},
  number={4},
  pages={6662-6669},
  doi={10.1109/LRA.2021.3093869}}

@ARTICLE{Lou_2022,
  author={Lou, Yunxia and Kumar, Anil and Xiang, Jiawei},
  journal={IEEE Transactions on Instrumentation and Measurement}, 
  title={{Machinery Fault Diagnosis Based on Domain Adaptation to Bridge the Gap Between Simulation and Measured Signals}}, 
  year={2022},
  volume={71},
  number={},
  pages={1-9},
  doi={10.1109/TIM.2022.3180416}}

@inproceedings{Hu_2022,
author = {Hu, Chuqing and Hudson, Sinclair and Ethier, Martin and Al-Sharman, Mohammad and Rayside, Derek and Melek, William},
title = {{Sim-to-Real Domain Adaptation for Lane Detection and Classification in Autonomous Driving}},
year = {2022},
publisher = {IEEE Press},
url = {https://doi.org/10.1109/IV51971.2022.9827450},
doi = {10.1109/IV51971.2022.9827450},
booktitle = {2022 IEEE Intelligent Vehicles Symposium (IV)},
pages = {457–463},
numpages = {7},
location = {Aachen, Germany}
}

@inproceedings{Sol_2024,
  title={{Sim-to-Real Domain Adaptation for Deformation Classification}},
  author={Sol, Joel and Fayyad, Jamil and Alijani, Shadi and Najjaran, Homayoun},
  booktitle={2024 IEEE International Conference on Systems, Man, and Cybernetics (SMC)},
  pages={2225--2231},
  year={2024},
  organization={IEEE},
  doi={10.1109/SMC54092.2024.10831103}
}

@article{Wang_2025,
title = {{Towards bridging the synthetic-to-real gap in quantitative photoacoustic tomography via unsupervised domain adaptation}},
journal = {Photoacoustics},
volume = {45},
pages = {100736},
year = {2025},
issn = {2213-5979},
doi = {https://doi.org/10.1016/j.pacs.2025.100736},
url = {https://www.sciencedirect.com/science/article/pii/S221359792500059X},
author = {Zeqi Wang and Wei Tao and Zhuang Zhang and Hui Zhao},
}

@InProceedings{Li_2025,
  title = 	 {{Evaluating Real-World Robot Manipulation Policies in Simulation}},
  author =       {Li, Xuanlin and Hsu, Kyle and Gu, Jiayuan and Mees, Oier and Pertsch, Karl and Walke, Homer Rich and Fu, Chuyuan and Lunawat, Ishikaa and Sieh, Isabel and Kirmani, Sean and Levine, Sergey and Wu, Jiajun and Finn, Chelsea and Su, Hao and Vuong, Quan and Xiao, Ted},
  booktitle = 	 {Proceedings of The 8th Conference on Robot Learning},
  pages = 	 {3705--3728},
  year = 	 {2025},
  editor = 	 {Agrawal, Pulkit and Kroemer, Oliver and Burgard, Wolfram},
  volume = 	 {270},
  series = 	 {Proceedings of Machine Learning Research},
  month = 	 {06--09 Nov},
  publisher =    {PMLR},
  url = 	 {https://proceedings.mlr.press/v270/li25c.html},
}

@article{Wiberg_2024,
title = {{Sim-to-real transfer of active suspension control using deep reinforcement learning}},
journal = {Robotics and Autonomous Systems},
volume = {179},
pages = {104731},
year = {2024},
issn = {0921-8890},
doi = {https://doi.org/10.1016/j.robot.2024.104731},
url = {https://www.sciencedirect.com/science/article/pii/S0921889024001155},
author = {Viktor Wiberg and Erik Wallin and Arvid Fälldin and Tobias Semberg and Morgan Rossander and Eddie Wadbro and Martin Servin},
}

@inproceedings{Tobin_2017,
  title={{Domain randomization for transferring deep neural networks from simulation to the real world}},
  author={Tobin, Josh and Fong, Rachel and Ray, Alex and Schneider, Jonas and Zaremba, Wojciech and Abbeel, Pieter},
  booktitle={2017 IEEE/RSJ international conference on intelligent robots and systems (IROS)},
  pages={23--30},
  year={2017},
  organization={IEEE},
  doi={10.1109/IROS.2017.8202133}
}

@inproceedings{Josifovski_2022,
  title={{Analysis of randomization effects on sim2real transfer in reinforcement learning for robotic manipulation tasks}},
  author={Josifovski, Josip and Malmir, Mohammadhossein and Klarmann, Noah and {\v{Z}}agar, Bare Luka and Navarro-Guerrero, Nicol{\'a}s and Knoll, Alois},
  booktitle={2022 IEEE/RSJ International Conference on Intelligent Robots and Systems (IROS)},
  pages={10193--10200},
  year={2022},
  organization={IEEE},
  doi={10.1109/IROS47612.2022.9981951}
}

@ARTICLE{Pan_2011,
  author={Pan, Sinno Jialin and Tsang, Ivor W. and Kwok, James T. and Yang, Qiang},
  journal={IEEE Transactions on Neural Networks}, 
  title={{Domain Adaptation via Transfer Component Analysis}}, 
  year={2011},
  volume={22},
  number={2},
  pages={199-210},
  doi={10.1109/TNN.2010.2091281}}

@InProceedings{Yan_2017,
author = {Yan, Hongliang and Ding, Yukang and Li, Peihua and Wang, Qilong and Xu, Yong and Zuo, Wangmeng},
title = {{Mind the Class Weight Bias: Weighted Maximum Mean Discrepancy for Unsupervised Domain Adaptation}},
booktitle = {Proceedings of the IEEE Conference on Computer Vision and Pattern Recognition (CVPR)},
month = {July},
year = {2017},
doi = {10.1109/CVPR.2017.107}
}

@inproceedings{Zhang_2018,
  title={{Importance weighted adversarial nets for partial domain adaptation}},
  author={Zhang, Jing and Ding, Zewei and Li, Wanqing and Ogunbona, Philip},
  booktitle={Proceedings of the IEEE conference on computer vision and pattern recognition (CVPR)},
  pages={8156--8164},
  year={2018},
  doi={10.1109/CVPR.2018.00851}
}

@inproceedings{Cao_2018,
author = {Cao, Zhangjie and Ma, Lijia and Long, Mingsheng and Wang, Jianmin},
title = {{Partial Adversarial Domain Adaptation}},
year = {2018},
isbn = {978-3-030-01236-6},
publisher = {Springer-Verlag},
address = {Berlin, Heidelberg},
url = {https://doi.org/10.1007/978-3-030-01237-3_9},
doi = {10.1007/978-3-030-01237-3_9},
booktitle = {Computer Vision – ECCV 2018: 15th European Conference, Munich, Germany, September 8-14, 2018, Proceedings, Part VIII},
pages = {139–155},
numpages = {17},
location = {Munich, Germany}
}

@inproceedings{Yosinski_2014,
author = {Yosinski, Jason and Clune, Jeff and Bengio, Yoshua and Lipson, Hod},
title = {{How transferable are features in deep neural networks?}},
year = {2014},
publisher = {MIT Press},
address = {Cambridge, MA, USA},
booktitle = {Proceedings of the 28th International Conference on Neural Information Processing Systems - Volume 2},
pages = {3320–3328},
numpages = {9},
location = {Montreal, Canada},
series = {NIPS'14}
}

@article{Chen_2020,
title = {{Transfer learning with deep neural networks for model predictive control of HVAC and natural ventilation in smart buildings}},
journal = {Journal of Cleaner Production},
volume = {254},
pages = {119866},
year = {2020},
issn = {0959-6526},
doi = {https://doi.org/10.1016/j.jclepro.2019.119866},
url = {https://www.sciencedirect.com/science/article/pii/S0959652619347365},
author = {Yujiao Chen and Zheming Tong and Yang Zheng and Holly Samuelson and Leslie Norford},
}

@inproceedings{Singh_2021,
 author = {Singh, Ankit},
 booktitle = {Advances in Neural Information Processing Systems},
 editor = {M. Ranzato and A. Beygelzimer and Y. Dauphin and P.S. Liang and J. Wortman Vaughan},
 pages = {5089--5101},
 publisher = {Curran Associates, Inc.},
 title = {{CLDA: Contrastive Learning for Semi-Supervised Domain Adaptation}},
 url = {https://proceedings.neurips.cc/paper_files/paper/2021/file/288cd2567953f06e460a33951f55daaf-Paper.pdf},
 volume = {34},
 year = {2021}
}

@inproceedings{Motiian_2017,
  title={{Unified deep supervised domain adaptation and generalization}},
  author={Motiian, Saeid and Piccirilli, Marco and Adjeroh, Donald A and Doretto, Gianfranco},
  booktitle={Proceedings of the IEEE international conference on computer vision},
  pages={5715--5725},
  year={2017},
  doi = {10.1109/ICCV.2017.609}
}

@article{Faroughi_2024,
    author = {Faroughi, Salah A. and Pawar, Nikhil M. and Fernandes, Célio and Raissi, Maziar and Das, Subasish and Kalantari, Nima K. and Kourosh Mahjour, Seyed},
    title = {{Physics-Guided, Physics-Informed, and Physics-Encoded Neural Networks and Operators in Scientific Computing: Fluid and Solid Mechanics}},
    journal = {Journal of Computing and Information Science in Engineering},
    volume = {24},
    number = {4},
    pages = {040802},
    year = {2024},
    month = {01},
    issn = {1530-9827},
    doi = {10.1115/1.4064449},
    url = {https://doi.org/10.1115/1.4064449},
    eprint = {https://asmedigitalcollection.asme.org/computingengineering/article-pdf/24/4/040802/7383247/jcise_24_4_040802.pdf},
}

@Article{Liu_2016,
journal="International Agrophysics",
issn="0236-8722",
volume="30",
number="4",
year="2016",
title="{{Ground penetrating radar for underground sensing in agriculture: a review}}",
author="Liu, Xiuwei
and Dong, Xuejun
and Leskovar, Daniel I.",
pages="533--543",
doi="10.1515/intag-2016-0010",
url="https://doi.org/10.1515/intag-2016-0010"
}

@article{Srivastava_2017,
  title={{Satellite soil moisture: Review of theory and applications in water resources}},
  author={Srivastava, Prashant K},
  journal={Water Resources Management},
  volume={31},
  number={10},
  pages={3161--3176},
  year={2017},
  publisher={Springer}, 
  DOI = {https://doi.org/10.1007/s11269-017-1722-6}
}

@Article{Ambadan_2020,
AUTHOR = {Thomas Ambadan, Jaison and Oja, Matilda and Gedalof, Ze’ev and Berg, Aaron A.},
TITLE = {{Satellite-Observed Soil Moisture as an Indicator of Wildfire Risk}},
JOURNAL = {Remote Sensing},
VOLUME = {12},
YEAR = {2020},
NUMBER = {10},
ARTICLE-NUMBER = {1543},
URL = {https://www.mdpi.com/2072-4292/12/10/1543},
ISSN = {2072-4292},
DOI = {10.3390/rs12101543}
}

@article{Jensen_2018,
doi = {10.1088/1748-9326/aa9853},
url = {https://doi.org/10.1088/1748-9326/aa9853},
year = {2018},
month = {jan},
publisher = {IOP Publishing},
volume = {13},
number = {1},
pages = {014021},
author = {Jensen, Daniel and Reager, John T and Zajic, Brittany and Rousseau, Nick and Rodell, Matthew and Hinkley, Everett},
title = {{The sensitivity of US wildfire occurrence to pre-season soil moisture conditions across ecosystems}},
journal = {Environmental Research Letters},
}

@article{Krueger_2015,
author = {Krueger, Erik S. and Ochsner, Tyson E. and Engle, David M. and Carlson, J.D. and Twidwell, Dirac and Fuhlendorf, Samuel D.},
title = {{Soil Moisture Affects Growing-Season Wildfire Size in the Southern Great Plains}},
journal = {Soil Science Society of America Journal},
volume = {79},
number = {6},
pages = {1567-1576},
doi = {https://doi.org/10.2136/sssaj2015.01.0041},
url = {https://acsess.onlinelibrary.wiley.com/doi/abs/10.2136/sssaj2015.01.0041},
eprint = {https://acsess.onlinelibrary.wiley.com/doi/pdf/10.2136/sssaj2015.01.0041},
year = {2015}
}

@article{Rao_2020,
title = {{SAR-enhanced mapping of live fuel moisture content}},
journal = {Remote Sensing of Environment},
volume = {245},
pages = {111797},
year = {2020},
issn = {0034-4257},
doi = {https://doi.org/10.1016/j.rse.2020.111797},
url = {https://www.sciencedirect.com/science/article/pii/S003442572030167X},
author = {Krishna Rao and A. Park Williams and Jacqueline Fortin Flefil and Alexandra G. Konings},
}

@Article{Alipour_2023,
AUTHOR = {Alipour, Mohamad and La Puma, Inga and Picotte, Joshua and Shamsaei, Kasra and Rowell, Eric and Watts, Adam and Kosovic, Branko and Ebrahimian, Hamed and Taciroglu, Ertugrul},
TITLE = {{A Multimodal Data Fusion and Deep Learning Framework for Large-Scale Wildfire Surface Fuel Mapping}},
JOURNAL = {Fire},
VOLUME = {6},
YEAR = {2023},
NUMBER = {2},
ARTICLE-NUMBER = {36},
URL = {https://www.mdpi.com/2571-6255/6/2/36},
ISSN = {2571-6255},
DOI = {10.3390/fire6020036}
}

@article{SU_2014,
title = {{A critical review of soil moisture measurement}},
journal = {Measurement},
volume = {54},
pages = {92-105},
year = {2014},
issn = {0263-2241},
doi = {https://doi.org/10.1016/j.measurement.2014.04.007},
url = {https://www.sciencedirect.com/science/article/pii/S0263224114001651},
author = {Susha Lekshmi S.U. and D.N. Singh and Maryam {Shojaei Baghini}},
}

@article{Bourgeau‐Chavez_2010,
    author = {Bourgeau‐Chavez, Laura L. and Garwood, Gordon C. and Riordan, Kevin and Koziol, Benjamin W. and Slawski, James},
    title = {{Development of calibration algorithms for selected water content reflectometry probes for burned and non‐burned organic soils of Alaska}},
    journal = {International Journal of Wildland Fire},
    volume = {19},
    number = {7},
    pages = {961-975},
    year = {2010},
    month = {11},
    issn = {1049-8001},
    doi = {10.1071/WF07175},
    url = {https://doi.org/10.1071/WF07175},
    eprint = {https://connectsci.au/wf/article-pdf/19/7/961/160936/wf07175.pdf},
}

@Article{Fragkos_2024,
AUTHOR = {Fragkos, Athanasios and Loukatos, Dimitrios and Kargas, Georgios and Arvanitis, Konstantinos G.},
TITLE = {{Response of the TEROS 12 Soil Moisture Sensor under Different Soils and Variable Electrical Conductivity}},
JOURNAL = {Sensors},
VOLUME = {24},
YEAR = {2024},
NUMBER = {7},
ARTICLE-NUMBER = {2206},
URL = {https://www.mdpi.com/1424-8220/24/7/2206},
PubMedID = {38610417},
ISSN = {1424-8220},
DOI = {10.3390/s24072206}
}

@article{Li_2021_soil,
title = {{Soil moisture retrieval from remote sensing measurements: Current knowledge and directions for the future}},
journal = {Earth-Science Reviews},
volume = {218},
pages = {103673},
year = {2021},
issn = {0012-8252},
doi = {https://doi.org/10.1016/j.earscirev.2021.103673},
url = {https://www.sciencedirect.com/science/article/pii/S0012825221001744},
author = {Zhao-Liang Li and Pei Leng and Chenghu Zhou and Kun-Shan Chen and Fang-Cheng Zhou and Guo-Fei Shang}
}

@article{Kim_2019,
  title={{A review of satellite-derived soil moisture and its usage for flood estimation}},
  author={Kim, Seokhyeon and Zhang, Runze and Pham, Hung and Sharma, Ashish},
  journal={Remote Sensing in Earth Systems Sciences},
  volume={2},
  number={4},
  pages={225--246},
  year={2019},
  publisher={Springer},
  doi={https://doi.org/10.1007/s41976-019-00025-7}
}

@article{Wu_2019,
title = {{A new drone-borne GPR for soil moisture mapping}},
journal = {Remote Sensing of Environment},
volume = {235},
pages = {111456},
year = {2019},
issn = {0034-4257},
doi = {https://doi.org/10.1016/j.rse.2019.111456},
url = {https://www.sciencedirect.com/science/article/pii/S0034425719304754},
author = {Kaijun Wu and Gabriela Arambulo Rodriguez and Marjana Zajc and Elodie Jacquemin and Michiels Clément and Albéric {De Coster} and Sébastien Lambot},
}

@article{Pathirana_2024,
title = {{Potential of ground-penetrating radar to calibrate electromagnetic induction for shallow soil water content estimation}},
journal = {Journal of Hydrology},
volume = {633},
pages = {130957},
year = {2024},
issn = {0022-1694},
doi = {https://doi.org/10.1016/j.jhydrol.2024.130957},
url = {https://www.sciencedirect.com/science/article/pii/S0022169424003512},
author = {Sashini Pathirana and Sébastien Lambot and Manokararajah Krishnapillai and Christina Smeaton and Mumtaz Cheema and Lakshman Galagedara},
}

@Article{Wu_2022,
AUTHOR = {Wu, Kaijun and Desesquelles, Henri and Cockenpot, Rodolphe and Guyard, Léon and Cuisiniez, Victor and Lambot, Sébastien},
TITLE = {{Ground-Penetrating Radar Full-Wave Inversion for Soil Moisture Mapping in Trench-Hill Potato Fields for Precise Irrigation}},
JOURNAL = {Remote Sensing},
VOLUME = {14},
YEAR = {2022},
NUMBER = {23},
ARTICLE-NUMBER = {6046},
URL = {https://www.mdpi.com/2072-4292/14/23/6046},
ISSN = {2072-4292},
DOI = {10.3390/rs14236046}
}

@INPROCEEDINGS{Aziz_2024,
  author={Aziz, Ishfaq and Soltanaghai, Elahe and Watts, Adam and Alipour, Mohamad},
  booktitle={IGARSS 2024 - 2024 IEEE International Geoscience and Remote Sensing Symposium}, 
  title={{Dual-Frequency Radar Wave-Inversion for Sub-Surface Material Characterization}}, 
  year={2024},
  volume={},
  number={},
  pages={5125-5129},
  doi={10.1109/IGARSS53475.2024.10640755}}

@INPROCEEDINGS{Sinchi_2023,
  author={Sinchi, Kurt Soncco and Calderon, Diego and Aziz, Ishfaq and Watts, Adam and Soltanaghai, Elahe and Alipour, Mohamad},
  booktitle={IGARSS 2023 - 2023 IEEE International Geoscience and Remote Sensing Symposium}, 
  title={{Under-Canopy Biomass Sensing using UAS-Mounted Radar: a Numerical Feasibility Analysis}}, 
  year={2023},
  volume={},
  number={},
  pages={3292-3295},
  doi={10.1109/IGARSS52108.2023.10282771}}

@article{Alam_2024,
title = {{Data-driven evaluation of building materials using Ground Penetrating Radar}},
journal = {Journal of Building Engineering},
volume = {95},
pages = {110188},
year = {2024},
issn = {2352-7102},
doi = {https://doi.org/10.1016/j.jobe.2024.110188},
url = {https://www.sciencedirect.com/science/article/pii/S235271022401756X},
author = {Ahmed Nirjhar Alam and Wesley F. Reinhart and Rebecca K. Napolitano},
}

@article{Hong_2024,
title = {{Research on reinforcement corrosion detection method based on the numerical simulation of ground-penetrating radar}},
journal = {Journal of Building Engineering},
volume = {85},
pages = {108760},
year = {2024},
issn = {2352-7102},
doi = {https://doi.org/10.1016/j.jobe.2024.108760},
url = {https://www.sciencedirect.com/science/article/pii/S2352710224003280},
author = {Shuxian Hong and Guanjin Mo and Shenyou Song and Daqian Li and Zuming Huang and Dongshuai Hou and Huanyong Chen and Xingquan Mao and Xingyu Lou and Biqin Dong},
}

@article{Kaplanvural_2023,
title = {{Volumetric water content estimation of concrete by particle swarm optimization of GPR data}},
journal = {Construction and Building Materials},
volume = {375},
pages = {130995},
year = {2023},
issn = {0950-0618},
doi = {https://doi.org/10.1016/j.conbuildmat.2023.130995},
url = {https://www.sciencedirect.com/science/article/pii/S0950061823007079},
author = {İsmail Kaplanvural},
}

@article{Kang_2025,
title = {{GPR-based depth estimation of ground interfaces in permafrost region: Electromagnetic method and cone penetration assessment}},
journal = {Measurement},
volume = {242},
pages = {116158},
year = {2025},
issn = {0263-2241},
doi = {https://doi.org/10.1016/j.measurement.2024.116158},
url = {https://www.sciencedirect.com/science/article/pii/S0263224124020438},
author = {Seonghun Kang and Geunwoo Park and Namsun Kim and Erol Tutumluer and Jong-Sub Lee},
}

@article{Feng_2019,
title = {{Improving reconstruction of tunnel lining defects from ground-penetrating radar profiles by multi-scale inversion and bi-parametric full-waveform inversion}},
journal = {Advanced Engineering Informatics},
volume = {41},
pages = {100931},
year = {2019},
issn = {1474-0346},
doi = {https://doi.org/10.1016/j.aei.2019.100931},
url = {https://www.sciencedirect.com/science/article/pii/S1474034618305895},
author = {Deshan Feng and Xun Wang and Bin Zhang},
}

@article{Haruzi_2022,
author = {Haruzi, P. and Schmäck, J. and Zhou, Z. and van der Kruk, J. and Vereecken, H. and Vanderborght, J. and Klotzsche, A.},
title = {{Detection of Tracer Plumes Using Full-Waveform Inversion of Time-Lapse Ground Penetrating Radar Data: A Numerical Study in a High-Resolution Aquifer Model}},
journal = {Water Resources Research},
volume = {58},
number = {5},
pages = {e2021WR030110},
doi = {https://doi.org/10.1029/2021WR030110},
url = {https://agupubs.onlinelibrary.wiley.com/doi/abs/10.1029/2021WR030110},
eprint = {https://agupubs.onlinelibrary.wiley.com/doi/pdf/10.1029/2021WR030110},
note = {e2021WR030110 2021WR030110},
year = {2022}
}

@ARTICLE{Feng_2023,
  author={Feng, Deshan and Liu, Yuxin and Wang, Xun and Zhang, Bin and Ding, Siyuan and Yu, Tianxiao and Li, Bingchao and Feng, Zheng},
  journal={IEEE Transactions on Geoscience and Remote Sensing}, 
  title={{Inspection and Imaging of Tree Trunk Defects Using GPR Multifrequency Full-Waveform Dual-Parameter Inversion}}, 
  year={2023},
  volume={61},
  number={},
  pages={1-15},
  doi={10.1109/TGRS.2023.3244946}}

@ARTICLE{Patsia_2023,
  author={Patsia, Ourania and Giannopoulos, Antonios and Giannakis, Iraklis},
  journal={IEEE Transactions on Geoscience and Remote Sensing}, 
  title={{GPR Full-Waveform Inversion With Deep-Learning Forward Modeling: A Case Study From Non-Destructive Testing}}, 
  year={2023},
  volume={61},
  number={},
  pages={1-10},
  doi={10.1109/TGRS.2023.3303683}}

@ARTICLE{Liu_2024,
  author={Liu, Yuxin and Feng, Deshan and Xiao, Yougan and Huang, Guoxing and Cai, Liqiong and Tai, Xiaoyong and Wang, Xun},
  journal={IEEE Transactions on Geoscience and Remote Sensing}, 
  title={{Full-Waveform Inversion of Multifrequency GPR Data Using a Multiscale Approach Based on Deep Learning}}, 
  year={2024},
  volume={62},
  number={},
  pages={1-12},
  doi={10.1109/TGRS.2024.3382331}}

@article{Xue_2024,
    author = {Xue, Jiyan and Huang, Qinghua and Wu, Sihong and Zhao, Li and Ma, Bowen},
    title = {{Real-time dual-parameter full-waveform inversion of GPR data based on robust deep learning}},
    journal = {Geophysical Journal International},
    volume = {238},
    number = {3},
    pages = {1755-1771},
    year = {2024},
    month = {07},
    issn = {1365-246X},
    doi = {10.1093/gji/ggae243},
    url = {https://doi.org/10.1093/gji/ggae243},
    eprint = {https://academic.oup.com/gji/article-pdf/238/3/1755/58690205/ggae243.pdf},
}

@ARTICLE{Zhang_2025,
  author={Zhang, Xiaowei and Zhao, Xuan and Lv, Shenghua and Xv, Lingfei and Lin, Chen and Wen, Jian},
  journal={IEEE Transactions on Geoscience and Remote Sensing}, 
  title={{Deep Learning Inversion of Ground-Penetrating Radar With Prior Physical Velocity Field for Complex Subsurface Root System}}, 
  year={2025},
  volume={63},
  number={},
  pages={1-15},
  doi={10.1109/TGRS.2025.3553688}}

@article{Aziz_2024_RSE,
title = {{Bayesian inversion of GPR waveforms for sub-surface material characterization: An uncertainty-aware retrieval of soil moisture and overlaying biomass properties}},
journal = {Remote Sensing of Environment},
volume = {313},
pages = {114351},
year = {2024},
issn = {0034-4257},
doi = {https://doi.org/10.1016/j.rse.2024.114351},
url = {https://www.sciencedirect.com/science/article/pii/S0034425724003778},
author = {Ishfaq Aziz and Elahe Soltanaghai and Adam Watts and Mohamad Alipour},
}

@article{Aziz_2025,
title = {{Rapid subsurface sensing via Bayesian-optimized FDTD modeling of ground penetrating radar}},
journal = {Journal of Building Engineering},
volume = {104},
pages = {112243},
year = {2025},
issn = {2352-7102},
doi = {https://doi.org/10.1016/j.jobe.2025.112243},
url = {https://www.sciencedirect.com/science/article/pii/S2352710225004802},
author = {Ishfaq Aziz and Mohamad Alipour},
}

@inproceedings{Bralich_2017,
  title={{Improving convolutional neural networks for buried target detection in ground penetrating radar using transfer learning via pretraining}},
  author={Bralich, John and Reichman, Dani{\"e}l and Collins, Leslie M and Malof, Jordan M},
  booktitle={Detection and Sensing of Mines, Explosive Objects, and Obscured Targets XXII},
  volume={10182},
  pages={198--208},
  year={2017},
  organization={SPIE},
  doi={10.1117/12.2263112}
}

@inproceedings{Reichman_2017,
  title={{Some good practices for applying convolutional neural networks to buried threat detection in Ground Penetrating Radar}},
  author={Reichman, Dani{\"e}l and Collins, Leslie M and Malof, Jordan M},
  booktitle={2017 9th International Workshop on Advanced Ground Penetrating Radar (IWAGPR)},
  pages={1--5},
  year={2017},
  organization={IEEE},
  doi={10.1109/IWAGPR.2017.7996100}
}

@inproceedings{Aydin_2019,
  title={{Transfer and multitask learning using convolutional neural networks for buried wire detection from ground penetrating radar data}},
  author={Aydin, Enver and Erdem, Seniha Esen Y{\"u}ksel},
  booktitle={Detection and Sensing of Mines, Explosive Objects, and Obscured Targets XXIV},
  volume={11012},
  pages={259--270},
  year={2019},
  organization={SPIE},
  doi={10.1117/12.2518875}
}

@article{Tong_2020,
title = {{Advances of deep learning applications in ground-penetrating radar: A survey}},
journal = {Construction and Building Materials},
volume = {258},
pages = {120371},
year = {2020},
issn = {0950-0618},
doi = {https://doi.org/10.1016/j.conbuildmat.2020.120371},
url = {https://www.sciencedirect.com/science/article/pii/S095006182032376X},
author = {Zheng Tong and Jie Gao and Dongdong Yuan},
}

@ARTICLE{Imai_2025,
  author={Imai, Takanori and Mizutani, Tsukasa and Iguchi, Tatsuya and Haneda, Toshihiro},
  journal={IEEE Transactions on Geoscience and Remote Sensing}, 
  title={{Enhancing Deep Learning-Based GPR Data Inversion With Unsupervised Domain Adaptation: Comparison of Domain Classifiers}}, 
  year={2025},
  volume={63},
  number={},
  pages={1-16},
  doi={10.1109/TGRS.2025.3568384}}

@article{Li_2026,
title = {{DDA-GPR: symmetric adversarial domain adaptation for few-shot underground cavity identification in urban roads}},
journal = {Measurement},
volume = {258},
pages = {119268},
year = {2026},
issn = {0263-2241},
doi = {https://doi.org/10.1016/j.measurement.2025.119268},
url = {https://www.sciencedirect.com/science/article/pii/S0263224125026272},
author = {Fanruo Li and Feng Yang and Wenxing Shi and Suping Peng and Maoxuan Xu and Lijianghan Qiao},
}

@INPROCEEDINGS{Huang_2023,
  author={Huang, Zhe and Jia, Yong and Fan, Xiaodong and Tang, Likun},
  booktitle={2023 8th International Conference on Signal and Image Processing (ICSIP)}, 
  title={{Domain Adaptive Recognition of Defects for Ground Penetrating Radar Images}}, 
  year={2023},
  volume={},
  number={},
  pages={409-413},
  doi={10.1109/ICSIP57908.2023.10271058}}

@article{Oturak_2022,
author = {Oturak, Mehmet and Yuksel, Seniha and Kucuk, Sefa},
year = {2022},
month = {12},
pages = {1-9},
title = {{Multi-source domain adaptation of GPR data for IED detection}},
volume = {17},
journal = {Signal, Image and Video Processing},
doi = {10.1007/s11760-022-02394-x}
}

@article{alipour2020big,
  title={A big data analytics strategy for scalable urban infrastructure condition assessment using semi-supervised multi-transform self-training},
  author={Alipour, Mohamad and Harris, Devin K},
  journal={Journal of Civil Structural Health Monitoring},
  volume={10},
  number={2},
  pages={313--332},
  year={2020},
  publisher={Springer}
}

@article{Fang_2025,
title = {Study on automated detection methods of shallow surface soil water content based on GPR signal level},
journal = {Remote Sensing of Environment},
volume = {331},
pages = {115003},
year = {2025},
issn = {0034-4257},
doi = {https://doi.org/10.1016/j.rse.2025.115003},
url = {https://www.sciencedirect.com/science/article/pii/S0034425725004079},
author = {Yunfeng Fang and Tianqing Hei and Zheng Tong and Tao Ma},
}

@article {Entekhabi_2010,
      author = "Dara  Entekhabi and Rolf H.  Reichle and Randal D.  Koster and Wade T.  Crow",
      title = "Performance Metrics for Soil Moisture Retrievals and Application Requirements",
      journal = "Journal of Hydrometeorology",
      year = "2010",
      publisher = "American Meteorological Society",
      address = "Boston MA, USA",
      volume = "11",
      number = "3",
      doi = "10.1175/2010JHM1223.1",
      pages=      "832 - 840",
      url = "https://journals.ametsoc.org/view/journals/hydr/11/3/2010jhm1223_1.xml"
}

\end{document}